\documentclass[preprintnumbers,showpacs,amsmath,amssymb,a4paper]{revtex4} 
\usepackage[]{epsfig} 
 
\newcommand{\bsig}{\boldsymbol{\sigma}}
\newcommand{\tw}{\ensuremath{t'}}

\newcommand{\ivec}{\ensuremath{\boldsymbol{i}}}
\newcommand{\jvec}{\ensuremath{\boldsymbol{j}}}
\newcommand{\ehat}[1]{\ensuremath{\boldsymbol{e}_{#1}}}
\newcommand{\ckt}[3][t]{\ensuremath{C^{(#2)}_{#3}(#1)}}
\newcommand{\ckl}[5][t]{\ensuremath{C^{(#2,#3)}_{#4,#5}(#1,\tw)}}
\newcommand{\rkl}[5][t]{\ensuremath{R^{(#2,#3)}_{#4,#5}(#1,\tw)}}
\newcommand{\Fkt}[3][t]{\ensuremath{F^{(#2)}_{#3}(#1)}}
\newcommand{\Rkt}[4][t]{\ensuremath{R^{(#2)}_{#3,#4}(#1)}}
\newcommand{\Ak}[2]{\ensuremath{A^{(#1)}_{#2}}}
\newcommand{\me}{\mathrm{e}}
\newcommand{\In}{\mathrm{I}}
\newcommand{\Hn}{\mathrm{H}}
\newcommand{\Phik}[3][t]{\ensuremath{\Phi^{(#2)}_{#3}(#1)}}

\newcommand{\Gk}[4]{\ensuremath{G^{(#1)}_{#2,#3}(#4)}}
\newcommand{\dd}{\mathrm{d}}
\newcommand{\Scal}{\mathcal{S}}
\newcommand{\taueq}{\tau_{\mathrm{eq}}}
\newcommand{\sign}[1][\pi]{(-1)^{#1}}

\newcommand{\eps}{\varepsilon}
\newcommand{\sgn}{\mathrm{sgn}}
\newcommand{\dt}{\Delta}
\newcommand{\even}{\mathrm{e}}
\newcommand{\odd}{\mathrm{o}}
\newcommand{\eq}{\mathrm{eq}}
\newcommand{\upi}{\mathrm{i}}

\begin{document} 
 
\title{General Solutions for Multispin Two-Time Correlation and 
Response Functions in the Glauber-Ising Chain} 
 
\author{Peter Mayer and Peter Sollich}

\affiliation{Department of Mathematics, King's College, Strand,
London, WC2R 2LS, UK}
 
\date{\today} 
 
\begin{abstract} 
The kinetic Glauber-Ising spin chain is one of the very few exactly
solvable models of non-equilibrium statistical
mechanics. Nevertheless, existing solutions do not yield tractable
expressions for two-time correlation and response functions of
observables involving products of more than one or two spins. We use a
new approach to solve explicitly the full hierarchy of differential
equations for the correlation and response functions. From this
general solution follow closed expressions for arbitrary multispin
two-time correlation and response functions, for the case where the
system is quenched from equilibrium at $T_\upi>0$ to some arbitrary $T
\geq 0$. By way of application, we give the results for two and
four-spin two-time correlation and response functions. From the
standard mapping, these also imply new exact results for two-time
particle correlation and response functions in one-dimensional
diffusion limited annihilation.
\end{abstract} 


\maketitle 

\section*{INTRODUCTION}

Glauber did some of his pioneering work in the field of nonequilibrium 
statistical mechanics when he introduced the kinetic version of 
the one-dimensional Ising model~\cite{Glauber63} about 40 years 
ago. He made a very careful choice for the model dynamics which, 
while resembling characteristic features of ferromagnets, also 
admits an analytic approach. Amongst other things Glauber actually 
derived the general solutions for one and two-spin correlation 
functions in his original paper. 

Since then the Glauber-Ising model has continuously appeared in the
literature in various contexts. For obvious reasons it has been
studied in much detail as a toy model for ferromagnets
\cite{Glauber63,BedShuOpp70,BrePra93b,BrePra96,PraBreSan97}.  A link
to one-dimensional diffusion-reaction processes has extended its
relevance further \cite{Spouge88,AmaFam90,Santos97}, and the model has
also been employed as an abstract metaphor for tapped granular media
\cite{BrePra02b}. The most recent occurrences in the literature relate
nonequilibrium fluctuation-dissipation relations and related questions
in higher-dimensional Ising models at criticality
\cite{GodLuc00,LipZan00,FDT}.

A large fraction of the publications that deal with the Glauber-Ising 
chain either make direct use of the results derived by Glauber
\cite{Glauber63}, or use facts which can be obtained directly using
his approach. This is rather
remarkable, seeing as (at least) three different versions of general solutions 
for arbitrary order one-time spin correlations have been derived in the 
mean time. The authors of \cite{BedShuOpp70} showed 
that a particular combination of correlation functions, denoted 
the $C_n$-functions, satisfy simpler dynamical equations than 
the correlations themselves. From the general solution given 
for the $C_n$-functions, arbitrary (one-time) correlations could be obtained. 
As pointed out by the authors, however, their general expressions are 
rather complicated already for the case of three-spin
correlations. 
In \cite{Felderhof71}, the rigorous general solution of the master 
equation was obtained via a Fermion mapping. There, arbitrary 
correlation functions can in principle be found by expanding the associated 
operator according to a given scheme. However, as pointed out in the 
article, for correlations of more than two spins the amount of 
algebra involved increases very rapidly, and the formalism is not
suitable for extracting two-time quantities. Finally,
a rigorous and exact solution for the generating function of 
correlators has been found more recently \cite{Aliev98} using Grassmann 
variables. The formalism is very compact and yields an explicit 
expression for the generating function. Naturally, correlation 
functions are encoded implicitly in this result and follow by 
taking appropriate derivatives, but multispin or two-time quantities
are again very difficult to obtain. The domain size distribution after 
quenching the system from a completely uncorrelated initial state 
to zero temperature is also briefly discussed in \cite{Aliev98}. The 
latter quantity was first obtained in \cite{DerZei96} using yet 
another method, namely the analogy between 
random walk problems and zero temperature dynamics of
one-dimensional Ising and Potts models. Via this route the authors of
\cite{DerZei96} also found
expressions for one-time correlation functions of arbitrary order,
valid however only for the dynamics after a quench from an
uncorrelated initial state.

In this paper we derive yet another version of the general solution
for arbitrary correlation functions in the finite Glauber-Ising model
subject to generic initial conditions. Compared to the existing
results, our solution is in the most explicit form and cannot be
simplified further for generic initial conditions. Like the authors of
Refs.~\cite{Glauber63,BedShuOpp70}, we start from the hierarchy of
differential equations for the correlation (or response) functions.
However, instead of trying to solve the hierarchy directly as a whole,
we judiciously split it into subsystems of inhomogeneous differential
equations.  These subsystems are first solved separately; arbitrary
correlation functions then follow in a recursive manner. Crucially,
this recursion can be solved explicitly, giving in the end a rather
compact solution of the full hierarchy. This result is of interest in
itself, but not directly useful for practical purposes since it
involves summations over a number of terms growing with the system
size $N$. However, in the bulk of the paper we show that if the system
is quenched from equilibrium at some initial temperature, then all
sums involving an extensive number of terms can be performed. This
gives the desired explicit expressions for arbitrary multispin
two-time correlation and response functions, involving sums of only a
finite number of terms.

In Sec.~\ref{sec:dynequ} we start with a recap of the definition of
the finite one-dimensional Glauber-Ising model, and give the
hierarchies of differential equations governing the evolution of
multispin one and two-time correlation (and response) functions.  The
general solution of these hierarchies, all of which have the same
structure, is then derived in Sec.~\ref{sec:general}. We briefly
discuss the procedure of taking the thermodynamic limit in our result,
which is in fact straightforward, and then proceed with the rigorous
analysis of the finite model. First we verify in
Sec.~\ref{sec:equilibrium} that the well-known equilibrium
correlations at $T>0$ are retrieved correctly. From these, one-time
multispin correlation functions after a quench from equilibrium at
$T_\upi > 0$ to arbitrary $T \geq 0$ follow, as shown in
Sec.~\ref{sec:correlquench}. Here we make contact with the results
given for the infinite chain given in \cite{BedShuOpp70}, in
\cite{DerZei96} (infinite chain, $T_\upi = \infty$ and $T=0$) and in
\cite{Aliev98} (finite chain, $T_\upi = \infty$). Using these one-time
correlations as initial conditions we finally obtain the two-time
multispin correlations (Sec.~\ref{sec:correl2quench}) and response
functions (Sec.~\ref{sec:responsequench}) after a quench; these are
the core new results of our study. In Sec.~\ref{sec:examples}, we
illustrate the procedure of extracting particular two-time functions
from these general results. We briefly discuss the spin-spin
correlation and response functions, already known from
\cite{GodLuc00,LipZan00}, and then turn to domain-wall correlations
and responses. There, as obvious applications, we give exact results
and scaling expansions for the time-dependent equilibrium domain-wall
autocorrelation at $T>0$ and for its two-time analogue for the
out-of-equilibrium dynamics at $T=0$. We also briefly discuss the
connection between the Glauber-Ising chain and diffusion-limited
reaction processes, indicating how e.g.\ multi-particle two-time
correlations in the latter processes may be obtained from our results.
We conclude in Sec.~\ref{sec:conclusions} with a summary and
discussion of our work. Useful representations and identities for the
functions appearing in our calculations are summarised in a number of
appendices.

Readers more interested in our results, rather than their derivation,
could omit Secs.~\ref{sec:general}--\ref{sec:responsequench} on a
first reading. They need to refer only to the end of
Secs.~\ref{sec:correl2quench} and~\ref{sec:responsequench}, where we
state explicitly the expressions for two and four-spin two-time
correlation and response functions. The applications described in
Sec.~\ref{sec:examples}, which are cross-referenced to the relevant
sections in the derivations, should then make it clear how our general
results are evaluated in practice.

\section{Dynamical Equations} 
\label{sec:dynequ}

The finite, one-dimensional Glauber-Ising model comprises a ring 
of $N$ Ising 
spins $\sigma_n \in \{-1,+1\}$ that evolve stochastically in time. We 
denote by $p(\bsig,t)$ the probability to find the system in state 
$\bsig=(\sigma_1,\ldots \sigma_N)$ at time $t$. Single spin-flip or 
heat-bath dynamics are described by the master equation
\begin{equation}
  \frac{\dd}{\dd t} \, p(\bsig,t) = \sum\limits_{n=1}^N \big[ w_n(F_n \bsig) 
  p(F_n \bsig,t) - w_n(\bsig) p(\bsig,t) \big],
  \label{equ:master}
\end{equation}
where $w_n(\bsig)$ is the rate for flipping the spin $\sigma_n$ in state 
$\bsig$, and $F_n$ denotes the spin-flip operator $F_n \bsig = (\sigma_1, 
\ldots -\sigma_n, \ldots \sigma_N)$. Glauber's choice $w_n(\bsig)=
\frac{1}{2} [1-\frac{\gamma}{2} \sigma_n (\sigma_{n-1} + \sigma_{n+1})]$ 
for the transition rates -- with $\gamma=\tanh(2J/T)$ -- is such that the 
equilibrium solution of (\ref{equ:master}) satisfies detailed balance and 
coincides with the Boltzmann distribution for the Ising Hamiltonian 
$\mathcal{H}=-J\sum_n \sigma_n \sigma_{n+1}$ at temperature $T$. 
The rates may equivalently be written as 
\begin{equation}
  w_n(\bsig)=\frac{1}{1+\me^{\Delta_n \mathcal{H}(\bsig)/T}},
  \label{equ:rates}
\end{equation}
where $\Delta_n\mathcal{H}(\bsig)=\mathcal{H}(F_n \bsig)-\mathcal{H}(\bsig)$ 
is the energy shift caused by flipping the $n^{\mathrm{th}}$ spin. We 
remark that whenever site-indices exceed the range $1,2,\ldots N$, as 
occurs for instance in the expression for the rates $w_1(\bsig)$ and 
$w_N(\bsig)$, we use $N$-periodicity of the ring, that is $\sigma_{N+1} 
= \sigma_1$ and $\sigma_0 = \sigma_N$.

\subsection{Correlation Functions}
\label{sec:correl}

We use the notation
\begin{equation}
\ckt{k}{\ivec} = \langle \sigma_{i_1}(t) \cdots \sigma_{i_k}(t) \rangle 
= \sum_{\bsig} \sigma_{i_1} \cdots \sigma_{i_k} \, p(\bsig,t)
\end{equation}
for the (disconnected) $k$-spin correlation function between 
sites $1 \leq i_1, \ldots i_k \leq N$ at time $t$. The summation 
$\sum_{\bsig}$ obviously runs over the $2^N$ states of the system. From 
the master equation (\ref{equ:master}) one readily obtains the evolution 
equation 
\begin{equation}
  \frac{\dd}{\dd t} \, \ckt{k}{\ivec} = - k \, \ckt{k}{\ivec} + 
  \frac{\gamma}{2} \sum\limits_{\eta=1}^{k} \left( 
  \ckt{k}{\ivec-\ehat{\eta}} + \ckt{k}{\ivec+\ehat{\eta}} \right),
  \label{equ:correl}
\end{equation}
{\em provided} that the index-vector $\ivec=(i_1,\ldots i_k)$ has pairwise 
distinct components. 
The $\ehat{\eta}$ appearing in (\ref{equ:correl}) are $k$-dimensional 
unit-vectors along the $\eta$-direction, that is $\ehat{1}=(1,0,\ldots 0)$, 
$\ehat{2}=(0,1,\ldots 0)$, \ldots $\ehat{k}=(0,0,\ldots 1)$. Due to 
the index-shifts $\ivec\pm\ehat{\eta}$, (\ref{equ:correl}) induces 
a set of coupled differential equations over all $k$-spin correlations. 
These equations are, however, not closed as for some $\ivec$ the 
index-shifts create a pair of equal components in $\ivec\pm\ehat{\eta}$. 
In this case, the order $k$ of the correlation function is reduced by two, 
since $\sigma_n^2=1$ for Ising spins; more precisely 
\begin{equation}
  \exists\ 1\le \mu < \nu \le k: i_\mu=i_\nu \Rightarrow 
  \ckt{k}{\ivec}=\ckt{k-2}{\ivec \setminus (i_\mu,i_\nu)}.
  \label{equ:links}
\end{equation}
Here we have introduced the short-hand $\ivec\setminus(i_\mu,i_\nu) = 
(i_1,\ldots i_{\mu-1},i_{\mu+1},\ldots i_{\nu-1},i_{\nu+1}, \ldots i_k)$ to 
indicate that the components $i_\mu,i_\nu$ are to be removed from $\ivec$. 
Since (\ref{equ:correl}) applies for any $k\geq 1$ and $\ivec$ with $1 \leq 
i_1, \ldots i_k \leq N$ pairwise distinct,
the dynamics of $(k-2)$-spin correlations are again governed by 
(\ref{equ:correl}) and in turn link to $(k-4)$-spin correlations, etc. 
For odd $k$ this hierarchy of differential equations, linked 
by the hierarchical property (\ref{equ:links}) of correlation functions, 
is closed at level $k=1$ where the formation of index-pairs becomes 
impossible. The equations on even levels, on the other hand, are closed 
when an index-pair occurs on level $k=2$, giving $\ckt{0}{}=\sum_{\bsig} 
\sigma_i^2 \, p(\bsig,t) = \sum_{\bsig} p(\bsig,t) =1$. 
Therefore the hierarchy of differential equations 
defined by (\ref{equ:correl}), (\ref{equ:links}) over all even (odd) 
levels $k'$ in the range $0 \leq k' \leq k$ constitutes a closed set 
of equations for the evolution of $k$-spin correlation functions with 
even (odd) $k$, respectively.

\subsection{Two-Time Correlations}
\label{sec:correl2}

We define multispin two-time correlation functions by
\begin{equation}
\ckl{k}{l}{\ivec}{\jvec}=\langle \sigma_{i_1}(t) \cdots 
\sigma_{i_k}(t) \sigma_{j_1}(\tw) \cdots \sigma_{j_l}(\tw)\rangle =
\sum_{\bsig,\bsig'} \sigma_{i_1} \cdots 
\sigma_{i_k} \, p(\bsig,t|\bsig',\tw) \, \sigma_{j_1}' \cdots \sigma_{j_l}' 
\, p(\bsig',\tw).
\label{equ:correlation_def}
\end{equation}
Here $p(\bsig,t|\bsig',\tw)$ denotes 
the conditional probability to find the system in state $\bsig$ at 
time $t \geq \tw$ given it was in state $\bsig'$ at $\tw$. 
This satisfies the master equation (\ref{equ:master}),
\begin{equation}
  \frac{\partial}{\partial t} \, p(\bsig,t|\bsig',\tw) = 
  \sum\limits_{n=1}^N \big[ w_n(F_n \bsig) p(F_n \bsig,t|\bsig',\tw) - 
  w_n(\bsig) p(\bsig,t|\bsig',\tw) \big],
  \label{equ:master2}
\end{equation}
with the initial condition
$p(\bsig,\tw|\bsig',\tw)=\delta_{\bsig,\bsig'}$; here
$\delta_{\bsig,\bsig'}=\prod_{n=1}^N \delta_{\sigma_n,\sigma_n'}$
is the Kronecker delta for states, which equals 1 for $\bsig=\bsig'$
and vanishes otherwise.
We deduce from (\ref{equ:master2}) the evolution 
equations for two-time correlations 
\begin{equation}
  \frac{\partial}{\partial t} \ckl{k}{l}{\ivec}{\jvec} = - k \, 
  \ckl{k}{l}{\ivec}{\jvec} + 
  \frac{\gamma}{2} \sum\limits_{\eta=1}^{k} \left( 
  \ckl{k}{l}{\ivec-\ehat{\eta}}{\jvec} + \ckl{k}{l}{\ivec+\ehat{\eta}}
  {\jvec} \right). 
  \label{equ:correl2}
\end{equation}
which again hold only for index vectors $\ivec$ with pairwise distinct 
components. It is clear from the definition that two-time correlations also 
exhibit hierarchical properties when index-pairs occur in $\ivec$ 
(or $\jvec$). Therefore (\ref{equ:correl2}) again induces a hierarchy 
of differential equations, linking $(k,l)$-spin to 
$(k-2,l)$-spin correlations etc. The hierarchies (\ref{equ:correl2}) 
and (\ref{equ:correl}) have the same structure and differ only in the
relevant initial values. For the two-time correlations these are
\begin{equation}
  \ckl[\tw]{k}{l}{\ivec}{\jvec} = \ckt[\tw]{k+l}{\ivec\,\cup\,\jvec}.
  \label{equ:correl2init}
\end{equation}
The abbreviation $\ivec \cup \jvec$ symbolises that the components of 
$\ivec$ and $\jvec$ are combined into a new vector $(i_1,\ldots i_k,
j_1,\ldots j_l)$. The initial values (\ref{equ:correl2init}) have to 
hold on each level $k$ of the hierarchy for all $\ivec$ with pairwise 
distinct components. To avoid trivial reductions in the order $l$ we 
restrict ourselves to vectors $\jvec$ with pairwise distinct components.

\subsection{Two-Time Response Functions}
\label{sec:response}

In order to define response functions, we consider a perturbation
$\delta\mathcal{H} =-h_B \, B(\bsig)$ to the Hamiltonian, with a
(generally time-dependent) field $h_B$ conjugate to the observable
$B(\bsig)$. The two-time response function $R_{A,B}(t,\tw)$ measures
the response of the observable 
$\langle A(t) \rangle = \sum_{\bsig} A(\bsig) \, p(\bsig,t)$
to a field impulse, where $h_B=h$ during a short time interval $[\tw,\tw+
\delta\tw]$ and $h_B=0$ at all other times.
Transition rates in the presence of a perturbation are defined via 
(\ref{equ:rates}). Regardless of whether the system has been perturbed 
or not we may write
\begin{equation}
  \langle A(t) \rangle = \sum\limits_{\bsig,\bsig'} A(\bsig) \, p(\bsig, 
  t|\bsig',\tw+\delta\tw) \, p(\bsig',\tw+\delta\tw).
  \label{equ:response1}
\end{equation}
From (\ref{equ:master2}) the conditional probability 
$p(\bsig,t|\bsig',\tw+\delta\tw)$ is unaffected by the perturbation 
since the field vanishes for $t > \tw+\delta\tw$.
To find the effect of the field on $p(\bsig',\tw+\delta\tw)$, one
expands to linear order in $\delta\tw$ to get
\begin{equation}
  p(\bsig',t'+\delta\tw) = p(\bsig',\tw) + 
\delta\tw \sum\limits_{n=1}^{N} \left[ 
  w_n(F_n \bsig') \, p(F_n \bsig',\tw) - 
  w_n(\bsig') \, p(\bsig',\tw) \right]. 
  \label{equ:response5}
\end{equation}
Inserting this into (\ref{equ:response1}) and taking the difference
between the perturbed and unperturbed cases gives the change $\delta
\langle A(t) \rangle$. Writing the expansion of the transition rates 
to linear order
in the field as $w_n(\bsig)=w_n(\bsig)|_{h_B=0}+ h \, 
w'_n(\bsig)$ then gives for the two-time response function
$R_{A,B}(t,\tw) = T \delta\langle A(t)\rangle /(h\delta\tw)$
\begin{equation}
  R_{A,B}(t,\tw) = T \sum\limits_{\bsig,\bsig'} A(\bsig) p(\bsig,t | 
  \bsig',\tw) \sum\limits_{n=1}^{N} \left[ 
  w'_n(F_n \bsig') \, p(F_n \bsig',\tw) - 
  w'_n(\bsig') \, p(\bsig',\tw) \right].
  \label{equ:response6}
\end{equation}
Here an extra factor of $T$ has been introduced into the definition of
the response function, for later convenience.
Since the derivation of equation (\ref{equ:response6}) uses only the master 
equation (\ref{equ:master}), it holds for any 
choice of $A(\bsig)$, $B(\bsig)$ and $w_n(\bsig)$. 
Comparing with (\ref{equ:correlation_def}) one sees that the
$t$-dependence arises in exactly the same way as for a correlation
function between $A(t)$ and some observable at time $\tw$. 
For multispin 
response functions, where $\langle A(t) \rangle = \langle \sigma_{i_1}(t) 
\cdots \sigma_{i_k}(t) \rangle$ and $\delta\mathcal{H}=-h_{\jvec}(\tw) \, 
\sigma_{j_1} \cdots \sigma_{j_l}$, one thus immediately has the
dynamical equations
\begin{equation}
  \frac{\partial}{\partial t} \, \rkl{k}{l}{\ivec}{\jvec} = - k \, 
  \rkl{k}{l}{\ivec}{\jvec} + 
  \frac{\gamma}{2} \sum\limits_{\eta=1}^{k} \left( 
  \rkl{k}{l}{\ivec-\ehat{\eta}}{\jvec} + \rkl{k}{l}{\ivec+\ehat{\eta}}
  {\jvec} \right).
  \label{equ:response}
\end{equation}
These are to be interpreted in exactly the same manner as 
(\ref{equ:correl}) and (\ref{equ:correl2}). To avoid a trivial 
reduction of the order $l$ of the perturbation we require, in analogy 
to two-time correlations, pairwise distinct components in $\jvec$. 
The initial conditions for (\ref{equ:response}) at $t=\tw$ also follow 
from (\ref{equ:response6}): working out $w'_n(\bsig)$ from 
(\ref{equ:rates}) gives
\begin{equation}
  w'_n(\bsig)=-\frac{1}{2T} \, \left[1-\frac{\gamma^2}
  {2} \, (1+\sigma_{n-1}\sigma_{n+1}) \right] \, \sigma_{j_1} \cdots 
  \sigma_{j_l} \, \sum\limits_{\nu=1}^l \delta_{n,j_\nu}. 
  \label{equ:response7}
\end{equation}
Substituting into (\ref{equ:response6}),
using $p(\bsig,\tw | \bsig',\tw)=\delta_{\bsig,\bsig'}$ and rearranging 
terms we then find
\begin{equation}
  \rkl[\tw]{k}{l}{\ivec}{\jvec}=\sum\limits_{\mu=1}^{k} \sum\limits_{\nu=1}
  ^{l} \delta_{i_\mu,j_\nu} \left[ \left( 1-\frac{\gamma^2}{2} \right) 
  \ckt[\tw]{k+l}{\ivec \, \cup \, \jvec} 
  - \frac{\gamma^2}{2} \ckt[\tw]{k+l+2}{\ivec \, \cup \, 
  \jvec^\nu} \right].
  \label{equ:responseinit}
\end{equation}
In (\ref{equ:responseinit}), $\jvec^\nu=(j_1,\ldots 
j_{\nu-1},j_\nu -1,j_\nu,j_\nu +1, j_{\nu+1},\ldots j_l)$ is the 
vector $\jvec$ with the additional components $j_\nu-1$ and $j_\nu+1$.
The structure of (\ref{equ:responseinit}) makes 
sense intuitively: an instantaneous response only occurs if the 
`observed' sites $i_i,\ldots i_k$ overlap with the `perturbed' sites 
$j_1,\ldots j_l$. At level $k=0$, the sum over $\mu$ in
(\ref{equ:responseinit}) is empty; as is done conventionally, we define such
empty sums as zero throughout. This makes sense: for $k=0$, the
observable is just a constant, $A(\bsig)=1$, and its value is
unaffected by any perturbation, giving a vanishing response. Equations
(\ref{equ:correl}) and (\ref{equ:correl2}) are to be read in the same
way for $k=0$.

\section{General Solution}
\label{sec:general}

We have seen in Sec.~\ref{sec:dynequ} that the dynamical equations 
describing the evolution of multispin one- and two-time correlation 
and two-time response functions all have the same structure. Hence we
obtain the  
various quantities from the same set of differential equations, 
e.g.\ (\ref{equ:correl}), by using the appropriate initial conditions. 
For this reason and in order to unify the analysis we use in this
section the symbol 
$F^{(k)}$ for the dynamical quantity and $A^{(k)}$ for its initial 
conditions. For the sake of readability we omit indices $\ivec$ in 
the text except when they are needed for clarity.

All dynamical functions $F^{(k)}$ discussed above are by definition 
$N$-periodic in the site indices $i_1, \ldots i_k$ and invariant under 
permutations of the latter. But it turns out that these symmetries are 
not directly helpful for solving the hierarchy (\ref{equ:correl}). We
therefore 
restrict the problem to functions $F^{(k)}$ with ordered  indices  
$\ivec \in N(k)$ where
\begin{equation}
  N(k)=\{ \ivec \in \mathbb{N}^k \, | \, 1 \leq i_1 < i_2 < \ldots 
  i_k \leq N \}.
  \label{equ:Nk}
\end{equation}
The set $N(k)$ is not closed under the index shifts $\ivec 
\mapsto \ivec-\ehat{\eta}$ and $\ivec \mapsto \ivec+\ehat{\eta}$ 
occurring on the r.h.s.\ of (\ref{equ:correl}), as shifted components 
might exceed the range $1,\ldots N$ or form pairs. Using $N$-periodicity, 
permutational symmetry and the hierarchical property, however, each 
function $F^{(k)}$ with a non-ordered index $\ivec \notin N(k)$ can be 
expressed in terms of another one, $F^{(k')}$, having an ordered index 
$\ivec' \in N(k')$. Hence we may formally rewrite (\ref{equ:correl}) on 
all levels $2 \leq k \leq N$ as
\begin{equation}
  \forall \ivec \in N(k): \, 
  \frac{\dd}{\dd t} \, \Fkt{k}{\ivec} \,\, = \!\! 
  \sum\limits_{\jvec \in N(k)} a^{(k)}_{\ivec,\jvec} \, 
  \Fkt{k}{\jvec} \,\,\,\, + \!\!\!\! \sum\limits_{\jvec' 
  \in N(k-2)} b^{(k)}_{\ivec,\jvec'} \, \Fkt{k-2}{\jvec'}.
  \label{equ:matrixform}
\end{equation}
Note that (\ref{equ:matrixform}) explicitly separates the terms on 
the r.h.s.\ of (\ref{equ:correl}) that link within level $k$ of the 
hierarchy from those connecting to level $k-2$ via the hierarchical 
property. Clearly the first term on the r.h.s.\ of (\ref{equ:correl}), 
linking within level $k$ for all $\ivec \in N(k)$, produces diagonal 
entries in $a^{(k)}$. Also, each function with an index 
$\ivec-\ehat{\eta} \in N(k)$ and $\ivec+\ehat{\eta} \in N(k)$ on the 
r.h.s.\ of (\ref{equ:correl}) yields corresponding off-diagonal entries 
in $a^{(k)}$. But when a shifted index is not contained in 
$N(k)$ several cases have to be distinguished. Here we only discuss 
the shifts $\ivec-\ehat{\eta}$ as a similar reasoning 
applies to $\ivec+\ehat{\eta}$. If $\ivec-\ehat{\eta} 
\notin N(k)$ and $2 \leq \eta \leq k$ the shift has created an index 
pair $i_{\eta-1}=i_\eta-1$. Hence the hierarchical property applies 
and the corresponding term on the r.h.s.\ of (\ref{equ:correl}) links 
to level $k-2$. Such terms are accounted for in $b^{(k)}$. On the
other hand, if
$\ivec-\ehat{\eta} \notin N(k)$ and $\eta=1$ then $i_1=1$.
$N$-periodicity and permutational symmetry of the underlying dynamical 
function $F^{(k)}$ allow us to replace the index $\ivec-\ehat{1}=(0,
i_2,\ldots i_k)$ by $(N,i_2,\ldots i_k)$ and then by $(i_2,\ldots i_k,N)$, 
respectively. If $(i_2,\ldots i_k,N) \in N(k)$ there is a link within 
level $k$ on the r.h.s.\ of (\ref{equ:correl}) and we add a corresponding 
entry to $a^{(k)}$ in the appropriate place $\ivec=(1,i_2,\ldots i_k)$, 
$\jvec=(i_2,\ldots i_k,N)$. Otherwise $i_k=N$ is an index pair and 
this term produces an entry in $b^{(k)}$. An explicit representation of 
$a^{(k)}$ as obtained by this construction is not necessary 
for our analysis; $b^{(k)}$ is given in (\ref{equ:bmatrix}) for later 
reference.

Rewriting (\ref{equ:correl}) in the form (\ref{equ:matrixform})
obviously does not change the basic problem of having to solve 
a hierarchy of differential equations. But (\ref{equ:matrixform}) 
suggests a somewhat unconventional approach for finding $F^{(k)}$: instead of 
simultaneously solving the full hierarchy induced by 
(\ref{equ:matrixform}) we can focus on a given level $k \geq 2$ 
and consider (\ref{equ:matrixform}) as an {\em inhomogeneous} system 
of differential equations in $F^{(k)}$. Hence we may write 
\begin{equation}
  \Fkt{k}{\ivec}=\Phik{k}{\ivec} \, +\int\limits_{0}^{t} \dd \tau 
  \sum\limits_{\jvec \in N(k)} \sum\limits_{\jvec' \in N(k-2)} 
  \Gk{k}{\ivec}{\jvec}{t-\tau} \, b^{(k)}_{\jvec,\jvec'} \, 
  \Fkt[\tau]{k-2}{\jvec'}, 
  \label{equ:Fsplit}
\end{equation}
where $\Phik{k}{\ivec}$ is the solution of the homogeneous equation
\begin{equation}
  \forall \ivec \in N(k): \, \frac{\dd}{\dd t} \, \Phik{k}{\ivec} \,\, = \!\!
  \sum\limits_{\jvec \in N(k)} a^{(k)}_{\ivec,\jvec} \, \Phik{k}{\jvec}, 
  \label{equ:homogeneous}
\end{equation}
satisfying the initial conditions $\forall \ivec \in N(k): 
\Phik[0]{k}{\ivec}=\Ak{k}{\ivec}$. The second term in (\ref{equ:Fsplit}), 
on the other hand, is a particular solution of (\ref{equ:matrixform}) 
obtained via the Green's function $G^{(k)}$ of (\ref{equ:homogeneous}). 
Thus we may first focus on solving the homogeneous equation 
(\ref{equ:homogeneous}) on the generic level $k$, obtaining
the Green's function on the way. Then (\ref{equ:Fsplit}) constitutes 
a recursion formula that allows us to successively express any solution 
on level $k$ in terms of solutions on lower levels $k-2$, $k-4$, etc. 
The recursion terminates either on level $0$ or $1$ of the hierarchy 
where the solutions are known: according to the discussion in 
Sec.~\ref{sec:dynequ}, for $k=0$ we actually have that $F^{(0)}(t)=A^{(0)}$ is 
constant in time. On level $k=1$, on the other hand, (\ref{equ:correl}) 
is homogeneous since no index pairs can be formed in a one-dimensional 
index vector. So $F^{(1)}(t)=\Phi^{(1)}(t)$ is just the 
solution of (\ref{equ:homogeneous}) on level $k=1$. 

Note that the applicability of this approach essentially relies on the 
fact that links between different levels of the hierarchy are unidirectional, 
i.e.\ only from $k$ to $k-2$. This is a peculiarity of the one-dimensional 
Glauber-Ising model.

\subsection{Homogeneous Solution}
\label{sec:homogeneous}

Solving the homogeneous equation (\ref{equ:homogeneous}) over ordered 
indices $\ivec \in N(k)$ is a non-trivial problem. In order to make 
progress we apply the method of images, in this context already used 
by Glauber \cite{Glauber63}, to extend (\ref{equ:homogeneous}) over 
index vectors $\ivec \in \{1, \ldots N \}^k$. This procedure is purely 
for mathematical convenience and only possible because we have 
reformulated (\ref{equ:correl}) over ordered $\ivec \in N(k)$ in the 
first place. As a generalisation of (\ref{equ:Nk}) we introduce the 
set of index vectors $N_\pi(k)$ that are ordered when permuted by $\pi$ 
\begin{equation}
  N_\pi(k)=\{ \ivec \in \mathbb{N}^k \, | \, 1 \leq i_{\pi(1)} < 
  i_{\pi(2)} < \ldots i_{\pi(k)} \leq N \}. 
  \label{equ:Nkpi} 
\end{equation}
Here, $\pi \in \Scal(k)$ denotes a permutation on $\{1,2,\ldots k\}$ and 
$\Scal(k)$ the set of all such permutations. By $(-1)^\pi$ we refer to 
the sign of a permutation, that is $(-1)^p$ where $p$ is the number of 
transpositions necessary to perform $\pi$. Now we define the 
homogeneous solutions over $\{1,2,\dots N\}^k$ by filling the 
hypercube with permutationally {\em antisymmetric} images according to
\begin{equation}
  \forall \ivec \in \{1,\dots N \}^k: \, 
  \bar{\Phi}^{(k)}_{\ivec}(t) = \left\{ 
    \begin{array}{cc}
      \sign \, \Phik{k}{(i_{\pi(1)},\ldots i_{\pi(k)})} & 
        \mbox{if}\ \exists \pi \in \Scal(k) : \ivec \in N_\pi(k) \\
      0 & \mbox{otherwise}
    \end{array}
  \right. .
  \label{equ:phibar}
\end{equation}
This equation states that for ordered indices $\ivec \in N(k)$ we have 
$\bar{\Phi}^{(k)}=\Phi^{(k)}$. Furthermore, when $\ivec$ has pairwise 
distinct components any transposition in $\ivec$ just changes the sign 
of $\bar{\Phi}^{(k)}$. Therefore, and since $\bar{\Phi}^{(k)}=0$ anyway for 
non-pairwise distinct indices, permuting the index $\ivec$ by 
$\pi_0 \in S(k)$ changes the sign of $\bar{\Phi}^{(k)}$ by 
$(-1)^{\pi_0}$ for any $\ivec \in \{1, \ldots N \}^k$. Setting $t=0$ 
in (\ref{equ:phibar}) defines the initial conditions $\bar{A}^{(k)}$ 
over the hypercube. As we explain below $\bar{\Phi}^{(k)}$ satisfies 
\begin{equation}
  \forall \ivec \in \{1,\dots N \}^k: \, 
  \frac{\dd}{\dd t} \, \bar{\Phi}^{(k)}_{\ivec}(t) = 
  -k \, \bar{\Phi}^{(k)}_{\ivec}(t) 
  + \frac{\gamma}{2} \sum\limits_{\eta=1}^{k} \left( 
  \bar{\Phi}^{(k)}_{\ivec-\ehat{\eta}}(t) + 
  \bar{\Phi}^{(k)}_{\ivec+\ehat{\eta}}(t) \right),  
  \label{equ:simplehomogeneous}
\end{equation}
if the functions $\bar{\Phi}^{(k)}$ are moreover extended 
$N$-(anti)periodically in each component of $\ivec$ for (even) odd 
$k$. We use the term $N$-antiperiodic in the sense that shifting 
any component of $\ivec$ by $\pm N$ changes the sign of $\bar{\Phi}^{(k)}$. 
Note that in contrast to (\ref{equ:correl}), which only holds for 
index vectors $\ivec$ with pairwise distinct components, 
(\ref{equ:simplehomogeneous}) applies for all $\ivec \in \{1, \ldots 
N \}^k$.

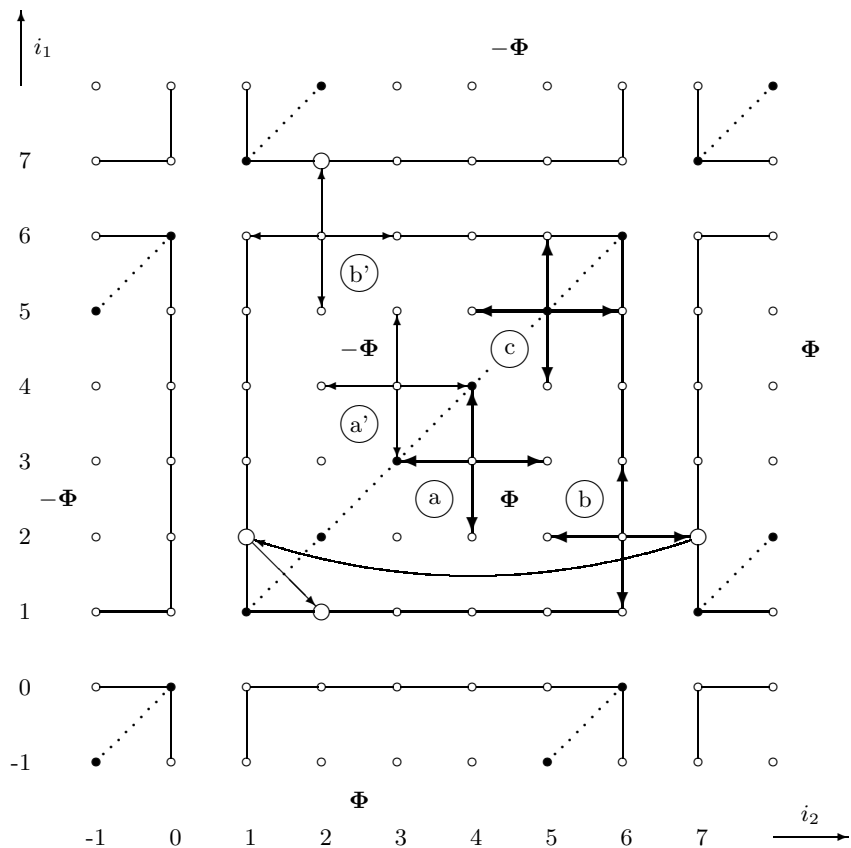
\begin{figure}[htb]
  \begin{picture}(11,11)
    \multiput(1, 1)(1,0){10}{\circle{0.1}}
    \multiput(1, 2)(1,0){10}{\circle{0.1}}
    \multiput(1, 3)(1,0){ 3}{\circle{0.1}}
    \put(4,3){\circle{0.2}}
    \multiput(5, 3)(1,0){ 6}{\circle{0.1}}
    \multiput(1, 4)(1,0){ 2}{\circle{0.1}}
    \put(3,4){\circle{0.2}}
    \multiput(4, 4)(1,0){ 5}{\circle{0.1}}
    \put(9,4){\circle{0.2}}
    \put(10,4){\circle{0.1}}
    \multiput(1, 5)(1,0){10}{\circle{0.1}}
    \multiput(1, 6)(1,0){10}{\circle{0.1}}
    \multiput(1, 7)(1,0){10}{\circle{0.1}}
    \multiput(1, 8)(1,0){10}{\circle{0.1}}
    \multiput(1, 9)(1,0){ 3}{\circle{0.1}}
    \put(4,9){\circle{0.2}}
    \multiput(5, 9)(1,0){ 6}{\circle{0.1}}
    \multiput(1,10)(1,0){10}{\circle{0.1}}
    \multiput(1, 1)(1,1){10}{\circle*{0.1}}
    \multiput(1, 7)(1,1){ 4}{\circle*{0.1}}
    \multiput(7, 1)(1,1){ 4}{\circle*{0.1}}
    \put(0.5,-0.5){\makebox(1,1){-1}}
    \put(1.5,-0.5){\makebox(1,1){ 0}}
    \put(2.5,-0.5){\makebox(1,1){ 1}}
    \put(3.5,-0.5){\makebox(1,1){ 2}}
    \put(4.5,-0.5){\makebox(1,1){ 3}}
    \put(5.5,-0.5){\makebox(1,1){ 4}}
    \put(6.5,-0.5){\makebox(1,1){ 5}}
    \put(7.5,-0.5){\makebox(1,1){ 6}}
    \put(8.5,-0.5){\makebox(1,1){ 7}}
    \put(10,0){\vector(1,0){1}}
    \put(10,0){\makebox(1,0.6){$i_2$}}
    \put(-0.5, 0.5){\makebox(1,1){-1}}
    \put(-0.5, 1.5){\makebox(1,1){ 0}}
    \put(-0.5, 2.5){\makebox(1,1){ 1}}
    \put(-0.5, 3.5){\makebox(1,1){ 2}}
    \put(-0.5, 4.5){\makebox(1,1){ 3}}
    \put(-0.5, 5.5){\makebox(1,1){ 4}}
    \put(-0.5, 6.5){\makebox(1,1){ 5}}
    \put(-0.5, 7.5){\makebox(1,1){ 6}}
    \put(-0.5, 8.5){\makebox(1,1){ 7}}
    \put(0,10){\vector(0,1){1}}
    \put(0,10){\makebox(0.6,1){$i_1$}}
    \thicklines
    \put(3.05,3){\line(1,0){0.85}}
    \put(4.10,3){\line(1,0){0.85}}
    \multiput(5.05,3)(1,0){3}{\line(1,0){0.9}}
    \multiput(8,5.05)(0,1){3}{\line(0,1){0.9}}
    \thinlines
    \multiput(3.05,8)(1,0){5}{\line(1,0){0.9}}
    \put(3,3.05){\line(0,1){0.85}}
    \put(3,4.10){\line(0,1){0.85}}
    \multiput(3,5.05)(0,1){3}{\line(0,1){0.9}}
    \multiput(3.05,2)(1,0){5}{\line(1,0){0.9}}
    \put(3.05,9){\line(1,0){0.85}}
    \put(4.10,9){\line(1,0){0.85}}
    \multiput(5.05,9)(1,0){3}{\line(1,0){0.9}}
    \multiput(2,3.05)(0,1){5}{\line(0,1){0.9}}
    \put(9,3.05){\line(0,1){0.85}}
    \put(9,4.10){\line(0,1){0.85}}
    \multiput(9,5.05)(0,1){3}{\line(0,1){0.9}}
    \put(1.05,9){\line(1,0){0.9}}
    \put(9.05,9){\line(1,0){0.9}}
    \put(1.05,8){\line(1,0){0.9}}
    \put(9.05,8){\line(1,0){0.9}}
    \put(1.05,3){\line(1,0){0.9}}
    \put(9.05,3){\line(1,0){0.9}}
    \put(1.05,2){\line(1,0){0.9}}
    \put(9.05,2){\line(1,0){0.9}}
    \put(2,1.05){\line(0,1){0.9}}
    \put(2,9.05){\line(0,1){0.9}}
    \put(3,1.05){\line(0,1){0.9}}
    \put(3,9.05){\line(0,1){0.9}}
    \put(8,1.05){\line(0,1){0.9}}
    \put(8,9.05){\line(0,1){0.9}}
    \put(9,1.05){\line(0,1){0.9}}
    \put(9,9.05){\line(0,1){0.9}}
    \thicklines
    \multiput(3,3)(1,1){3}{\multiput(0,0)(0.1,0.1){10}{\circle*{0.05}}}
    \multiput(6,6)(0.1,0.1){ 3}{\circle*{0.05}}
    \multiput(6.8,6.8)(0.1,0.1){ 3}{\circle*{0.05}}
    \multiput(7,7)(0.1,0.1){10}{\circle*{0.05}}
    \thinlines
    \multiput(9,9)(0.1,0.1){10}{\circle*{0.05}}
    \multiput(1,7)(0.1,0.1){10}{\circle*{0.05}}
    \multiput(1,1)(0.1,0.1){10}{\circle*{0.05}}
    \multiput(7,1)(0.1,0.1){10}{\circle*{0.05}}
    \multiput(9,3)(0.1,0.1){10}{\circle*{0.05}}
    \multiput(3,9)(0.1,0.1){10}{\circle*{0.05}}
    \put( 6, 4){\makebox(1,1){$\boldsymbol{\Phi}$}}
    \put( 4, 0){\makebox(1,1){$\boldsymbol{\Phi}$}}
    \put(10, 6){\makebox(1,1){$\boldsymbol{\Phi}$}}
    \put( 4, 6){\makebox(1,1){$-\boldsymbol{\Phi}$}}
    \put( 0, 4){\makebox(1,1){$-\boldsymbol{\Phi}$}}
    \put( 6,10){\makebox(1,1){$-\boldsymbol{\Phi}$}}
    \thicklines
    \put(8.05,4){\vector( 1, 0){0.85}}
    \put(7.95,4){\vector(-1, 0){0.9}}
    \put(8,4.05){\vector( 0, 1){0.9}}
    \put(8,3.95){\vector( 0,-1){0.9}}
    \put(7.05,7){\vector( 1, 0){0.9}}
    \put(6.95,7){\vector(-1, 0){0.9}}
    \put(7,7.05){\vector( 0, 1){0.9}}
    \put(7,6.95){\vector( 0,-1){0.9}}
    \put(6.05,5){\vector( 1, 0){0.9}}
    \put(5.95,5){\vector(-1, 0){0.9}}
    \put(6,5.05){\vector( 0, 1){0.9}}
    \put(6,4.95){\vector( 0,-1){0.9}}
    \thinlines
    \put(6.5,6.5){\circle{0.5}}
    \put(6.25,6.25){\makebox(0.5,0.5){c}}
    \put(5.5,4.5){\circle{0.5}}
    \put(5.25,4.25){\makebox(0.5,0.5){a}}
    \put(7.5,4.5){\circle{0.5}}
    \put(7.25,4.25){\makebox(0.5,0.5){b}}
    \put(4.05,8){\vector( 1, 0){0.9}}
    \put(3.95,8){\vector(-1, 0){0.9}}
    \put(4,8.05){\vector( 0, 1){0.85}}
    \put(4,7.95){\vector( 0,-1){0.9}}
    \put(5.05,6){\vector( 1, 0){0.9}}
    \put(4.95,6){\vector(-1, 0){0.9}}
    \put(5,6.05){\vector( 0, 1){0.9}}
    \put(5,5.95){\vector( 0,-1){0.9}}
    \put(4.5,5.5){\circle{0.5}}
    \put(4.25,5.25){\makebox(0.5,0.5){a'}}
    \put(4.5,7.5){\circle{0.5}}
    \put(4.25,7.25){\makebox(0.5,0.5){b'}}
    \qbezier(3.10,3.95)(6,3)(8.90,3.95)
    \put(3.40,3.85){\vector(-3,1){0.3}}
    \put(3.08,3.92){\vector(1,-1){0.84}}
  \end{picture}
  \caption{\label{fig:images} Illustration of the domain extension for 
    homogeneous solutions in a system of size $N=6$ and at level $k=2$ 
    of the hierarchy. The square in the center of the figure represents 
    the region $\{ 1,\ldots 6 \}^2$ consisting of $N(2)$ (lower triangle) 
    and its image (upper triangle). Associated with each circle in the 
    lower triangle is a homogeneous solution $\bar{\Phi}=\Phi$. The 
    upper triangle contains the permutationally antisymmetric image 
    $\bar{\Phi}=-\Phi$. 
    Along the dotted diagonal the index vector $\ivec=(i_1,i_2)$ is 
    not pairwise distinct and hence $\bar{\Phi}=0$ (full circles). The 
    surrounding sections of squares represent the $N$-antiperiodic 
    extension of $\bar{\Phi}$ over $\mathbb{Z}^2$. Finally, the 
    cross of arrows 
    over a circle in position $\ivec$ points to the shifted index vectors 
    $\ivec \pm \ehat{1}$ and $\ivec \pm \ehat{2}$ in correspondence 
    with the terms on the r.h.s.\ of (\ref{equ:simplehomogeneous}). 
    Based on this simple example we consider the following 
    special cases discussed in the text: (a) For $\ivec=(3,4) \in N(2)$
    the shifted index vectors $\ivec+\ehat{1}=(4,4)$ and 
    $\ivec-\ehat{2}=(3,3)$ are 
    not in $N(2)$ and must therefore contain an index pair 
    (dots on the diagonal). The associated $\bar{\Phi}$'s are zero 
    and do not contribute in (\ref{equ:simplehomogeneous}). 
    (b) For $\ivec=(2,6) \in N(2)$ we have 
    $\ivec + \ehat{2} = (2,7) \not\in N(2)$ (big circle). 
    Using $N$-antiperiodicity we replace 
    $\bar{\Phi}_{2,7}$ by $-\bar{\Phi}_{2,1}$ which in turn satisfies 
    $\bar{\Phi}_{2,1}=-\Phi_{1,2}$. Hence, this arrow effectively 
    points to $(1,2) \in N(2)$ and produces the correct term on the 
    r.h.s.\ of (\ref{equ:simplehomogeneous}). (a',b') If 
    (\ref{equ:simplehomogeneous}) satisfies the cases (a) and 
    (b) in $N(2)$ it is immediately clear that it also holds for (a') 
    and (b') in the image of $N(2)$. (c) For $\ivec=(5,5)$ we have 
    e.g.\ $\ivec-\ehat{1}=(4,5)$ and $\ivec-\ehat{2}=(5,4)$. Since 
    $\bar{\Phi}_{5,4}=-\Phi_{4,5}$ these terms cancel in 
    (\ref{equ:simplehomogeneous}). Based on this illustration it is 
    straightforward to prove cases like e.g.\ $\ivec=(6,6)$ that are 
    not discussed in detail in the text.}
\end{figure}
As a first step in proving that $\bar{\Phi}^{(k)}$ satisfies 
(\ref{equ:simplehomogeneous}) we consider ordered indices 
$\ivec \in N(k)$. According to (\ref{equ:phibar}) we have 
$\bar{\Phi}^{(k)}=\Phi^{(k)}$ for such $\ivec$. Hence 
(\ref{equ:simplehomogeneous}) should reproduce the homogeneous 
equation (\ref{equ:homogeneous}). Equivalently, by expressing 
$\bar{\Phi}^{(k)}$ in terms of $\Phi^{(k)}$ on the r.h.s.\ of 
(\ref{equ:simplehomogeneous}) we must recover the matrix $a^{(k)}$. 
In the first term on the r.h.s.\ of (\ref{equ:simplehomogeneous}),
and in all terms with shifted indices $\ivec-\ehat{\eta} \in N(k)$ and 
$\ivec+\ehat{\eta} \in N(k)$, we have $\bar{\Phi}^{(k)}=\Phi^{(k)}$. 
So these terms directly reproduce the matrix entries discussed in 
the text below (\ref{equ:matrixform}). Now we focus on the case 
where $\ivec-\ehat{\eta} \notin N(k)$; as before, 
$\ivec+\ehat{\eta} \notin N(k)$ is covered by a similar reasoning. 
If $\ivec-\ehat{\eta} \notin N(k)$ and $2 \leq \eta \leq k$ the 
shift has created an index pair $i_{\eta-1}=i_\eta-1$. According 
to (\ref{equ:phibar}) the corresponding $\bar{\Phi}^{(k)}$ is 
zero as required (case (a) in Fig.~\ref{fig:images}). Alternatively, 
when $\ivec-\ehat{\eta} \notin N(k)$ 
and $\eta=1$ we must have $i_1=1$ and therefore $\ivec-\ehat{1}=(0,
i_2,\ldots i_k) \notin \{1, \ldots N \}^k$. Using $N$-(anti)periodicity 
we replace this index by $(N,i_2,\ldots i_k) \in \{1, \ldots N \}^k$ 
and modify the sign of $\bar{\Phi}^{(k)}$ by $(-1)^{k-1}$. Next, 
applying a cyclic permutation $\pi_\mathrm{c}$ turns this index into 
$(i_2,\ldots i_k,N)$ and changes the sign of $\bar{\Phi}^{(k)}$ according 
to (\ref{equ:phibar}) once more, by $(-1)^{\pi_\mathrm{c}}$. But 
$(-1)^{\pi_\mathrm{c}}=(-1)^{k-1}$ in fact cancels the sign-change 
caused by $N$-(anti)periodicity. So, all in all, if $(i_2,\ldots i_k,N)
\in N(k)$ we recover the correct matrix element of $a^{(k)}$ between 
$\ivec=(1,i_2,\ldots i_k)$, $\jvec=(i_2,\ldots i_k,N)$ (case (b) in 
Fig.~\ref{fig:images}). Otherwise $i_k=N$ and 
hence $\bar{\Phi}^{(k)}=0$ as required.

Next consider indices $\ivec \in N_\pi(k)$ for some given $\pi \in S(k)$ 
that are ordered when permuted by $\pi$. 
Here we use the fact that $\bar{\Phi}^{(k)}$ is antisymmetric under 
permutations for all $\ivec \in \{ 1,\ldots N \}^k$. So, regardless of 
whether a shifted index on the r.h.s.\ of (\ref{equ:simplehomogeneous}) 
is also contained in $N_\pi(k)$ or not, we permute it by $\pi$ and change 
the sign of $\bar{\Phi}^{(k)}$ by $(-1)^\pi$. In cases where a shifted 
index lies outside $\{1,\ldots N\}^k$ we first use $N$-(anti)periodicity, 
then permute by $\pi$ and finally apply $\pi_\mathrm{c}$. As discussed 
above, this again just gives a sign change $(-1)^\pi$. Due to linearity of 
(\ref{equ:simplehomogeneous}) the overall factor of $(-1)^\pi$ drops out.
Therefore $\pi$ maps each sector $\ivec \in N_\pi(k)$ of 
(\ref{equ:simplehomogeneous}) onto $\ivec \in N(k)$. Since 
(\ref{equ:simplehomogeneous}) holds on $\ivec \in N(k)$ this implies 
that (\ref{equ:simplehomogeneous}) is also satisfied over $\ivec \in 
N_\pi(k)$ for any $\pi \in S(k)$ (cases (a',b') in Fig.~\ref{fig:images}).

We complete the proof that $\bar{\Phi}^{(k)}$ satisfies 
(\ref{equ:simplehomogeneous}) over $\ivec \in \{1,\ldots N\}^k$ by 
showing that it holds for non-pairwise distinct indices, too. In 
this case there must be at least one index pair, say $i_\mu=i_\nu$, 
in $\ivec$. Thus the l.h.s.\ of (\ref{equ:simplehomogeneous}), the 
first term on the r.h.s.\ and each term with a shifted index 
$\eta \neq \mu, \nu$ contain this pair and are identically zero 
according to (\ref{equ:phibar}). We also know that swapping the 
$\mu^\mathrm{th}$ and $\nu^\mathrm{th}$ component of $\ivec-\ehat{\mu}$ 
or $\ivec+\ehat{\mu}$ turns it into $\ivec-\ehat{\nu}$ or 
$\ivec+\ehat{\nu}$, respectively, since $i_\mu=i_\nu$. But it takes an 
odd number of transpositions to swap index-components, consequently the 
associated $\bar{\Phi}^{(k)}$'s have opposite signs and cancel each 
other in (\ref{equ:simplehomogeneous}) (case (c) in Fig.~\ref{fig:images}). 
When index shifts exceed the 
hypercube $\{1,\ldots N\}^k$ we use $N$-(anti)periodicity and the 
same argument applies. 

Summarising so far, by appropriately extending the definition of
$\Phi^{(k)}$ beyond the  
physically relevant range $\ivec \in N(k)$ we have managed to rewrite 
the homogeneous equation (\ref{equ:homogeneous}) in the much simpler 
form (\ref{equ:simplehomogeneous}). A comparison of 
(\ref{equ:simplehomogeneous}) with the original hierarchy 
(\ref{equ:correl}) shows that {\em antisymmetry} has naturally 
eliminated all links between different levels $k$ of the hierarchy. 
Now we can solve (\ref{equ:simplehomogeneous}) over the space 
of $N$-(anti)\-periodic functions. For this purpose we introduce the 
discrete Fourier transform $f_q=\mathcal{F}\{f_n\}$ as 
\begin{eqnarray}
  \forall q \in Q: \, f_q & = & \sum\limits_{n=1}^N f_n \, \me^{-i n q} 
  \label{equ:fourier} \\
  \forall n \in \mathbb{Z}: \, f_n & = & \frac{1}{N} \sum\limits_{q \in Q} 
  f_q \, \me^{+i n q} .
  \label{equ:fourierinv}
\end{eqnarray}
If one restricts $1 \leq n \leq N$ in (\ref{equ:fourierinv}), then 
(\ref{equ:fourier}), (\ref{equ:fourierinv}) form a transform pair 
when $Q$ is any equidistant $N$-partitioning of the interval $[0,2 \pi]$. 
But in order to retrieve $N$-periodic or $N$-antiperiodic functions from 
(\ref{equ:fourierinv}) the particular choice $Q=Q_\mathrm{o}$ or 
$Q=Q_\mathrm{e}$ has to be made, respectively, where
\begin{eqnarray}
  Q_\mathrm{e} & = & \big\{ \makebox[1.5cm]{$\frac{\pi}{N} + 
    \frac{2 \pi}{N} \, n$} \, \big| \, n=0, \ldots N-1 \big\} = 
    \big\{ {\textstyle \frac{\pi}{N}, \frac{3 \pi}{N}, \ldots 
    \frac{(2N-1)\pi}{N}} \big\}
    \label{equ:Qeven} \\
  Q_\mathrm{o}  & = & \big\{ \makebox[1.5cm]{$\frac{2 \pi}{N} \, n$}
    \, \big| \, n=0, \ldots N-1 \big\} = 
    \big\{ {\textstyle \,0\; , \frac{2 \pi}{N}, \ldots 
    \frac{(2N-2)\pi}{N}} \big\}. 
    \label{equ:Qodd}
\end{eqnarray}
So on even levels $k$ of the hierarchy we set $Q=Q_\mathrm{e}$ as 
we are dealing with $N$-antiperiodic functions while $Q=Q_\mathrm{o}$ 
on odd levels corresponds to $N$-periodic functions. In either case 
the transforms (\ref{equ:fourier}), (\ref{equ:fourierinv}) satisfy the 
usual shifting and convolution properties of Fourier transforms. Hence 
the $k$-fold transform of (\ref{equ:simplehomogeneous}) follows 
immediately as 
\begin{equation}
  \forall \boldsymbol{q} \in Q^k: \, 
  \frac{\dd}{\dd t}\,\bar{\Phi}^{(k)}_{\boldsymbol{q}}(t) 
  =\bar{\Phi}^{(k)}_{\boldsymbol{q}}(t) \left( -k + \gamma 
  \sum\limits_{\eta=1}^k \cos q_{\eta} \right).
  \label{equ:homogeneousfourier}
\end{equation}
Here $\boldsymbol{q}=(q_1,\ldots q_k)$ are the Fourier variables 
corresponding to $\ivec = (i_1,\ldots i_k)$. Note that we use the 
subscript to distinguish $\bar{\Phi}^{(k)}_{\ivec}$ from its transform
$\bar{\Phi}^{(k)}_{\boldsymbol{q}}$.
In Fourier space the evolution equations (\ref{equ:homogeneousfourier}) 
for $\bar{\Phi}^{(k)}$ are decoupled and can therefore be integrated 
easily. Then, by inverting the Fourier transform and using the initial
condition $\bar{\Phi}^{(k)}_{\ivec}(0)= \bar{A}^{(k)}_{\ivec}$ we obtain 
\begin{equation}
  \bar{\Phi}^{(k)}_{\ivec}(t) \,\, = \!\! \sum\limits_{j_1,\ldots j_k} 
  \prod\limits_{\eta=1}^k \me^{-t} \, I_{i_{\eta}-j_{\eta}}(\gamma t) \,
  \bar{A}^{(k)}_{\jvec}, 
  \label{equ:solutionphibar}
\end{equation}
where the sum runs over $\jvec \in \{1,\ldots N\}^k$. An expression for 
$I_n(x)$ is given in (\ref{equ:I}) in Appendix~\ref{sec:bessel}, where
we also discuss the properties of $I_n(x)$ that are relevant for our analysis.
Finally, by expressing the initial conditions 
$\bar{A}^{(k)}$ in terms of $A^{(k)}$ via (\ref{equ:phibar}) and after
some 
rearranging, we obtain the solution $\Phi^{(k)}$ of the homogeneous 
equation (\ref{equ:homogeneous}) on level $1 \leq k \leq N$ 
\begin{equation}
  \forall \ivec \in N(k): \, \Phik{k}{\ivec} \,\, = \!\! 
  \sum\limits_{\jvec \in N(k)} 
  \Gk{k}{\ivec}{\jvec}{t} \Ak{k}{\jvec}
  \label{equ:homogeneoussolution}
\end{equation}
with the Green's function
\begin{equation}
  \forall t>0: \, 
  \Gk{k}{\ivec}{\jvec}{t} \,\, = \!\! \sum\limits_{\pi 
  \in \mathcal{S}(k)} \sign \prod\limits_{\eta=1}^{k} \me^{-t} \,
  I_{i_\eta-j_{\pi(\eta)}} (\gamma t).
  \label{equ:greensfunction}
\end{equation}
The Green's function as such is defined over the index range
$\ivec, \jvec \in N(k)$. Expression (\ref{equ:greensfunction}), 
however, carries the antisymmetry of the images and is therefore 
well defined over $\ivec,\jvec \in \mathbb{Z}^k$. For later 
convenience we note that (\ref{equ:greensfunction}) is 
in fact permuationally antisymmetric in $\ivec$ and $\jvec$.
To see the antisymmetry in $\jvec$, consider
the sum over $\pi \in \Scal(k)$. If, instead of $\jvec$, we 
substitute a version permuted by $\pi_0$, we may use $\pi \circ \pi_0 
\in \Scal(k)$ as the summation variable since the permutations form a
group. Hence the replacement of $\jvec$ by $(j_{\pi_0(1)},\ldots 
j_{\pi_0(k)})$ just   
changes the sign of the Green's function by $(-1)^{\pi_0}$, proving
permutational antisymmetry. Antisymmetry over $\ivec$ can be shown
similarly and is an immediate consequence of the images.

\subsection{The Recursion Formula}
\label{sec:recursion}

In this section we use the recursion formula (\ref{equ:Fsplit}) to 
construct the general solution of the hierarchy (\ref{equ:matrixform}). 
It is convenient to express the homogeneous solution $\Phi^{(k)}$ 
contained in (\ref{equ:Fsplit}) in terms of the Green's function $G^{(k)}$ 
via (\ref{equ:homogeneoussolution}). This gives 
\begin{equation}
  \Fkt{k}{\ivec} \,\, = \!\! \sum\limits_{\jvec \in N(k)} 
  \Gk{k}{\ivec}{\jvec}{t} \, \Ak{k}{\jvec} + \int\limits_0^t \dd \tau 
  \!\! \sum\limits_{\jvec \in N(k)} \sum\limits_{\jvec' \in N(k-2)} 
  \Gk{k}{\ivec}{\jvec}{t-\tau} \, b^{(k)}_{\jvec,\jvec'} \, 
  \Fkt[\tau]{k-2}{\jvec'}.
  \label{equ:recursion}
\end{equation}
Now we show that a particular sequence of simplifications, essentially 
based on the similarity of different levels of the hierarchy, can be 
applied when (\ref{equ:recursion}) is iterated. Our proof is inductive, 
hence we start with proposing a structure for the result. 
Working out the first few iterations of (\ref{equ:recursion}) suggests 
that $L+1\leq \lfloor \frac{k}{2} \rfloor$ iterations should give the
following expression
\begin{equation}
  \Fkt{k}{\ivec} = \sum\limits_{l=0}^L 
  \sum\limits_{\pi\in\mathcal{P}(l,k)} \sign \prod\limits_{\lambda
  =1}^l H_{i_{\pi(2\lambda)}-i_{\pi(2\lambda-1)}}(2t) \; \Phik{k-2l}
  {(i_{\pi(2l+1)},\ldots i_{\pi(k)})} + \Rkt{k}{\ivec}{L}.
  \label{equ:Literations}  
\end{equation}
Here $\mathcal{P}(l,k)$ denotes the set of permutations that correspond 
to choosing $0 \leq l \leq \lfloor \frac{k}{2} \rfloor$ ordered pairs from 
the numbers $\{1,2,\ldots k \}$ and keeping the remaining $k-2l$ numbers 
in ascending order, i.e.\
\begin{equation}
  \begin{array}{rcl}
    \mathcal{P}(l,k) & = & \big\{\pi \in \Scal(k) \, \big| \, 
    \pi(1)<\pi(2),\quad \pi(3)<\pi(4), \ldots \quad \pi(2l-1)<\pi(2l), \\
    & & \pi(1)<\pi(3)<\ldots \pi(2l-1), \quad 
    \pi(2l+1)<\pi(2l+2)<\ldots \pi(k) \big\}.
  \end{array}
  \label{equ:Plk}
\end{equation}
An expression for $H_n(t)$ is given in (\ref{equ:H}). We also discuss 
relevant features of this function in Appendix \ref{sec:Hn}. 
The $\Phi^{(k)}$ appearing in (\ref{equ:Literations})  
are the homogeneous solutions (\ref{equ:homogeneoussolution}). Finally, 
$R^{(k)}$ denotes the remainder term of the recursion that links to level 
$k-2L-2$, and is of the form
\begin{eqnarray}
  \Rkt{k}{\ivec}{L} & = & \int\limits_0^t \dd\tau \left\{ 
    \sum\limits_{\pi\in\mathcal{P}(L,k)} 
    \sign \prod\limits_{\lambda=1}^L H_{i_{\pi(2\lambda)}-
    i_{\pi(2\lambda-1)}}(2(t-\tau)) \right.  
  \label{equ:Rn} \\ 
  & \times & \left. \sum\limits_{\jvec \in N(k-2L)} \;
    \sum\limits_{\jvec' \in N(k-2L-2)} \Gk{k-2L}{(i_{\pi(2L+1)},
    \ldots i_{\pi(k)})}{\jvec}{t-\tau} \, b^{(k-2L)}_{\jvec,\jvec'} \, 
    \Fkt[\tau]{k-2L-2}{\jvec'} \right\}.
  \nonumber
\end{eqnarray}

The first step in our proof that (\ref{equ:Literations}) is 
the product of iterating (\ref{equ:recursion}) is to consider $L=0$.
The two sums in (\ref{equ:Literations}) collapse to a single term, 
that is $l=0$ and $\mathcal{P}(0,k)=\{\mathrm{Id}\}$. This obviously 
also applies to the first sum in (\ref{equ:Rn}). Defining the value of 
empty products in (\ref{equ:Literations}) and (\ref{equ:Rn}) as one --
a convention we use throughout -- then
reduces (\ref{equ:Literations}) to (\ref{equ:recursion}) for $L=0$. Now 
assume that (\ref{equ:Literations}) is true for some $L \geq 0$. 
Consequently, for (\ref{equ:Literations}) to apply also for
$L \rightarrow L+1$, the remainder has to satisfy
\begin{equation}
  \Rkt{k}{\ivec}{L} = \Rkt{k}{\ivec}{L+1} \;\; + \!\!\!\sum\limits_{\pi \in 
  \mathcal{P}(L+1,k)} \!\! \sign \prod\limits_{\lambda=1}^{L+1} 
  H_{i_{\pi(2 \lambda)} - i_{\pi(2 \lambda -1)}}(2 t) \; \Phik{k-2(L+1)}
  {(i_{\pi(2(L+1)+1)},\ldots i_{\pi(k)})}.
  \label{equ:Rninduction}
\end{equation}
Of course we need $L+1 \leq \lfloor \frac{k}{2} \rfloor -1$ since 
(\ref{equ:recursion}) only holds on levels $k \geq 2$. We show below 
that the remainder (\ref{equ:Rn}) satisfies (\ref{equ:Rninduction}) 
which proves that (\ref{equ:Literations}) is the result of 
iterating (\ref{equ:recursion}).

First we evaluate the matrix product $G^{(k-2L)} \cdot b^{(k-2L)}$ 
in (\ref{equ:Rn}); we show in Appendix~\ref{sec:greensproject} that 
this can be expressed in terms of Green's functions 
$G^{(k-2L-2)}$ and the functions $H_n(t)$.
Substituting (\ref{equ:Gb}) into (\ref{equ:Rn}) gives 
\begin{eqnarray}
  \Rkt{k}{\ivec}{L} & = & \int\limits_0^t \dd\tau \left\{ 
    \sum\limits_{\pi\in\mathcal{P}(L,k)} 
    \sign \sum_{1 \leq \mu < \nu \leq k-2L} (-1)^{\nu-\mu-1} \right.  
    \nonumber \\
  & \times & \left( -\frac{\partial}{\partial \tau} \, 
    H_{i_{\pi(2L+\nu)}-i_{\pi(2L+\mu)}}
    (2(t-\tau)) \right) \prod\limits_{\lambda=1}^L H_{i_{\pi(2\lambda)}-
    i_{\pi(2\lambda-1)}}(2(t-\tau))  \label{equ:Rnstep1} \\
  & \times & \left. \sum\limits_{\jvec \in N(k-2L-2)} \Gk{k-2L-2}
    {(i_{\pi(2L+1)}, \ldots i_{\pi(k)}) \setminus (i_{\pi(2L+\mu)},
    i_{\pi(2L+\nu)})}{\jvec}{t-\tau} \Fkt[\tau]{k-2L-2}{\jvec} \right\}
  \nonumber
\end{eqnarray}

Then we express $F^{(k-2L-2)}$ in (\ref{equ:Rnstep1}) via the 
recursion formula (\ref{equ:recursion}) and use the general
identity for our Green's function 
\begin{equation}
  \forall \tau,\tau'>0: \, 
  \sum\limits_{\jvec \in N(k)} \Gk{k}{\ivec}{\jvec}{\tau} \, 
  \Gk{k}{\jvec}{\ivec'}{\tau'}=\Gk{k}{\ivec}{\ivec'}{\tau+\tau'}. 
  \label{equ:greensconvolution}
\end{equation}
This gives 
\begin{eqnarray}
  \Rkt{k}{\ivec}{L} & = & \int\limits_0^t \dd\tau \left\{ 
    \sum\limits_{\pi\in\mathcal{P}(L,k)} 
    \sign \sum_{1 \leq \mu < \nu \leq k-2L} (-1)^{\nu-\mu-1} \right.  
    \nonumber \\
  & \times & \left( -\frac{\partial}{\partial \tau} \, 
    H_{i_{\pi(2L+\nu)}-i_{\pi(2L+\mu)}}
    (2(t-\tau)) \right) \prod\limits_{\lambda=1}^L H_{i_{\pi(2\lambda)}-
    i_{\pi(2\lambda-1)}}(2(t-\tau)) \nonumber \\
  & \times & \left[ \sum\limits_{\jvec \in N(k-2L-2)} 
    \Gk{k-2L-2}{(i_{\pi(2L+1)}, \ldots i_{\pi(k)}) \setminus (i_{\pi(2L+\mu)},
    i_{\pi(2L+\nu)})}{\jvec}{t} \, \Ak{k-2L-2}{\jvec}  \right.
    \label{equ:Rnstep2} \\
  & + & \int\limits_0^{\tau} \dd\tau' \sum\limits_{\jvec \in N(k-2L-2)}
    \; \sum\limits_{\jvec' \in N(k-2L-4)} 
    \nonumber \\
  & & \Gk{k-2L-2}
    {(i_{\pi(2L+1)}, \ldots i_{\pi(k)}) \setminus (i_{\pi(2L+\mu)},
    i_{\pi(2L+\nu)})}{\jvec}{t-\tau'} \, b^{(k-2L-2)}_{\jvec,\jvec'} \, 
    \Fkt[\tau']{k-2L-4}{\jvec'}  {\displaystyle \Bigg] \Bigg\} }. \nonumber
  \nonumber
\end{eqnarray}
In (\ref{equ:Rnstep2}) $\tau$-dependencies via the Green's functions have 
dropped out.

Next, we rewrite the permuations in (\ref{equ:Rnstep2}) in terms of
those from the set $\mathcal{P}(L+1,k)$. 
Consider the case $L \geq 1$ first. According to (\ref{equ:Plk}) 
the first $2L$ values of $\pi \in \mathcal{P}(L,k)$ correspond to 
$L$ ordered pairs chosen from $\{1,\dots k\}$ whereas $\pi(2L+1), 
\ldots \pi(k)$ are the remaining numbers in ascending order. Now the $\mu,\nu$ 
sum chooses another ordered pair $\pi(2L+\mu)<\pi(2L+\nu)$ since $\mu<\nu$ 
and $\pi$ is ordered in that range. In contrast to the $L$ pairs 
selected by $\pi$ the new pair does not obey the 
mutual ordering between pairs. As both sums run over all possible choices 
we get each set of $L+1$ pairs $L+1$ times, with $\pi(2L+\mu)$ in all 
mutual orderings from $\pi(2L+\mu)<\pi(1)<\ldots \pi(2L-1)$ to 
$\pi(1)<\ldots \pi(2L-1)<\pi(2L+\mu)$. But this ordering is only relevant 
for the second line in (\ref{equ:Rnstep2}). Hence we may restrict the 
choice of pairs to those contained in $\mathcal{P}(L+1,k)$ and confine 
the sum over the $L+1$ ways of selecting a ``special'' pair to the 
second line in (\ref{equ:Rnstep2}) -- which thereby turns into the 
derivative of the product of $L+1$ terms 
\begin{eqnarray}
  \Rkt{k}{\ivec}{L} & = & \int\limits_0^t \dd\tau \left\{ 
    \sum\limits_{\pi\in\mathcal{P}(L+1,k)} 
    \sign \left( -\frac{\partial}{\partial \tau}
    \prod\limits_{\lambda=1}^{L+1} H_{i_{\pi(2\lambda)}-
    i_{\pi(2\lambda-1)}}(2(t-\tau)) \right) \right. \nonumber \\
  & \times & \left[ \sum\limits_{\jvec \in N(k-2(L+1))} 
    \Gk{k-2(L+1)}{(i_{\pi(2(L+1)+1)}, \ldots i_{\pi(k)})}{\jvec}{t} \, 
    \Ak{k-2(L+1)}{\jvec}  \right.
    \label{equ:Rnstep3} \\
  & + & \int\limits_0^{\tau} \dd\tau' \sum\limits_{\jvec \in N(k-2(L+1))} 
    \; \sum\limits_{\jvec' \in N(k-2(L+1)-2)} 
    \nonumber \\
  & & \Gk{k-2(L+1)}
    {(i_{\pi(2(L+1)+1)}, \ldots i_{\pi(k)})}{\jvec}{t-\tau'}
    \, b^{(k-2(L+1))}_{\jvec,\jvec'} \, 
    \Fkt[\tau']{k-2(L+1)-2}{\jvec'}  {\displaystyle \Bigg] \Bigg\} }. \nonumber
  \nonumber
\end{eqnarray}

Now we convince ourselves that the sign of the 
permutation in (\ref{equ:Rnstep3}) matches the sign-factor in 
(\ref{equ:Rnstep2}): in (\ref{equ:Rnstep2}) we start from the ordered 
vector $(1,2,\ldots k)$. Applying $\pi \in \mathcal{P}(L,k)$ permutes 
this vector as described, with the sign of the permutation being 
$(-1)^\pi$. The sum over $\mu,\nu$ then chooses components $2L+\mu$, $2L+\nu$ 
as pair number $L+1$. Now it takes $\nu-\mu-1$ transpositions to move 
those components next to each other and an even number of further ones 
-- thus irrelevant -- to move this pair around. Altogether, $(-1)^\pi 
\; (-1)^{\nu-\mu-1}$ in (\ref{equ:Rnstep2}) 
is the sign of the corresponding permutation $\pi \in
\mathcal{P}(L+1,k)$ in (\ref{equ:Rnstep3}).

In the $L=0$ 
case, equivalence of (\ref{equ:Rnstep2}), (\ref{equ:Rnstep3}) is obvious: 
since $\mathcal{P}(0,k)=\{\mathrm{Id}\}$ we can remove this sum and all 
$\pi$'s from (\ref{equ:Rnstep2}); the empty product over $\lambda$ is
one by definition. So (\ref{equ:Rnstep2}) is just the 
explicit formulation of the sum over $\pi \in \mathcal{P}(1,k)$ in
(\ref{equ:Rnstep3}).

Finally an integration by parts in $\tau$ turns (\ref{equ:Rnstep3}) 
into (\ref{equ:Rninduction}), using $H_n(0)=0$. This completes 
the proof that 
(\ref{equ:Literations}) is the result of iterating (\ref{equ:recursion}).

Having proved (\ref{equ:Literations}), we can now set $L=\lfloor 
\frac{k}{2}\rfloor -1$. The remainder term in (\ref{equ:Literations})
then links  to level $0$ or $1$ of the hierarchy for $k$ even or odd,
respectively. 
But, as discussed below (\ref{equ:homogeneous}), we know 
the functions $F^{(0)}=A^{(0)}$ and $F^{(1)}=\Phi^{(1)}$ appearing in 
the remainder. In fact it can be simplified in full analogy to 
the above calculation for $L<\lfloor \frac{k}{2}\rfloor -1$ and this yields
the {\em general},  
{\em explicit} solution of the hierarchy (\ref{equ:matrixform}) as 
\begin{equation}
  \Fkt{k}{\ivec}=\sum\limits_{l=0}^{\lfloor \frac{k}{2} \rfloor} 
    \sum\limits_{\pi \in 
    \mathcal{P}(l,k)} \sign \prod\limits_{\lambda=1}^{l} H_{i_{\pi
    (2\lambda)}-i_{\pi(2\lambda-1)}}(2t) \; \Phik{k-2l}{(i_{\pi(2l+1)},\ldots 
    i_{\pi(k)})}.
  \label{equ:gsolution}
\end{equation}
The homogeneous solutions $\Phi^{(k)}$ appearing in (\ref{equ:gsolution}) 
are given by (\ref{equ:homogeneoussolution}) for $1 \leq k \leq N$.
We have incorporated the link to level $0$ in (\ref{equ:gsolution}), 
which occurs for even $k$, by defining $\Phi^{(0)}=A^{(0)}$.
The definition of the sets $\mathcal{P}(l,k)$ may be found in 
(\ref{equ:Plk}) and expressions for the functions $I_n(x)$ --- 
contained in the $\Phi^{(k)}$ --- and $H_n(t)$ are given in 
(\ref{equ:I}) and (\ref{equ:H}), respectively. According to 
(\ref{equ:gsolution}) all functions $I_n(x)$ and $H_n(t)$ 
appearing in $F^{(k)}$ are associated with (even) odd levels 
for (even) odd $k$ and consequently the appropriate sets 
$Q=Q_\even$ or $Q=Q_\odd$ as given by Eqs.~(\ref{equ:Qeven},\ref{equ:Qodd}) 
must be substituted in their representations.
Explicit versions of (\ref{equ:gsolution}) for the first
few levels read, bearing in mind that the product over $\lambda$
equals one for $l=0$,
\begin{eqnarray*}
  \Fkt{1}{i}     & = & \Phik{1}{i} \\[1ex]
  \Fkt{2}{\ivec} & = & \Phik{2}{\ivec} + 
                       H_{i_2-i_1}(2t) \, \Phi^{(0)} \\[1ex] 
  \Fkt{3}{\ivec} & = & \Phik{3}{\ivec} + 
                       H_{i_2-i_1}(2t) \, \Phik{1}{i_3} - 
                       H_{i_3-i_1}(2t) \, \Phik{1}{i_2} +
                       H_{i_3-i_2}(2t) \, \Phik{1}{i_1} \\[1ex] 
  \Fkt{4}{\ivec} & = & \Phik{4}{\ivec} \\
    & + & 
  \makebox[3.1cm][l]{$H_{i_2-i_1}(2t) \, \Phik{2}{(i_3,i_4)}$} - 
  \makebox[3.1cm][l]{$H_{i_3-i_1}(2t) \, \Phik{2}{(i_2,i_4)}$} +
  \makebox[3.1cm][l]{$H_{i_4-i_1}(2t) \, \Phik{2}{(i_2,i_3)}$} \\
    & + & 
  \makebox[3.1cm][l]{$H_{i_3-i_2}(2t) \, \Phik{2}{(i_1,i_4)}$} - 
  \makebox[3.1cm][l]{$H_{i_4-i_2}(2t) \, \Phik{2}{(i_1,i_3)}$} +
  \makebox[3.1cm][l]{$H_{i_4-i_3}(2t) \, \Phik{2}{(i_1,i_2)}$} \\
    & + & 
  \makebox[4cm][l]{$H_{i_2-i_1}(2t) \, H_{i_4-i_3}(2t) \, \Phi^{(0)}$} -
  \makebox[4cm][l]{$H_{i_3-i_1}(2t) \, H_{i_4-i_2}(2t) \, \Phi^{(0)}$} +
  \makebox[4cm][l]{$H_{i_4-i_1}(2t) \, H_{i_3-i_2}(2t) \, \Phi^{(0)}$}
\end{eqnarray*}
We remind the reader that the functions $F^{(k)}$ only have direct 
physical meaning if the indices $\ivec$ are chosen from $N(k)$. 
Extended over all indices $\ivec \in \{1,\ldots N\}^k$ they are 
permutationally antisymmetric; this follows from (\ref{equ:recursion}) 
and the corresponding property of the Green's function. Therefore the 
$F^{(k)}$ at any time $t$ can be used directly as antisymmetrized 
initial conditions $\bar{A}^{(k)}$ in (\ref{equ:solutionphibar}); this  
property will be useful below.  

Equation (\ref{equ:gsolution}) is evidently the simplest possible 
representation of $F^{(k)}$ for generic initial conditions $A^{(k)}$ 
as encoded in the $\Phi^{(k)}$. While for the generic case the complexity 
of $F^{(k)}$ grows rapidly with the level $k$, we will see below that in
physically interesting cases significant simplifications occur.

\subsection{The Thermodynamic Limit}
\label{sec:Ntoinfty}

The entire analysis presented in Sec.~\ref{sec:general} can be 
repeated in full analogy for an infinite spin chain. There, one 
takes the thermodynamic limit in (\ref{equ:correl}), which thereby 
becomes an infinite hierarchy of differential equations. The 
individual levels $k$ may again be rewritten in the matrix form 
(\ref{equ:matrixform}) over ordered indices. For the infinite 
chain, however, one chooses $-\infty < i_1 < i_2 < \ldots i_k < 
+\infty$ to obtain complete equations. The infinite matrices 
$a^{(k)}$ and $b^{(k)}$ have a slightly simpler structure since 
the shifts $\ivec-\ehat{1}$ and $\ivec+\ehat{k}$ can never produce 
an index pair. In other words $N$-periodicity of the finite ring 
is lost. Nevertheless (\ref{equ:matrixform}) may be treated as 
an inhomogeneous system of differential equations which can be 
solved according to the steps presented in Sec.~\ref{sec:homogeneous} 
and Sec.~\ref{sec:recursion}. Rather than the discrete 
$N$-(anti)\-periodic Fourier transforms (\ref{equ:fourier}), 
(\ref{equ:fourierinv}) the standard Fourier series expansion 
is useful for solving the homogeneous equations. The structure 
of the solutions (\ref{equ:gsolution}) of the hierarchy and the 
homogeneous solutions (\ref{equ:homogeneoussolution}) remains 
completely unaffected. But instead of the functions $I_n(x)$, 
$H_n(t)$ their $N \to \infty$ limits $\In_n(x)$, $\Hn_n(t)$ 
discussed in Appendix \ref{sec:bessel}, \ref{sec:Hn} appear in 
(\ref{equ:homogeneoussolution}), (\ref{equ:gsolution}). The 
sum over the initial conditions in (\ref{equ:homogeneoussolution}) 
obviously runs over $-\infty < j_1 < \ldots j_k < \infty$ for 
an infinite chain.

Alternatively, the rigorous solutions (\ref{equ:homogeneoussolution}), 
(\ref{equ:gsolution}) for the finite ring also produce the same 
$N \to \infty$ limit. This is evident for (\ref{equ:gsolution}) as 
it comprises a finite number of functions, the limit of each of which 
exists. In (\ref{equ:homogeneoussolution}), however, one has to keep 
in mind that the functions $I_n(x)$ are $N$-(anti)\-periodic. 
Therefore all summation indices $\jvec$ that are close to $\ivec$ 
{\em modulo} $N$ contribute to the sum in (\ref{equ:homogeneoussolution}); 
the site $N$ for instance is a neighbour of site $1$.  
So a meaningful $N \to \infty$ limit can only be taken if we 
first shift the index range of the finite ring. An appropriate 
replacement for $N(k)$ is for instance $-\lfloor \frac{N-1}{2} \rfloor 
\leq i_1 < \ldots i_k \leq \lfloor \frac{N}{2} \rfloor$. Shifting 
the index range on the ring changes the domains of the solutions 
(\ref{equ:homogeneoussolution}), (\ref{equ:gsolution}) but leaves 
them unaffected otherwise. Convergence for $N\to\infty$ of the sum in 
(\ref{equ:homogeneoussolution}) over the symmetric index range can
then be proved \cite{Thesis}, for any fixed index vector $\ivec$. 

In the following we continue to analyse the finite ring of spins. 
But according to the above discussion the thermodynamic limit may be 
taken at any stage. It just amounts to replacing the functions 
$I_n(x)$, $H_n(t)$ by their $N \to \infty$ limits $\In_n(x)$, 
$\Hn_n(t)$ and extending summation ranges. From now on we 
omit the limits for summations that are taken over the ring, as 
for instance in (\ref{equ:solutionphibar}). In a finite system 
these summations run over $1,\ldots N$ while in the thermodynamic 
limit they are to be evaluated over $-\infty \ldots +\infty$.

\section{Equilibrium correlations}
\label{sec:equilibrium}

As discussed in the beginning of Sec.~\ref{sec:general} the functions 
$F^{(k)}(t)=C^{(k)}(t)$ describe the evolution of correlations if we 
use the initial conditions $A^{(k)}=C^{(k)}(0)$. Here we show that our 
solution (\ref{equ:gsolution}) reproduces the well known equilibrium 
correlation functions for $T>0$. 

Equilibrium correlations are obtained from the dynamical solutions 
(\ref{equ:gsolution}) by taking the limit $t \to \infty$. Let us 
first focus on the homogeneous solutions 
$\Phi^{(k)}$. It is convenient to consider the representation 
(\ref{equ:solutionphibar}) instead of (\ref{equ:homogeneoussolution}); 
recall that for $\ivec \in N(k)$ we have $\bar{\Phi}^{(k)}=\Phi^{(k)}$. 
When taking the modulus of (\ref{equ:solutionphibar}) and using the 
triangular inequality we may drop the initial conditions 
$|\bar{A}^{(k)}| \leq |C^{(k)}(0)| \leq 1$. This gives the first 
inequality in 
\begin{equation}
  \left| \Phik{k}{\ivec} \right| \leq 
  \sum\limits_{j_1,\ldots j_k} 
  \prod\limits_{\eta=1}^k \me^{-t} 
  \left| I_{i_{\eta}-j_{\eta}}(\gamma t) \right| = 
  \prod\limits_{\eta=1}^k \me^{-t} \sum\limits_{j_\eta} 
  \left| I_{i_{\eta}-j_{\eta}}(\gamma t) \right| \leq 
  \me^{-k \, t/\taueq}.
  \label{equ:Phieq}
\end{equation}
As indicated, the $k$-dimensional sum in 
(\ref{equ:Phieq}) factorizes once the initial condition terms are removed. 
Then, using the 
bound (\ref{equ:Isum}) on $\sum_n |I_n(x)|$ gives the second 
inequality in (\ref{equ:Phieq}) for any system size $N$. Here
$1/\taueq=1-\gamma$ is  
the smallest eigenvalue of the master operator 
\cite{Felderhof71} corresponding to (\ref{equ:master}). Since 
$T>0$ implies $\gamma = \tanh(2J/T) < 1$ the equilibration 
time $\taueq$ is finite. Therefore at any level $k \geq 1$ 
of the hierarchy the homogeneous solutions vanish in equilibrium. 
At level $k=0$, however, we have $\Phi^{(0)}(t)=A^{(0)}=
C^{(0)}=1$ at all times.

Next consider the $t \to \infty$ limit of the functions $H_n(t)$. 
Based on the representation (\ref{equ:Halt}) it is straightforward 
to show that for $T>0$ and arbitrary system size $N$ 
we are left with the sum 
\begin{equation}
  H_{n,\mathrm{eq}}=\lim\limits_{t\rightarrow\infty} H_n(t)=
  \frac{1}{N} \sum_{q \in Q} \, \sin(n \, q) \, 
  \frac{\gamma \sin q}{1-\gamma \cos q}. 
  \label{equ:Hneq}
\end{equation}
By combining the results (\ref{equ:Phieq}) and (\ref{equ:Hneq}) we now 
obtain the equilibrium correlations from (\ref{equ:gsolution}). Any 
correlation function $C^{(2k+1)}$ containing an odd number of spins 
is expressed in terms of $\Phi^{(1)}, \Phi^{(3)},\ldots \Phi^{(2k+1)}$. 
But all 
these homogeneous solutions vanish in equilibrium and hence so do the
correlations of odd order, as expected. For equilibrium correlations
of even order 
$C^{(2k)}$, on the other hand, the only nonzero terms in 
(\ref{equ:gsolution}) are those having a factor $\Phi^{(0)}=1$. 
This gives 
\begin{equation}
  C^{(2k)}_{\ivec,\mathrm{eq}} = \sum\limits_{\pi \in \mathcal{P}(k)} 
  \sign \prod\limits_{\lambda=1}^k 
  H_{i_{\pi(2\lambda)}-i_{\pi(2\lambda-1)},\mathrm{eq}}, 
  \label{equ:ceqredundant}
\end{equation}
where we have introduced the notation $\mathcal{P}(k)=\mathcal{P}(k,2k)$ 
for the usual set of ordered pairings. Since the functions $H_{n,\mathrm{eq}}$ 
in (\ref{equ:ceqredundant}) are associated with even levels we need to
set $Q=Q_\mathrm{e}$ when substituting (\ref{equ:Hneq}). For ordered indices 
$\ivec \in N(2k)$ the result (\ref{equ:ceqredundant}) describes multispin 
equilibrium correlations in the finite Glauber-Ising model at $T>0$. Over 
non-ordered indices $\ivec \in \{1,\ldots N \}^{2k}$, however, 
(\ref{equ:ceqredundant}) carries the permutational antisymmetry of 
(\ref{equ:gsolution}); this property will be useful below. 

The link between (\ref{equ:ceqredundant}) and the result obtained 
from a transfer-matrix calculation is established by realizing that 
for $1 \leq n \leq N-1$ we have for (\ref{equ:Hneq}) the identity 
\begin{equation}
  \frac{1}{N} \sum_{q \in Q_\mathrm{e}} \, \sin(n \, q) \, 
  \frac{\gamma \sin q}{1-\gamma \cos q} 
  = 
  \frac{\left[ \tanh (J/T) \right]^n + \left[ \tanh (J/T) \right]^{N-n}}
  {1+\left[ \tanh (J/T) \right]^N}. 
  \label{equ:Hneqtransfer}
\end{equation}
The l.h.s.\ of (\ref{equ:Hneqtransfer}) is identically zero 
for $n=0,N$ and $N$-antiperiodic, however. To prove (\ref{equ:Hneqtransfer}) 
one notes that the sum may be rewritten in the form of the inverse Fourier 
transform (\ref{equ:fourierinv}). So we may equivalently prove the 
converse statement (\ref{equ:fourier}) for $q \in Q_\mathrm{e}$. The 
latter just involves a geometric summation and is straightforward. 
We finally remark that when the representation (\ref{equ:Hneqtransfer}) 
is used for $H_{n,\mathrm{eq}}$, the permutational antisymmetry of 
(\ref{equ:ceqredundant}) is lost. Only for ordered index vectors 
$\ivec \in N(2k)$ are all combinations $n=i_{\pi(2\lambda)}-
i_{\pi(2\lambda-1)}$ in (\ref{equ:ceqredundant}) in the range where 
(\ref{equ:Hneqtransfer}) applies.

\section{Correlations after a quench}
\label{sec:correlquench}

Next we derive dynamical correlation functions 
$C^{(k)}(t)$ from our general solution (\ref{equ:gsolution}). We 
consider the example where the system is quenched at $t=0$ from 
an equilibrium state at $\gamma_\mathrm{i} < 1$, corresponding to
$T_\mathrm{i}> 0$, to some arbitrary temperature $\gamma$. As in 
Sec.~\ref{sec:equilibrium} we may identify $F^{(k)}(t)=C^{(k)}(t)$ 
if we use equilibrium correlations as the initial conditions 
$A^{(k)}$. 

On odd levels of the hierarchy we have $A^{(2k+1)}=0$ 
since equilibrium correlations between any odd number of spins 
vanish. Consequently $\Phi^{(2k+1)}=0$ according to 
(\ref{equ:homogeneoussolution}) and via (\ref{equ:gsolution}) 
also $C^{(2k+1)}(t)=0$ at all times. This conclusion is trivial 
because the master equation (\ref{equ:master}) and the equilibrium 
initial state are invariant under spin inversion $\bsig \to -\bsig$. 
On even levels $2k \geq 2$ of the hierarchy we express 
$A^{(2k)}$ via (\ref{equ:Hneq}) and (\ref{equ:ceqredundant}) as 
\begin{equation}
  \bar{A}^{(2k)}_{\ivec} = \sum\limits_{\pi \in \mathcal{P}(k)} 
  \sign \prod\limits_{\lambda=1}^k \, \frac{1}{N} \sum_{q \in Q_\mathrm{e}} 
  \sin\left[ q \, (i_{\pi(2\lambda)}-i_{\pi(2\lambda-1)}) \right]  
  \frac{\gamma_\mathrm{i} \sin q}{1-\gamma_\mathrm{i} \cos q}.
  \label{equ:corinitbar}
\end{equation}
As pointed out below (\ref{equ:ceqredundant}),
this representation for the equilibrium correlation functions is 
permutationally antisymmetric and consequently zero for non-pairwise 
distinct indices $\ivec$. This is why we have been able to write
$\bar{A}$ instead of  
$A$ in (\ref{equ:corinitbar}). Knowing $\bar{A}$ is convenient 
as it allows us to derive the homogeneous solutions from 
(\ref{equ:solutionphibar}) instead of the more complex expression 
(\ref{equ:homogeneoussolution}). In fact, since (\ref{equ:corinitbar}) 
is factorized over index pairs, the sum in (\ref{equ:solutionphibar}) 
also factorizes into two-dimensional sums
\begin{equation}
  \Phi^{(2k)}_{\ivec}(t) \,\, = \!\! 
  \sum\limits_{\pi \in \mathcal{P}(k)} 
  \! \sign \! \prod\limits_{\lambda=1}^k \left[
  \sum\limits_{m,n} 
  \me^{-2t} \, I_{i_{\pi(2\lambda)}-m}(\gamma t) \,
  I_{i_{\pi(2\lambda-1)}-n}(\gamma t) \, 
  \frac{1}{N} \sum_{q \in Q_\mathrm{e}} 
  \sin\left[ q \, (m-n) \right]  
  \frac{\gamma_\mathrm{i} \sin q}{1-\gamma_\mathrm{i} \cos q}
  \right].
  \label{equ:corphi}
\end{equation}
Using the properties (\ref{equ:Icossum}), (\ref{equ:Iconv}) of the 
functions $I_n(x)$, the sum $\sum_{m,n}$ in (\ref{equ:corphi}) can 
easily be evaluated. Here it is essential that the functions $I_n(x)$ 
in (\ref{equ:corphi}) are associated with the even levels $2k$ and that 
in the argument of $\sin[q\, (m-n)]$ we have $q \in Q_\mathrm{e}$ 
too, as (\ref{equ:Icossum}) would not apply otherwise. We obtain 
\begin{equation}
  \Phi^{(2k)}_{\ivec}(t) = \sum\limits_{\pi \in \mathcal{P}(k)} 
  \sign \prod\limits_{\lambda=1}^k H_{i_{\pi(2\lambda)}-
  i_{\pi(2\lambda-1)}}'(2t),
  \label{equ:corphisolution}
\end{equation}
where $H_n'(t)$ is the result of the summation over $m,n$ and 
may be written as
\begin{equation}
  H_n'(t) = \frac{1}{N} \sum_{q \in Q_\mathrm{e}} 
  \sin (n \, q) \, 
  \frac{\gamma_\mathrm{i} \sin q}{1-\gamma_\mathrm{i} \cos q} \, 
  \me^{-t (1-\gamma \cos q)}.
  \label{equ:Hnprime}
\end{equation}
We remark that this straightforward derivation of the homogeneous 
solutions heavily rests upon the permutational antisymmetry of our 
expression (\ref{equ:ceqredundant}) for equilibrium correlations. 
When Glauber derived the dynamics of two-spin correlations 
\cite{Glauber63} he expressed the equilibrium values via 
(\ref{equ:Hneqtransfer}). He introduced images at this stage of 
the calculation to make (\ref{equ:Hneqtransfer}) permutationally 
antisymmetric.

Having worked out the homogeneous solutions (\ref{equ:corphisolution}) 
that correspond to our problem, we now obtain the correlations after the 
quench from (\ref{equ:gsolution}) 
\begin{equation}
  \ckt{2k}{\ivec} = \sum\limits_{l=0}^k \sum\limits_{\pi \in 
  \mathcal{P}(l,2k)} 
  \!\!\! (-1)^{\pi} \!\!\!\!\!\! \sum\limits_{\hat{\pi} \in\mathcal{P}(k-l)} 
  \!\!\!\! (-1)^{\hat{\pi}} 
  \prod\limits_{\lambda=1}^l H_{i_{\pi(2\lambda)}-
  i_{\pi(2\lambda-1)}}(2t) 
  \prod\limits_{\lambda'=1}^{k-l} H_{i_{\pi(2l+\hat{\pi}(2\lambda'))}-
  i_{\pi(2l+\hat{\pi}(2\lambda'-1))}}'(2t). 
  \label{equ:corquenchbig}
\end{equation}
For the term with $l=k$ to correctly give the contribution from
$\Phi^{(k-l)}=C^{(0)}=1$ one has to define $\mathcal{P}(0)=\{
\mathrm{Id} \}$ here. To simplify (\ref{equ:corquenchbig}), we now focus on 
the combination of pairings that occur
for some fixed $l$: each $\pi \in \mathcal{P}(l,2k)$ draws $l$ 
ordered pairs from $\{1,\ldots 2k\}$. These define the indices
of the factors $H_n(2t)$ in (\ref{equ:corquenchbig}). Out of 
the remaining, ordered $2k-2l$ numbers $\pi(2l+1), \ldots \pi(2k)$, 
the permutation $\hat{\pi} \in P(k-l)$ selects another $k-l$ ordered 
pairs, which give the indices of the factors $H_n'(2t)$. If we denote 
by $\pi_l \in \Scal(2k)$ the overall permutation on 
$\{1,\dots 2k\}$, then all $\pi_l$ satisfy 
\begin{equation}
  \begin{array}{c}
  \pi_l(1) < \pi_l(2), \, \pi_l(3) < \pi_l(4), \ldots \,
  \pi_l(2k-1) < \pi_l(2k) \\[1ex] 
  \pi_l(1) < \pi_l(3) < \ldots \pi_l(2l-1), \, 
  \pi_l(2l+1) < \pi_l(2l+3) < \ldots \pi_l(2k-1). 
  \end{array}
  \label{equ:semiordered}
\end{equation}
By ordering the first $l$ pairs against the subsequent $k-l$ ones, 
each $\pi_l \in \Scal(2k)$ can be mapped onto an ordered pairing 
$\bar{\pi} \in \mathcal{P}(k)$. This leaves the sign of the permutation 
unaffected as it takes an even number of transpositions to reorder 
pairs, i.e.\ $(-1)^{\bar{\pi}} = (-1)^{\pi_l} = (-1)^\pi \, 
(-1)^{\hat{\pi}}$. Conversely, we have that for each $\bar{\pi} \in 
\mathcal{P}(k)$ there are ${k \choose l}$ semi-ordered pairings 
$\pi_l$ of the form (\ref{equ:semiordered}). These correspond to 
the different ways of distributing $k-l$ primes among the functions 
$H_n(2t)$ in (\ref{equ:corquenchbig}) when we choose the index pairs
according to $\bar{\pi}$. Together with the summation over $l$ we can
thus factorize (\ref{equ:corquenchbig}) for each $\bar{\pi}$, giving 
\begin{equation}
  \ckt{2k}{\ivec} = \sum\limits_{\bar{\pi}\in\mathcal{P}(k)} 
  (-1)^{\bar{\pi}} \prod\limits_{\lambda=1}^k H_{i_{\bar{\pi}(2\lambda)}-
  i_{\bar{\pi}(2\lambda-1)}}''(2t), 
  \label{equ:corquench}
\end{equation}
where $H_n''(t) = H_n(t) + H_n'(t)$. Since the functions $H_n(t)$ in 
(\ref{equ:corquenchbig}) are associated with even levels $2k$ we need to set 
$Q=Q_\mathrm{e}$ in the representation (\ref{equ:Halt}). Together 
with (\ref{equ:Hnprime}) this yields 
\begin{equation}
  H_n''(t) = \frac{1}{N} \sum_{q \in Q_\mathrm{e}} 
  \sin (n \, q) \left\{ 
  \frac{\gamma_\mathrm{i} \sin q}{1-\gamma_\mathrm{i} \cos q} \, 
  \me^{-t (1-\gamma \cos q)}+\frac{\gamma \sin q}{1-\gamma \cos q} 
  \left[ 1-\me^{-t (1-\gamma \cos q)} \right] \right\}.
  \label{equ:Hn2prime}
\end{equation}
At $t=0$ our solution for dynamical correlations (\ref{equ:corquench}), 
(\ref{equ:Hn2prime}) obviously reproduces the equilibrium values 
(\ref{equ:Hneq}), (\ref{equ:ceqredundant}) for the initial temperature 
$\gamma_\mathrm{i}$. Note that for a quench from $T_\mathrm{i} =
\infty$ ($\gamma_\mathrm{i} = 0$), i.e.\ a random initial 
configuration, (\ref{equ:Hn2prime}) reduces to $H_n''(t) = H_n(t)$. 
It is equally clear that we recover the equilibrium 
values at $\gamma$ in the limit $t \to \infty$. The spectrum of relaxation 
times for correlations is encoded in the exponents in (\ref{equ:Hn2prime}) 
and depends on the final temperature $\gamma$ after the quench and the 
system size via $Q_\mathrm{e}$ as given in (\ref{equ:Qeven}). In the 
thermodynamic limit $N \to \infty$ we can
replace $\frac{1}{N} \sum_q$ by 
$\frac{1}{2\pi} \int \dd q$ in (\ref{equ:Hn2prime}) and, for consistency, 
then use the symbol $\Hn_n''(t)$. Further properties 
of $H_n''(t)$ are discussed in Appendix~\ref{sec:Hn}. An equivalent 
formulation of the results (\ref{equ:corquench}), (\ref{equ:Hn2prime}) 
for the infinite spin chain was derived in \cite{BedShuOpp70}.

It is worthwhile to note that the general solution (\ref{equ:gsolution}) 
preserves the structure of ordered pairings. In the case at hand, for 
instance, we used the initial conditions (\ref{equ:ceqredundant}) and 
arrived at the result (\ref{equ:corquench}). Both expressions have 
the same structure and (\ref{equ:corquench}) again carries the 
permutational antisymmetry of (\ref{equ:gsolution}). Therefore the 
dynamical correlations after an arbitrary sequence of temperature 
shifts are still of the form (\ref{equ:corquench}) and could be written 
down immediately as long as the system is initially in an 
equilibrium state. Examples of two and four-spin correlation functions 
as obtained from (\ref{equ:corquench}) read
\begin{eqnarray*}
  \ckt{2}{\ivec} & = & H_{i_2-i_1}''(2t) \\
  \ckt{4}{\ivec} & = & H_{i_2-i_1}''(2t) \, H_{i_4-i_3}''(2t) - 
                       H_{i_3-i_1}''(2t) \, H_{i_4-i_2}''(2t) + 
                       H_{i_4-i_1}''(2t) \, H_{i_3-i_2}''(2t).  
\end{eqnarray*}
The number of terms grows as the number of permutations contained in
$\mathcal{P}(k)$, i.e., $(2k-1)!! = 1 \cdot 3 \cdots (2k-1)$, giving 
$15$ terms for $C^{(6)}$, $105$ terms for $C^{(8)}$ and so on.

\section{Two-time correlations after a quench}
\label{sec:correl2quench}

As discussed above the evolution equations (\ref{equ:correl2}) 
of two-time correlations $C^{(k,l)}(t,t')$ may be written in the 
form (\ref{equ:matrixform}). Hence these functions again follow 
from the general solution (\ref{equ:gsolution}) if we use 
the equal time initial conditions (\ref{equ:correl2init}) 
at $t=t'$; obviously time arguments $t$ in (\ref{equ:gsolution}) 
have to be replaced by $\Delta=t-t'$ in this setup. We 
consider the same quench as in Sec.~\ref{sec:correlquench} 
because this allows us to express the initial conditions 
(\ref{equ:correl2init}) in terms of (\ref{equ:corquench}) for 
even $k+l$. On all levels $k$ where $k+l$ is odd, that is 
all even levels $k$ for odd $l$ and vice versa,
we have $C^{(k,l)}(t',t')=C^{(k+l)}(t')=0$. 
Hence $C^{(k,l)}(t,t')=0$ at all times 
from (\ref{equ:homogeneoussolution}), 
(\ref{equ:gsolution}); this is again just a consequence of 
the spin inversion symmetry $\bsig \to -\bsig$. So from now 
on we assume that $k+l$ is even.

In analogy to Sec.~\ref{sec:correlquench} an expression 
$\bar{A}$ for the equal time initial conditions 
(\ref{equ:correl2init}) that is permutationally antisymmetric 
in $\ivec$ would be desirable. Without loss of generality 
we require that the vector $\jvec$ --- only appearing as a 
parameter --- is ordered, $\jvec \in N(l)$. Now the following 
problems have to be handled. First, if 
$\ivec \in N(k)$ is also ordered, our expression for $\bar{A}$ 
has to produce the correlation function between the $k+l$ 
sites $\ivec \cup \jvec$. But the combined vector 
is not necessarily ordered and hence the associated 
correlation cannot be expressed directly via (\ref{equ:corquench}). 
Second, there could still be pairs of equal components between $\ivec$ 
and $\jvec$. While the correlation over $\ivec \cup \jvec$ then reduces 
its order via the hierarchical property (\ref{equ:links}), 
the expression (\ref{equ:corquench}) is identically zero.
A systematic construction of $\bar{A}$ based on 
(\ref{equ:links}) and (\ref{equ:corquench}) is possible but 
rather cumbersome. Let us therefore immediately state the 
result for $\bar{A}$ and prove its correctness instead:
\begin{equation}
  \bar{A}^{(k,l)}_{\ivec,\jvec}(t') = \sum_{\pi\in \mathcal{P}(\frac{k+l}{2})} 
  \sign \prod_{\lambda=1}^{\frac{k+l}{2}} 
  \mathcal{U}_{(\ivec \cup \jvec)_{\pi(2\lambda-1)}, 
  (\ivec \cup \jvec)_{\pi(2\lambda)}}^{\,\jvec}(t')
  \label{equ:cor2initbar}
\end{equation}
where
\begin{equation}
  \mathcal{U}_{a,b}^{\jvec}(t') = \left\{
    \begin{array}{rccc}
      & H_{j_\nu-j_\mu}''(2 t') & & (a,b)=(j_\mu,j_\nu) \\[1.5ex]
      (-1)^{\nu-1} \, \delta_{i_\eps,j_\nu} + 
      \left[ \prod\limits_{\lambda=1}^l \sgn(j_\lambda-i_\eps) \right]
      & H_{j_\nu-i_\eps}''(2 t') & \quad \mbox{for} \quad & 
      (a,b)=(i_\eps,j_\nu)\\
      \left[ \prod\limits_{\lambda=1}^l \sgn(j_\lambda-i_\eps) \, 
      \sgn(j_\lambda-i_\delta) \right] & H_{i_\delta-i_\eps}''(2 t') & & 
      (a,b)=(i_\eps,i_\delta)
    \end{array}
  \right.
  \label{equ:Ucal}
\end{equation}
The $\sgn(n)$'s in (\ref{equ:Ucal}) are the standard sign functions, with
$\sgn(n)=-\sgn(-n)=1$ for $n>0$ and $\sgn(0)=0$. To illustrate the
notation of (\ref{equ:cor2initbar}) and (\ref{equ:Ucal}) we consider 
as a simple example the 6-spin two-time correlation with $k=l=3$. 
In this case the sum in (\ref{equ:cor2initbar}) runs over the 15 
ordered pairings contained in $\mathcal{P}(3)$. Among these focus 
on the particular pairing $\pi$ that gives $(1,3,2,4,5,6)$. 
Here the indices $(a,b)$ of the 3 factors $\mathcal{U}$ in 
(\ref{equ:cor2initbar}) are $((\ivec \cup \jvec)_1,
(\ivec \cup \jvec)_3)=(i_1,i_3)$, $((\ivec \cup \jvec)_2,
(\ivec \cup \jvec)_4)=(i_2,j_1)$ and $((\ivec \cup \jvec)_5,
(\ivec \cup \jvec)_6)=(j_2,j_3)$. According to (\ref{equ:Ucal}) 
we substitute $\prod_\lambda \sgn(j_\lambda-i_1) \sgn(j_\lambda-i_3) 
H_{i_3-i_1}''(2t')$ for the first factor $\mathcal{U}_{i_1,i_3}$ 
as both indices are drawn from $\ivec$. The second factor 
$\mathcal{U}_{i_2,j_1}$ comprises components from $\ivec$ as 
well as $\jvec$ and must therefore be expressed as $(-1)^{1-1} 
\delta_{i_2,j_1} + \prod_\lambda \sgn(j_\lambda-i_2) 
H_{j_1-i_2}''(2t')$. The third factor, with both indices from $\jvec$,
is simply $\mathcal{U}_{j_2,j_3}=H_{j_3-j_2}''(2t')$. 
The remaining terms in the sum over $\pi \in \mathcal{P}(3)$ in
(\ref{equ:cor2initbar}) are worked out similarly.

A few further comments on (\ref{equ:Ucal}) are in order. First, the
case $(a,b)=(j_\nu,i_\eps)$ cannot occur in (\ref{equ:Ucal}) because
all pairs in $\pi$ are by definition ordered. Second, the
$\mathcal{U}$'s carry the superscript $\jvec$ since the products in
(\ref{equ:Ucal}) run over the full vector $\jvec$, so that
$\mathcal{U}$ depends on all components of $\jvec$ except in the first
case of (\ref{equ:Ucal}). Finally, the factor $(-1)^{\nu-1}$ in
(\ref{equ:Ucal}) depends on the actual number $\nu$ of the component
$j_\nu$ drawn from $\jvec$.

Having clarified the interpretation of (\ref{equ:cor2initbar}) and 
(\ref{equ:Ucal}) we now prove that these equations are indeed the 
initial conditions $\bar{A}$ corresponding to (\ref{equ:correl2init}). 
Let us first show that our expressions are antisymmetric under 
permutations in $\ivec \in \{ 1,\ldots N \}^k$. It is
sufficient to prove antisymmetry under transposition of any two 
adjacent components of $\ivec$, say $i_\mu,i_{\mu+1}$, since all 
permutations may be decomposed into such transpositions. Now, given 
$i_\mu,i_{\mu+1}$, we split the pairings $\mathcal{P}$ in 
(\ref{equ:cor2initbar}) into those $\mathcal{P}'$ containing the 
pair $(i_\mu,i_{\mu+1})$ and the remaining pairings $\mathcal{P}''$ 
where $i_\mu$ and $i_{\mu+1}$ are distributed over separate pairs. 
For each $\pi \in \mathcal{P}'$ there is a factor 
$\mathcal{U}_{i_\mu,i_{\mu+1}}$ in (\ref{equ:cor2initbar}) which, 
according to (\ref{equ:Ucal}), is of the form $\prod_\lambda 
\sgn(j_\lambda- i_\mu) \sgn(j_\lambda-i_{\mu+1}) 
H_{i_{\mu+1}-i_\mu}''(2t')$. With $H_{-n}''(t)=-H_n''(t)$ this
immediately shows that 
$\mathcal{U}_{i_{\mu+1},i_\mu}=-\mathcal{U}_{i_\mu,i_{\mu+1}}$;
therefore each term in 
(\ref{equ:cor2initbar}) with $\pi \in \mathcal{P}'$ is antisymmetric 
under $i_\mu \leftrightarrow i_{\mu+1}$. For $\pi \in 
\mathcal{P}''$ we have to apply a different reasoning. 
Suppose, for instance, that a particular pairing $\pi \in \mathcal{P}''$ 
contains the pairs $(i_\eps,i_\mu)$ and $(i_\delta,i_{\mu+1})$. 
As the pairs are ordered we have $\eps < \mu$, $\delta < 
\mu+1$ and the mutual ordering $\eps < \delta$. But $\delta \neq 
\mu$ implies $\delta < \mu$ and hence the conditions $\eps < \mu+1$, 
$\delta < \mu$ and $\eps < \delta$ also hold. So there is another 
pairing $\pi' \in \mathcal{P}''$ that contains $(i_\eps,i_{\mu+1})$ 
and $(i_\delta,i_\mu)$ but is identical to $\pi$ otherwise. Since 
$\pi$ and $\pi'$ have opposite signs and because the transposition 
$i_\mu \leftrightarrow i_{\mu+1}$ maps the pairings corresponding to 
$\pi$ and $\pi'$ onto each other, the sum of the two associated 
terms in (\ref{equ:cor2initbar}) is antisymmetric under 
$i_\mu \leftrightarrow i_{\mu+1}$. This argument is readily extended 
to all $\pi \in \mathcal{P}''$ by considering the remaining cases; 
these are pairs of the form $(i_\eps,i_\mu),(i_{\mu+1},x)$ where 
$\eps < \mu$ and $x \in \{i_{\mu+2},\ldots i_k,j_1,\ldots j_l \}$ 
or $(i_\mu,x),(i_{\mu+1},y)$ with $x,y \in \{i_{\mu+2},\ldots j_l \}$.

Now we turn to ordered index vectors $\ivec \in N(k)$. Then 
(\ref{equ:cor2initbar}) must produce the correlation function 
over the sites $\ivec \cup \jvec$ according to the initial 
condition (\ref{equ:correl2init}). Let us for the moment focus 
on those $\ivec \in N(k), \jvec \in N(l)$ for which the combined 
vector $\ivec \cup \jvec$ has pairwise distinct components. 
This allows us to ignore the Kronecker delta in (\ref{equ:Ucal}). 
Thus $\mathcal{U}_{a,b}^{\jvec}$ contains a factor $\prod_\lambda 
\sgn(j_\lambda-i_\eps)$ whenever the index $a$ is drawn from $\ivec$,
i.e.\ when $a=i_\eps$, and similarly for $b$. Since all components of
$\ivec$ occur 
exactly once in the product over $\lambda$ in (\ref{equ:cor2initbar}) we 
can therefore rewrite (\ref{equ:cor2initbar}) as 
\begin{equation}
  \bar{A}^{(k,l)}_{\ivec,\jvec}(t') = 
  \prod\limits_{\mu=1}^k \prod\limits_{\nu=1}^l 
  \sgn(j_\nu-i_\mu)
  \sum_{\pi\in \mathcal{P}(\frac{k+l}{2})} 
  \sign \prod_{\lambda=1}^{\frac{k+l}{2}} 
  H_{(\ivec \cup \jvec)_{\pi(2\lambda)}-
  (\ivec \cup \jvec)_{\pi(2\lambda-1)}}''(2t').
  \label{equ:cor2initbarpwd}
\end{equation}
Apart from the products over the $\sgn$'s this
corresponds to our result 
(\ref{equ:corquench}) for correlation functions. But 
(\ref{equ:corquench}) is antisymmetric under index permutation. Hence, 
if $\bar\pi \in \Scal(k+l)$ is the permutation that orders
$\ivec\cup\jvec$, i.e.\ $\ivec \cup \jvec \in N_{\bar{\pi}}(k+l)$, 
then the pairing sum in (\ref{equ:cor2initbarpwd}) must give $(-1)^{\bar\pi}$ 
times the correlation over $\ivec \cup \jvec$. Thus the products 
over the $\sgn$'s must coincide with $(-1)^{\bar\pi}$ in order to cancel 
the sign of $\bar\pi$. We can convince ourselves that this is indeed 
the case. Keep $\mu$ fixed and evaluate the 
product over $\nu$. For each $\nu$ with $j_\nu < i_\mu$ we have 
$\sgn(j_\nu-i_\mu)=-1$ while $i_\mu < j_\nu$ gives 
$\sgn(j_\nu-i_\mu)=+1$ and is irrelevant for the product. Altogether 
this gives $(-1)^{n_\mu}$ where $j_{n_\mu} < i_\mu < j_{n_\mu+1}$. 
Since both $\ivec \in N(k)$ and $\jvec \in N(l)$ are ordered we 
have $n_1 \leq n_2 \leq \ldots n_k$. Now if we start ordering 
$\ivec \cup \jvec$ beginning with $i_k$, it takes $n_k$ transpositions 
to turn $\ivec \cup \jvec$ into $(i_1,\ldots i_{k-1}, j_1,\ldots 
j_{n_k}, i_k, j_{n_k+1}, \ldots j_l)$, a further $n_{k-1}$ 
transpositions to turn this into $(i_1,\ldots i_{k-2}, j_1,\ldots 
j_{n_{k-1}}, i_{k-1}, j_{n_{k-1}+1}, \ldots j_{n_k}, i_k, j_{n_k+1}, 
\ldots j_l)$ and so on. Therefore the products over the $\sgn$'s give $(-1)$ 
to the power of the number of transpositions necessary to order 
$\ivec \cup \jvec$, which is equal to $(-1)^{\bar\pi}$.

Finally we lift the restriction that $\ivec \cup \jvec$ must have 
pairwise distinct components; $\ivec \in N(k)$ and $\jvec \in N(l)$ 
is of course still required. Assume that there are $p$ pairs 
$i_{\eps_1}=j_{\delta_1}, i_{\eps_2}=j_{\delta_2}, \ldots 
i_{\eps_p}=j_{\delta_p}$ between $\ivec$ and $\jvec$. We 
write $k'=k-p$, $l'=l-p$, $\ivec'=\ivec \setminus (i_{\eps_1}, 
\ldots i_{\eps_p})$ and $\jvec'=\jvec \setminus (j_{\delta_1}, \ldots 
j_{\delta_p})$. The combined vector $\ivec' \cup \jvec'$ then
contains no further component pairs. According to the hierarchical 
property (\ref{equ:links}) the equations (\ref{equ:cor2initbar}) and 
(\ref{equ:Ucal}) must then reduce to (\ref{equ:cor2initbarpwd}) with 
$k,l,\ivec,\jvec$ replaced by $k',l',\ivec',\jvec'$, respectively. 
The first step towards showing that this is true consists 
in splitting the pairings in (\ref{equ:cor2initbar}) again. Now, 
however, we denote by $\mathcal{P}'$ the pairings that contain 
all the pairs $(i_{\eps_q},j_{\delta_q})$ with $q=1,\ldots p$ and 
by $\mathcal{P}''$ the remaining pairings.
Let us focus on $\pi \in \mathcal{P}''$ first. For each 
such $\pi$ at least one pair, say $(i_{\eps_q},j_{\delta_q})$, is not 
formed. Instead $i_{\eps_q}$ may be paired up with $(i_u,i_{\eps_q})$, 
$(i_{\eps_q},i_v)$ or $(i_{\eps_q},j_w)$ where $w \neq \delta_q$. In 
either case there is a factor $\mathcal{U}=0$ in the product in 
(\ref{equ:cor2initbar}) since $\mathcal{U}_{i_u,i_{\eps_q}}$, 
$\mathcal{U}_{i_{\eps_q},i_v}$ and $\mathcal{U}_{i_{\eps_q},j_w}$ 
all contain a factor $\sgn(j_{\delta_q}-i_{\eps_q})=0$ according to 
(\ref{equ:Ucal}); for $\mathcal{U}_{i_{\eps_q},j_w}$ the Kronecker 
delta is also zero since $\jvec \in N(l)$ is ordered and hence 
$j_w \neq j_{\delta_q} = i_{\eps_q}$ for $w \neq \delta_q$. So 
each term in the sum over $\pi$ 
in (\ref{equ:cor2initbar}) is zero for $\pi \in \mathcal{P}''$.
Next we turn to the pairings 
$\pi \in \mathcal{P}'$. Each such $\pi$ may be constructed as 
follows. Starting from $\ivec \cup \jvec = (i_1, \ldots i_{\eps_1} 
\ldots i_{\eps_2} \ldots i_{\eps_p} \ldots i_k, j_1 \ldots 
j_{\delta_1} \ldots j_{\delta_2} \ldots j_{\delta_p} \ldots j_l)$, 
we first group all pairs $(i_{\eps_q},j_{\delta_q})$ 
with $q=1,\ldots p$ into the lowest components of the combined vector. 
It takes $k-\eps_1+\delta_1-1$ transpositions to move $j_{\delta_1}$ 
next to $i_{\eps_1}$ and an even (irrelevant) number of further
transpositions to get $(i_{\eps_1},j_{\delta_1},i_1, \ldots i_{\eps_2} 
\ldots i_{\eps_3} \ldots i_{\eps_p} \ldots i_k, j_1 \ldots 
j_{\delta_2} \ldots j_{\delta_3} \ldots j_{\delta_p} \ldots j_l)$. 
Moving $j_{\delta_2}$ next to $i_{\eps_2}$ then only takes 
$k-\eps_2+\delta_2-2$ transpositions because component $j_{\delta_1}$ 
is no longer contained between $i_{\eps_2}$ and $j_{\delta_2}$
in the latter vector. Clearly, regrouping $\ivec \cup \jvec$ into
the form $\ivec_\mathrm{rg}=(i_{\eps_1},j_{\delta_1},
i_{\eps_2},j_{\delta_2}, \ldots 
i_{\eps_p},j_{\delta_p}) \cup \ivec' \cup \jvec'$ then requires a 
permutation $\pi_{\mathrm{rg}}$ of sign
\begin{equation}
  (-1)^{\pi_{\mathrm{rg}}} = \prod\limits_{q=1}^p (-1)^{k-\eps_q+\delta_q-q}.
  \label{equ:pirg}
\end{equation}
Now by arranging the top $k'+l'$ components $\ivec' \cup \jvec'$ of 
$\ivec_\mathrm{rg}$ according to $\hat{\pi} \in \mathcal{P}((k'+l')/2)$ 
and ordering the pairs $(i_{\eps_q},j_{\delta_q})$ against 
$((\ivec' \cup \jvec')_{\hat{\pi}(2 \lambda-1)}, 
(\ivec' \cup \jvec')_{\hat{\pi}(2 \lambda)})$, any pairing $\pi \in 
\mathcal{P}'((k+l)/2)$ of $\ivec \cup \jvec$ may be obtained. 
Also, since rearranging pairs takes an even number of transpositions, 
$(-1)^\pi = (-1)^{\pi_{\mathrm{rg}}} (-1)^{\hat{\pi}}$. Using this 
representation for the pairings $\pi \in \mathcal{P}'$ we may now 
rewrite (\ref{equ:cor2initbar}). First we substitute the factorisation 
for $(-1)^\pi$ just given. Second, since all pairings $\pi$ contain the pairs 
$(i_{\eps_q},j_{\delta_q})$ with $q=1,\ldots p$ each product in 
(\ref{equ:cor2initbar}) has the factors 
$\mathcal{U}_{i_{\eps_q},j_{\delta_q}}$ which just give 
$(-1)^{\delta_q-1}$ according to (\ref{equ:Ucal}). The remaining 
factors are over the pairings $\hat{\pi}$ of $\ivec' \cup \jvec'$. 
Since $\ivec' \cup \jvec'$ has pairwise distinct components we 
may apply a similar reasoning as in (\ref{equ:cor2initbarpwd}) 
and thus obtain
\begin{eqnarray}
  \bar{A}^{(k,l)}_{\ivec,\jvec}(t') & = & (-1)^{\pi_{\mathrm{rg}}} 
  \prod\limits_{q=1}^p (-1)^{\delta_q-1} 
  \prod\limits_{
  \stackrel{{\scriptstyle \mu=1}}{\mu \neq \eps_1,\ldots \eps_p}}^k 
  \prod\limits_{\nu=1}^l \sgn(j_\nu-i_\mu) \nonumber \\
  & \times & \sum_{\pi\in \mathcal{P}(\frac{k'+l'}{2})} 
  \sign \prod_{\lambda=1}^{\frac{k'+l'}{2}} 
  H_{(\ivec' \cup \jvec')_{\pi(2\lambda)}-
  (\ivec' \cup \jvec')_{\pi(2\lambda-1)}}''(2t').
  \label{equ:cor2initbarppairs}
\end{eqnarray}
From (\ref{equ:pirg}) we see that equation
(\ref{equ:cor2initbarppairs}) reduces to  
(\ref{equ:cor2initbarpwd}) and therefore implements the 
hierarchical property (\ref{equ:links}) for pairs between 
$\ivec$ and $\jvec$ if the identity 
\begin{equation}
  1 = \prod\limits_{q=1}^p \left[(-1)^{k-\eps_q-q-1} 
  \prod\limits_{
  \stackrel{{\scriptstyle \mu=1}}{\mu \neq \eps_1,\ldots \eps_p}}^k
  \sgn(j_{\delta_q}-i_\mu)\right]
  \label{equ:identity} 
\end{equation}
holds. The product over $\mu$ is easily evaluated when using 
that $\ivec \in N(k)$ and $\jvec \in N(l)$ are ordered and 
$i_{\eps_q}=j_{\delta_q}$. Hence $\sgn(j_{\delta_q}-i_\mu)=-1$ 
for $\mu=\eps_q+1,\ldots k$. Since $\mu=\eps_{q+1},\ldots \eps_p$ 
are omitted, the product over $\mu$ gives $(-1)^{k-\eps_q-p+q}$. 
So the factors in the product over $q$ become $(-1)^{-1-p}$,
yielding the overall product $(-1)^{-p(p+1)}=1$ since $p (p+1)$ is 
even for any integer $p$. This completes our proof that 
equations (\ref{equ:cor2initbar}) and (\ref{equ:Ucal}) are 
the initial conditions $\bar{A}$ corresponding to 
(\ref{equ:correl2init}). 

Based on the expression (\ref{equ:cor2initbar}) for $\bar{A}$ it is 
straightforward to write down the corresponding homogeneous solutions 
$\Phi$ via (\ref{equ:solutionphibar}). Every component of $\ivec$ in 
(\ref{equ:cor2initbar}) --- which becomes a summation variable in 
(\ref{equ:solutionphibar}) --- occurs in exactly one of the factors of
the product over $\lambda$
in (\ref{equ:cor2initbar}), paired up with (depending on the
particular pairing $\pi$) either a different component of $\ivec$
or with a component of the fixed vector $\jvec$. Once the 
$k$-dimensional summation over $\ivec$ from (\ref{equ:solutionphibar}) 
is exchanged with the summation over $\pi$ from 
(\ref{equ:cor2initbar}), it thus factorizes into one- and 
two-dimensional sums. This procedure yields 
\begin{equation}
  \Phi^{(k,l)}_{\ivec,\jvec}(t,t') = \sum_{\pi\in \mathcal{P}(\frac{k+l}{2})} 
  \sign \prod_{\lambda=1}^{\frac{k+l}{2}} 
  \mathcal{V}_{(\ivec \cup \jvec)_{\pi(2\lambda-1)}, 
  (\ivec \cup \jvec)_{\pi(2\lambda)}}^{\,\jvec}(t,t'),
  \label{equ:cor2phisolution}
\end{equation}
where $\mathcal{V}$ denotes the case-dependent sums over $\mathcal{U}$
\begin{equation}
  \mathcal{V}_{a,b}^{\, \jvec}(t,t') = \left\{
    \begin{array}{rcc}
      H_{j_\nu-j_\mu}''(2 t') & & (a,b)=(j_\mu,j_\nu) \\[1.5ex]
      (-1)^{\nu-1} \, \me^{-\dt} \, I_{i_\eps-j_\nu}(\gamma \dt) + 
      \mathcal{E}_{i_\eps,j_\nu}^{\,\jvec}(t,t')
      & \quad \mbox{for} \quad & (a,b)=(i_\eps,j_\nu)\\[1.5ex]
      \mathcal{F}_{i_\eps,i_\delta}^{\,\jvec}(t,t')
       & & (a,b)=(i_\eps,i_\delta)
    \end{array}
  \right.
  \label{equ:Vcal}
\end{equation}
and $\mathcal{E}$, $\mathcal{F}$ are, respectively, the one 
and two-dimensional sums 
\begin{eqnarray}
  \mathcal{E}_{i_\eps,j_\nu}^{\,\jvec}(t,t') & = & 
  \sum_m \prod_{\lambda=1}^{\mathrm{dim}(\jvec)} \sgn(j_\lambda-m) \, 
  \me^{-\dt} \, I_{i_\eps-m}(\gamma \dt) \,
  H_{j_\nu-m}''(2t') 
  \label{equ:Esum} \\
  \mathcal{F}_{i_\eps,i_\delta}^{\,\jvec}(t,t') & = & 
  \sum_{m,n} \prod_{\lambda=1}^{\mathrm{dim}(\jvec)} \sgn(j_\lambda-m) \, 
  \sgn(j_\lambda-n) \, 
  \me^{-2 \dt} \, I_{i_\eps-m}(\gamma \dt) \, I_{i_\delta-n}(\gamma \dt) \,
  H_{n-m}''(2t').
  \label{equ:Fsum}
\end{eqnarray}
Equations (\ref{equ:cor2phisolution}) and (\ref{equ:Vcal}) are to 
be read in exactly the same manner as (\ref{equ:cor2initbar}) 
and (\ref{equ:Ucal}). The sums over $m$ and $n$ in 
(\ref{equ:Esum}) and (\ref{equ:Fsum}) each run over $N$ different
values, so that these expressions are not practical for large system
sizes. We show in
Appendix~\ref{sec:EF}, however, that both sums can be reduced to ones
involving only a finite number of terms, so that the limit $N\to\infty$
can be taken without problems. In
(\ref{equ:Esum}) and (\ref{equ:Fsum}) we  
have replaced the dimension $l$ of $\jvec$, being the upper limit 
of the products, by the generic expression $\dim(\jvec)$. This is 
for later convenience. Note that according to 
(\ref{equ:cor2phisolution}) and (\ref{equ:Vcal}) the homogeneous 
solutions for {\em arbitrary} order two-time multispin correlation 
functions can be expressed purely in terms of the functions $I$, $H''$, 
$\mathcal{E}$ and $\mathcal{F}$. Examples of low order homogeneous 
solutions read 
\begin{eqnarray}
  \Phi_{\, ,\jvec}^{(0,2)}(t,t') & = & H_{j_2-j_1}''(2t') 
  \label{equ:phicor02} \\[1ex]
  \Phi_{i,j}^{(1,1)}(t,t') & = & \me^{-\dt} \, I_{i-j}(\gamma \dt) + 
  \mathcal{E}_{i,j}^{\,j}(t,t') 
  \label{equ:phicor11} \\[1ex]
  \Phi_{\ivec,\jvec}^{(2,2)}(t,t') & = & 
  \mathcal{F}_{i_1,i_2}^{\,\jvec}(t,t') \, H_{j_2-j_1}''(2t') 
  \label{equ:phicor22} \\
  & - & 
  \big[+\me^{-\dt} \, I_{i_1-j_1}(\gamma \dt) + 
    \mathcal{E}_{i_1,j_1}^{\,\jvec}(t,t') \big] 
  \big[-\me^{-\dt} \, I_{i_2-j_2}(\gamma \dt) + 
    \mathcal{E}_{i_2,j_2}^{\,\jvec}(t,t') \big] \nonumber \\
  & + & 
  \big[-\me^{-\dt} \, I_{i_1-j_2}(\gamma \dt) + 
    \mathcal{E}_{i_1,j_2}^{\,\jvec}(t,t') \big] 
  \big[+\me^{-\dt} \, I_{i_2-j_1}(\gamma \dt) + 
    \mathcal{E}_{i_2,j_1}^{\,\jvec}(t,t') \big] \nonumber
\end{eqnarray}

Substituting the result (\ref{equ:cor2phisolution}) into 
(\ref{equ:gsolution}) finally delivers analytic expressions 
for the desired two-time correlation functions. We 
have worked out $C^{(1,1)}(t,t'), C^{(2,2)}(t,t'), 
C^{(1,3)}(t,t'), C^{(3,1)}(t,t')$ and $C^{(3,3)}(t,t')$ 
and it appears that a factorisation similar 
to the one which simplified (\ref{equ:corquenchbig}) to
(\ref{equ:corquench}) is generally possible. 
We do not have a proof for this claim, however, and therefore 
take (\ref{equ:gsolution}) together with (\ref{equ:cor2phisolution}) 
as our final result, which expresses an arbitrary multispin two-time
correlation function as a sum over a finite number of terms. Examples
of low order multispin two-time 
correlations, expressed in terms of homogeneous solutions 
via (\ref{equ:gsolution}), read 
\begin{eqnarray}
  C_{i,j}^{(1,1)}(t,t') & = & \Phi_{i,j}^{(1,1)}(t,t') 
  \label{equ:cor11} \\
  C_{\ivec,\jvec}^{(2,2)}(t,t') & = & \Phi_{\ivec,\jvec}^{(2,2)}(t,t') + 
    H_{i_2-i_1}(2\dt) \, \Phi_{\, ,\jvec}^{(0,2)}(t,t') 
  \label{equ:cor22} 
\end{eqnarray}
In Sec.~\ref{sec:examples} below we present a discussion the 
correlation functions $C^{(1,1)}(t,t')$ and $C^{(2,2)}(t,t')$. 
In particular, we demonstrate how the expressions 
(\ref{equ:cor11}) and (\ref{equ:cor22}) simplify when using 
the results of Appendix~\ref{sec:EF} to rewrite the sums (\ref{equ:Esum}) 
and (\ref{equ:Fsum}).

\section{Two-time responses after a quench}
\label{sec:responsequench}

Quite analogously to the case of two-time correlations discussed
above, two-time multispin
response functions $R^{(k,l)}(t,t')$ follow from the 
general solution (\ref{equ:gsolution}) with time 
arguments replaced by $\Delta=t-t'$. The equal 
time initial conditions $R^{(k,l)}(t',t')$ are given in 
terms of correlation functions in (\ref{equ:responseinit}). 
We consider again a quench from equilibrium as introduced in 
Sec.~\ref{sec:correlquench}. Consequently, on all 
levels $k$ with odd $k+l$ the initial conditions 
(\ref{equ:responseinit}) vanish and so do the $R^{(k,l)}(t,t')$ 
for all $t \geq t'$. Thus we immediately focus on 
$k+l$ even. Without loss of generality we restrict 
the vector $\jvec$, labelling the sites to which the 
perturbation couples, to ordered $\jvec \in N(l)$. 

Equation (\ref{equ:responseinit}) comprises a correlation 
function over the sites $\ivec \cup \jvec^\nu$, where 
$\jvec^\nu$ is the index vector $\jvec$ with the additional 
components $j_\nu-1$ and $j_\nu+1$ as defined below 
(\ref{equ:responseinit}). Since $\jvec$ is fixed we can
immediately use the hierarchical property (\ref{equ:links}) 
of correlation functions to remove component pairs from 
$\jvec^\nu$ in cases where $\jvec^\nu \notin N(l+2)$. 
More precisely this corresponds to replacing $\jvec^\nu$ in 
(\ref{equ:responseinit}) by the ``squashed'' index vector 
\begin{equation}
  \jvec^{\nu,\mathrm{s}} = \left\{
  \begin{array}{lcccr}
    (j_1,\ldots j_{\nu-1} &,j_\nu-1 &,j_\nu &,j_\nu+1 &,j_{\nu+1},\ldots j_l)\\
    (j_1,\ldots j_{\nu-2} &         &,j_\nu &,j_\nu+1 &,j_{\nu+1},\ldots j_l)\\
    (j_1,\ldots j_{\nu-1} &,j_\nu-1 &,j_\nu &         &,j_{\nu+2},\ldots j_l)\\
    (j_1,\ldots j_{\nu-2} &         &,j_\nu &         &,j_{\nu+2},\ldots j_l) 
  \end{array}
  \right.
  \quad \!\! \mbox{for} \quad 
  \begin{array}{c}
    j_{\nu-1}<j_\nu-1, j_\nu+1<j_{\nu+1} \\
    j_{\nu-1}=j_\nu-1, j_\nu+1<j_{\nu+1} \\
    j_{\nu-1}<j_\nu-1, j_\nu+1=j_{\nu+1} \\
    j_{\nu-1}=j_\nu-1, j_\nu+1=j_{\nu+1} 
  \end{array}
  \label{equ:jnus}
\end{equation}
We denote the dimension of $\jvec^{\nu,\mathrm{s}}$ by 
$l^{\nu,\mathrm{s}} \in \{ l-2,l,l+2\}$. Having made this 
modification in (\ref{equ:responseinit}), which guarantees 
that $\jvec^{\nu,\mathrm{s}} \in N(l^{\nu,\mathrm{s}})$ is 
ordered, we may now use (\ref{equ:cor2initbar}) to write 
down the initial condition $\bar{A}$ corresponding to 
(\ref{equ:responseinit}); recall that (\ref{equ:cor2initbar}) 
is the permutationally antisymmetric version of 
(\ref{equ:correl2init}). This gives for the permutationally
antisymmetrized initial condition for $R_{\ivec,\jvec}^{(k,l)}(t,t')$  
\begin{eqnarray}
  \bar{A}_{\ivec,\jvec}^{(k,l)}(t') & = &\sum_{\mu=1}^k \sum_{\nu=1}^l 
  \delta_{i_\mu,j_\nu} \left[ \left(1-\frac{\gamma^2}{2}\right) 
  \sum_{\pi \in \mathcal{P}(\frac{k+l}{2})} \sign 
  \prod_{\lambda=1}^{\frac{k+l}{2}} 
  \mathcal{U}_{(\ivec \cup \jvec)_{\pi(2\lambda-1)},
  (\ivec \cup \jvec)_{\pi(2\lambda)}}^{\, \jvec}(t') \right. \nonumber \\
  & - & \left. \frac{\gamma^2}{2}   
  \sum_{\pi \in \mathcal{P}(\frac{k+l^{\nu,\mathrm{s}}}{2})} \sign 
  \prod_{\lambda=1}^{\frac{k+l^{\nu,\mathrm{s}}}{2}} 
  \mathcal{U}_{(\ivec \cup \jvec^{\nu,\mathrm{s}})_{\pi(2\lambda-1)},
  (\ivec \cup \jvec^{\nu,\mathrm{s}})_{\pi(2\lambda)}}^{\, 
  \jvec^{\nu,\mathrm{s}}}(t') \right]. 
  \label{equ:responseinitbar1}
\end{eqnarray}
The fact that the Kronecker deltas in (\ref{equ:responseinitbar1}) 
contain components of $\ivec$ does not affect permutational 
antisymmetry: the only occurrence of the summation variable 
$\mu$ in (\ref{equ:responseinitbar1}) is in $\delta_{i_\mu,j_\nu}$. 
But finite sums are independent of the order in which they are 
taken, hence permutational antisymmetry is unaffected. 

We now 
note that for each $\mu,\nu$ the Kronecker delta in 
(\ref{equ:responseinitbar1}), enforcing $i_\mu = j_\nu$, 
allows us to simplify the expressions inside the square brackets. As 
discussed in the text below Eqs.~(\ref{equ:cor2initbar},
\ref{equ:Ucal}), the products over $\lambda$ are zero for 
all pairings that do not contain the pair $(i_\mu,j_\nu)$. So for 
each $\mu,\nu$ it is sufficient to restrict the sums over the 
ordered pairings in (\ref{equ:responseinitbar1}) to those that do 
contain the pair $(i_\mu,j_\nu)$. For the pairings drawn from 
$\ivec \cup \jvec$, the corresponding factor $\mathcal{U}$ with 
index $(a,b)=(i_\mu,j_\nu)$ is $(-1)^{\nu-1}$ according to 
(\ref{equ:Ucal}). But for the pairings drawn from $\ivec \cup 
\jvec^{\nu,\mathrm{s}}$ this factor is $-(-1)^{\nu-1}$ since 
$j_\nu$ is component number $\nu+1$ or $\nu-1$ of 
$\jvec^{\nu,\mathrm{s}}$ as can be seen from (\ref{equ:jnus}). 
Thus we may rewrite (\ref{equ:responseinitbar1}) as 
\begin{eqnarray}
  \bar{A}_{\ivec,\jvec}^{(k,l)}(t') & = &\sum_{\mu=1}^k \sum_{\nu=1}^l 
  (-1)^{\nu-1} \, \delta_{i_\mu,j_\nu} \left[\left(1-\frac{\gamma^2}{2}\right) 
  \sum_{\pi \in \mathcal{P'}(\frac{k+l}{2})} \sign 
  \prod_{\lambda=1}^{\frac{k+l}{2}} 
  {\mathcal{U}'}_{(\ivec \cup \jvec)_{\pi(2\lambda-1)},
  (\ivec \cup \jvec)_{\pi(2\lambda)}}^{\, \jvec}(t') \right. \nonumber \\
  & + & \left. \frac{\gamma^2}{2}   
  \sum_{\pi \in \mathcal{P'}(\frac{k+l^{\nu,\mathrm{s}}}{2})} \sign 
  \prod_{\lambda=1}^{\frac{k+l^{\nu,\mathrm{s}}}{2}} 
  {\mathcal{U}'}_{(\ivec \cup \jvec^{\nu,\mathrm{s}})_{\pi(2\lambda-1)},
  (\ivec \cup \jvec^{\nu,\mathrm{s}})_{\pi(2\lambda)}}^{\, 
  \jvec^{\nu,\mathrm{s}}}(t') \right]. 
  \label{equ:responseinitbar}
\end{eqnarray}
Here $\mathcal{P}'$ emphasises that we only sum over the pairings 
that contain $(i_\mu,j_\nu)$ for each given $\mu,\nu$. Since we have 
pulled the factors $\pm (-1)^{\nu-1}$ --- contained in each product in 
(\ref{equ:responseinitbar}) --- out of the square brackets we have to 
replace $\mathcal{U}$ by $1$ for $(a,b)=(i_\mu,j_\nu)$, and indicated
this by writing $\mathcal{U}'$ instead of $\mathcal{U}$. 

From the permutationally antisymmetric initial condition 
(\ref{equ:responseinitbar}) that corresponds to (\ref{equ:responseinit}) 
one easily obtains the homogeneous solutions for two-time multispin 
response functions via (\ref{equ:solutionphibar}). We use 
the fact that according to our construction of (\ref{equ:responseinitbar}) 
the entire expression inside the square brackets is independent of $i_\mu$ 
for each fixed $\mu$. So the summation over this index is trivial. The 
sum over the remaining indices $\ivec \setminus (i_\mu)$ follows in 
full analogy to the derivation of (\ref{equ:cor2phisolution}) and 
consequently we obtain 
\begin{eqnarray}
  \Phi_{\ivec,\jvec}^{(k,l)}(t,t') & = & \sum_{\mu=1}^k \sum_{\nu=1}^l 
  (-1)^{\nu-1} \, \me^{-\dt} \, I_{i_\mu-j_\nu}(\gamma \dt) \, \nonumber \\
  & \times &
  \left[\left(1-\frac{\gamma^2}{2}\right) 
  \sum_{\pi \in \mathcal{P'}(\frac{k+l}{2})} \sign 
  \prod_{\lambda=1}^{\frac{k+l}{2}} 
  {\mathcal{V}'}_{(\ivec \cup \jvec)_{\pi(2\lambda-1)},
  (\ivec \cup \jvec)_{\pi(2\lambda)}}^{\, \jvec}(t,t') \right. \nonumber \\
  & + & \left. \frac{\gamma^2}{2}   
  \sum_{\pi \in \mathcal{P'}(\frac{k+l^{\nu,\mathrm{s}}}{2})} \sign 
  \prod_{\lambda=1}^{\frac{k+l^{\nu,\mathrm{s}}}{2}} 
  {\mathcal{V}'}_{(\ivec \cup \jvec^{\nu,\mathrm{s}})_{\pi(2\lambda-1)},
  (\ivec \cup \jvec^{\nu,\mathrm{s}})_{\pi(2\lambda)}}^{\, 
  \jvec^{\nu,\mathrm{s}}}(t,t') \right].
  \label{equ:responsephisolution} 
\end{eqnarray}
Here $\mathcal{V}'=1$ for indices $(a,b)=(i_\mu,j_\nu)$ and 
$\mathcal{V}'=\mathcal{V}$ as given by (\ref{equ:Vcal}) otherwise. 
As in the case of two-time correlations we see that the homogeneous solutions 
for {\em arbitrary} two-time multispin response functions can be expressed 
entirely in terms of $I,H'',\mathcal{E}$ and $\mathcal{F}$ according 
to (\ref{equ:responsephisolution}). At level $k=0$, 
the sum over $\mu$ in (\ref{equ:responsephisolution}) is empty and so
we set 
$\Phi^{(0,l)}=A^{(0,l)}=0$ as discussed at the end of 
Sec.~\ref{sec:response}. In Appendix~\ref{sec:res1122}
we explicitly demonstrate how the first nontrivial 
homogeneous solutions (\ref{equ:phires11}) and (\ref{equ:phires22}) 
are obtained from the general result (\ref{equ:responsephisolution}): 
\begin{eqnarray}
  \Phi_{\, , \jvec}^{(0,2)} \!\! & = & 0 
  \label{equ:phires02} \\[1ex]
  \Phi_{i,j}^{(1,1)} \!\! & = & 
  \me^{-\dt} \, I_{i-j} 
  \left[ {\textstyle \left( 1-\frac{\gamma^2}{2} \right)} - 
  {\textstyle\frac{\gamma^2}{2}} H_{2}''(2t') \right] 
  \label{equ:phires11} \\[1ex]
  \Phi_{\ivec,\jvec}^{(2,2)} \!\! & = & \me^{-\dt} \, 
  I_{i_1-j_1} \left[ - {\textstyle \left(1-\frac{\gamma^2}{2} \right) }
  \left( - \me^{-\dt} \, I_{i_2 -j_2} + 
  \mathcal{E}_{i_2,j_2}^{\,(j_1,j_2)} \right) + 
  {\textstyle \frac{\gamma^2}{2} }
  \left( + \me^{-\dt} \, I_{i_2 -j_1+1} + 
  \mathcal{E}_{i_2,j_1-1}^{\,(j_1-1,j_1)} \right)
  \right] \nonumber \\ 
  & + & \me^{-\dt} \, 
  I_{i_2-j_1} \left[ + {\textstyle \left(1-\frac{\gamma^2}{2} \right) }
  \left( - \me^{-\dt} \, I_{i_1 -j_2} + 
  \mathcal{E}_{i_1,j_2}^{\,(j_1,j_2)} \right)  - 
  {\textstyle \frac{\gamma^2}{2} }
  \left( + \me^{-\dt} \, I_{i_1 -j_1+1} + 
  \mathcal{E}_{i_1,j_1-1}^{\,(j_1-1,j_1)} \right)
  \right] \nonumber \\ 
  & - & \me^{-\dt} \, 
  I_{i_1-j_2} \left[ + {\textstyle \left(1-\frac{\gamma^2}{2} \right) }
  \left( + \me^{-\dt} \, I_{i_2 -j_1} + 
  \mathcal{E}_{i_2,j_1}^{\,(j_1,j_2)} \right) - 
  {\textstyle \frac{\gamma^2}{2} }
  \left( - \me^{-\dt} \, I_{i_2 -j_2-1} + 
  \mathcal{E}_{i_2,j_2+1}^{\,(j_2,j_2+1)} \right)
  \right] \nonumber \\ 
  & - & \me^{-\dt} \, 
  I_{i_2-j_2} \left[ - {\textstyle \left(1-\frac{\gamma^2}{2} \right)} 
  \left( + \me^{-\dt} \, I_{i_1 -j_1} + 
  \mathcal{E}_{i_1,j_1}^{\,(j_1,j_2)} \right) + 
  {\textstyle \frac{\gamma^2}{2} }
  \left( - \me^{-\dt} \, I_{i_1 -j_2-1} + 
  \mathcal{E}_{i_1,j_2+1}^{\,(j_2,j_2+1)} \right)
  \right] 
  \label{equ:phires22}
\end{eqnarray}
To save space we have omitted the arguments of 
the functions $I$ and $\mathcal{E}$ which are obviously 
$\gamma \Delta$ and $(t,t')$, respectively. The expression 
(\ref{equ:phires22}) for $\Phi^{(2,2)}$ only holds if $\jvec$ 
is of the form $\jvec=(j,j+1)$; for $j_1+1<j_2$
the result (\ref{equ:responsephisolution}) gives a more complicated form.

As usual, actual two-time response functions may be expressed 
in terms of homogeneous solutions (\ref{equ:responsephisolution}) 
via (\ref{equ:gsolution}). It is not clear to us whether a 
factorization of the combined expression (\ref{equ:gsolution}), 
(\ref{equ:responsephisolution}) is possible. So we state our 
results explicitly in terms of homogeneous solutions, as for instance 
\begin{eqnarray}
  R_{i,j}^{(1,1)}(t,t') & = & \Phi_{i,j}^{(1,1)}(t,t') 
  \label{equ:res11} \\
  R_{\ivec,\jvec}^{(2,2)}(t,t') & = & \Phi_{\ivec,\jvec}^{(2,2)}(t,t')  + 
    H_{i_2-i_1}(2\dt) \, \Phi_{\, ,\jvec}^{(0,2)}(t,t') 
  \label{equ:res22}
\end{eqnarray}
The responses $R^{(1,1)}(t,t')$ and $R^{(2,2)}(t,t')$ are discussed 
further in the subsequent section. In particular, we show that the 
identities for $\mathcal{E}$ given in Appendix~\ref{sec:EF} yield 
a simplification of the expression (\ref{equ:res22}).

\section{Simple Examples of Two-Time Functions}
\label{sec:examples}

In this section we illustrate the procedure of extracting 
explicit expressions for multispin two-time correlation and 
response functions from the general solutions derived in 
Secs.~\ref{sec:correl2quench} and~\ref{sec:responsequench}. 
For some of the examples we also give a discussion of their
implication for the physics of the Glauber-Ising chain.

The simplest possible two-time correlation functions comprise 
only one spin at each time. By combining (\ref{equ:phicor11}), 
(\ref{equ:cor11}) and rewriting the sum $\mathcal{E}$ using 
(\ref{equ:Eodd}) we obtain 
\begin{equation}
  C_{i,j}^{(1,1)}(t,t') = \langle \sigma_i(t) \, \sigma_j(t') \rangle = 
  \me^{-\dt} \, I_{i-j}(\gamma \dt) + 
  \widetilde{H}_{j-i}''(\dt,2t'). 
  \label{equ:cor11general}
\end{equation}
Expressions for the functions $I_n(x)$, $\tilde{H}_n''(t_1,t_2)$ 
are given in (\ref{equ:I}), (\ref{equ:H2prime2ttilde}). Since the 
two-time correlations (\ref{equ:cor11general}) are the solutions 
of (\ref{equ:correl2}) at level $k=1$, the set $Q=Q_{\odd}$ 
must be used in (\ref{equ:I}).
The result (\ref{equ:cor11general}) applies for the finite 1$d$ 
Glauber-Ising model quenched from an equilibrium state at some 
$T_{\upi}>0$ to an arbitrary temperature $T \geq 0$ at time $t=0$. 
According to the discussion of the functions $I$ and $H$ in 
Appendices~\ref{sec:bessel} and~\ref{sec:Hn}, taking the thermodynamic 
limit $N \to \infty$ and quenching from a random initial 
configuration $T_{\upi} = \infty$ just amounts to replacing 
$I_{j-i}$ and $\tilde{H}_{j-i}''$ in (\ref{equ:cor11general}) 
with $\In_{j-i}$ and $\tilde{\Hn}_{j-i}$, respectively. 
Writing out $\tilde{\Hn}_{j-i}(\dt,2t')$ explicitly 
using (\ref{equ:H2prime2ttilde}) we thus have 
\begin{equation}
  C_{i,j}^{(1,1)}(t,t') = \me^{-\dt} \, \In_{i-j}(\gamma \dt) + 
  \sgn(j-i) \, \Hn_{j-i}(\dt, 
  2t') + \gamma \int_0^{\dt} \!\!  \dd \tau \, \me^{-\tau} \, 
  \In_{j-i}(\gamma \tau) \, \Hn_1(\dt-\tau,2t'). 
  \label{equ:cor11gamma}
\end{equation}
If we additionally quench the system to $T=0$ then (\ref{equ:cor11general}) 
together with the zero temperature formula (\ref{equ:H2ttildezero}) for 
$\tilde{\Hn}_n(t_1,t_2)$ simply yields 
\begin{equation}
  C_{i,j}^{(1,1)}(t,t') = \me^{-(t+t')} \left\{ \In_{i-j}(t+t') + 
  \!\! \int_0^{2t'} \!\!\!\! \dd \tau \, \In_{i-j}(t+t'-\tau) \, 
  \left[ \In_0 + \In_1 \right](\tau) \right\}.
  \label{equ:cor11zero}
\end{equation}
Here we have introduced the short hand $[ \, \cdot \, ](x)$ to 
indicate that all functions contained in the square bracket have the 
same argument $x$. The functions $\In_n(x)$ in (\ref{equ:cor11gamma}), 
(\ref{equ:cor11zero}) are the modified Bessel functions (\ref{equ:In}).

Similarly we obtain the two-time spin response functions to a 
local magnetic field $h_j$ in the non-equilibrium state after 
the quench -- scaled by $T$ according to our definition in 
Sec.~\ref{sec:response} -- by putting together 
(\ref{equ:phires11}), (\ref{equ:res11}) 
\begin{equation}
  R_{i,j}^{(1,1)}(t,t') = T \left. \frac{\delta \, \langle \sigma_i(t) \rangle}
  {\delta \, h_j(t')} \right|_{h_j=0} = \me^{-\dt} \, I_{i-j}(\gamma \dt)  
  \left[ 1-\frac{\gamma^2}{2} \left(1 + H_{2}''(2t') \right) \right]. 
  \label{equ:res11general}
\end{equation}
$I_n(x)$ and $H_n''(t)$ are given in (\ref{equ:I}) (with $Q=Q_{\odd}$)
and (\ref{equ:Hn2prime}). Taking the thermodynamic limit $N \to \infty$ 
and letting $T_{\upi} = \infty$ turns (\ref{equ:res11general}) into 
\begin{equation}
  R_{i,j}^{(1,1)}(t,t') = \me^{-\dt} \, \In_{i-j}(\gamma \dt)  
  \left[ 1-\frac{\gamma^2}{2} \left(1 + \Hn_{2}(2t') \right) \right].  
  \label{equ:res11gamma}
\end{equation}
For the zero temperature quench (\ref{equ:res11gamma}) may be 
simplified further by using (\ref{equ:H2tzero}), giving 
\begin{equation}
  R_{i,j}^{(1,1)}(t,t') = \frac{1}{2} \, \me^{-(t+t')} \, \In_{i-j}(\dt)  
  \left[\In_0 + 2\In_1 + \In_2 \right](2t'). 
  \label{equ:res11zero}
\end{equation}

The exact solutions (\ref{equ:cor11gamma}), 
(\ref{equ:res11gamma}) agree, as they should, with their temporal
Laplace (with respect to $\Delta,t'$) and spatial Fourier (with
respect to $n=i-j$) transforms as derived in 
\cite{GodLuc00}. While the latter are surprisingly simple for the 
functions (\ref{equ:cor11gamma}) and (\ref{equ:res11gamma}), 
a corresponding transformation of higher order functions is 
virtually impossible. The correlations (\ref{equ:cor11gamma}) 
and a modified version of the responses (\ref{equ:res11gamma}), 
slightly differing in the way the transition rates in the presence of
a field are defined, were also studied in \cite{LipZan00}; the 
long-time scaling form of the correlation functions can equivalently 
be found using a picture of annihilating random walkers 
\cite{FonIsoNewStein01}.

Both publications \cite{GodLuc00, LipZan00} considered the 
non-equilibrium fluctuation-dissipation relations associated 
with the auto-correlation and response functions. For any pair of
correlation and response functions one can define a
fluctuation-dissipation theorem (FDT) ``violation'' factor $X$ via
\begin{equation}
R(t,\tw) = X(t,\tw) \frac{\partial}{\partial\tw} C(t,\tw)
\label{equ:Xdef}
\end{equation}
so that $X(t,\tw)=1$ corresponds to the usual equilibrium FDT (recall
that our response function contains an extra factor of $T$; in the
conventional definition, (\ref{equ:Xdef}) would have $X/T$ on the
r.h.s.\ instead of $X$). Under appropriate conditions, $T/X$ can then
be interpreted as an effective temperature $T_{\rm eff}$; this
interpretation is well-established for mean-field spin glass models
but its extension away from mean-field remains open and has prompted
much recent work~\cite{CugKur93,CugKurPel97,CriRit03}.

\begin{figure}[htb]
  \begin{picture}(15,7)
    \put(0,0){\epsfig{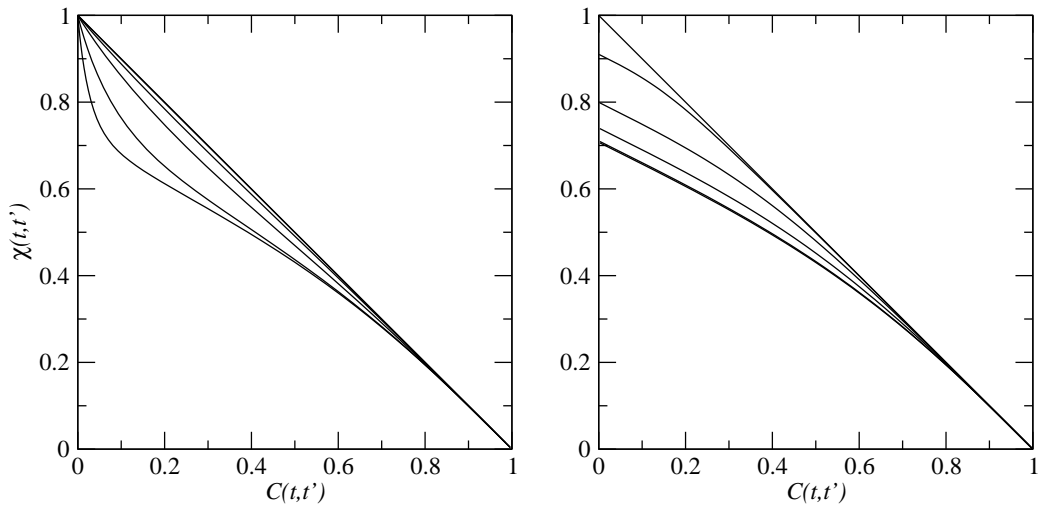}}
  \end{picture}
  \caption{\label{fig:FDT} FD plots of the local spin-spin correlation
    and response functions, for fixed $\tw$ (left) and fixed $t$ 
    (right). The values of $\tw/\taueq$ are $\tw/\taueq = 10^{-3},
    10^{-2}, 10^{-1},10^{-1/2},10^0,10^1$ from bottom to top on the 
    left; the same values are used for $t/\taueq$ on the right. See 
    text for discussion.}
\end{figure}

In \cite{FDT} we have given a detailed discussion of the FDT
violations that result for non-local correlation and response
functions, extending the work of \cite{GodLuc00, LipZan00} for the
local quantities. Here we therefore only illustrate one aspect of our
results, by considering the local quantities obtained for $i=j$ from
(\ref{equ:cor11gamma}), (\ref{equ:res11gamma}) for a quench to
nonzero temperature. When all times $t,\tw$ and $\taueq=1/(1-\gamma)$
are large, one expects results that depend only on the time ratios
$t/\tw$ and $\tw/\taueq$; this is indeed what we find,
in agreement with the results given in \cite{LipZan00}. 
A convenient way of representing the results is a parametric
fluctuation-dissipation (FD) plot of the integrated response function
against the correlation function; the former is defined as
\begin{equation}
\chi(t,\tw) = \int_{\tw}^t \dd\tau \, R(t,\tau)\ ,
\end{equation}
and gives the response to a field switched on at $\tw$ and held
constant after that. One notes from
(\ref{equ:Xdef}) and $R(t,\tw) = -(\partial/\partial\tw) \chi(t,\tw)$,
that the negative slope of such a plot is guaranteed to give
$X(t,\tw)$ only if $\tw$ is used as the curve parameter while $t$ is held
fixed~\cite{SolFieMay02}. This gives the FD plots in
Fig.~\ref{fig:FDT}(right); on the left we show the FD plot reported in
\cite{LipZan00}, which was produced with the opposite convention of
keeping $\tw$ fixed and varying $t$ along the curves.
We see that the plot on the right produces a
physically far more plausible representation of the crossover from the
aging regime 
$\tw\ll\taueq$ to equilibrium $\tw\gg\taueq$; in particular, one reads
off that the FDT violation factor $X(t,\tw)$ is always $\leq 1$ and
its value in the limit $C\to 0$ evolves smoothly from 1/2 to 1 as the
ratio $t/\taueq$ grows from 0 to $\infty$.

Among the higher order correlation functions, one of primary 
interest is the domain wall correlation. 
The domain wall indicator (or defect variable) $n_i=\frac{1}{2} \, 
(1-\sigma_i \, \sigma_{i+1}) \in \{0,1\}$ measures the 
presence ($n_i=1$) of a domain wall between sites $i$ and $i+1$. 
It is a well known fact \cite{Spouge88,AmaFam90,Santos97} 
that Glauber dynamics of the (unmagnetised) spin system 
$\bsig$ corresponds to the diffusion limited reaction process 
$A+A \rightleftharpoons \emptyset$ 
in terms of $\boldsymbol{n}=(n_1,\ldots n_N)$; here we interpret 
$n_i$ as the particle occupation number of site $i$. The relevant 
rates follow immediately from (\ref{equ:rates}): the transitions 
$\ldots \uparrow \downarrow \downarrow \ldots \rightleftharpoons 
\ldots \uparrow \uparrow \downarrow \ldots$ for spins, where 
$\uparrow$ and $\downarrow$ symbolise $\sigma_i=+1$ and $\sigma_i=-1$, 
respectively, occur with rate $1/2$ and correspond to diffusion 
$\ldots  1 0  \ldots \rightleftharpoons \ldots 0 1 \ldots$ of 
particles. The process $\ldots \uparrow \downarrow \uparrow 
\ldots \rightleftharpoons \ldots \uparrow \uparrow \uparrow \ldots$,
on the other hand, which maps onto
$\ldots 1 1 \ldots \rightleftharpoons \ldots 0 0 \ldots$, 
gives the rate $(1+\gamma)/2$ for particle annihilation and the 
$(1-\gamma)/2$ for pair creation. 
For zero temperature dynamics, where $\gamma=1$, this obviously reduces 
to diffusion limited pair annihilation (DLPA). In the context of the 
Glauber-Ising model, understanding the dynamics of domain walls or 
defects is crucial for e.g.\ explaining the coarsening process after 
a quench. Here our general solutions for multispin two-time correlation 
and response functions given in the preceding sections are the key 
for a comprehensive analysis. Due to the mapping to diffusion 
reaction processes, our results simulatenously describe multi-particle 
two-time correlation and response functions in the latter systems.

As a simple example we now consider the two-time connected correlation 
function between two domain walls, or equivalently particles,  
$C_{i-j}(t,t')= 4 [ \langle n_i(t) \, n_j(t') \rangle - 
\langle n_i(t) \rangle \, \langle n_j(t') \rangle ]$. 
In terms of spins this gives 
\begin{equation}
  C_{i-j}(t,t') = \langle \sigma_i(t) \, \sigma_{i+1}(t) \, 
  \sigma_j(t') \, \sigma_{j+1}(t') \rangle - 
  \langle \sigma_i(t) \, \sigma_{i+1}(t) \rangle \, 
  \langle \sigma_j(t') \, \sigma_{j+1}(t') \rangle. 
  \label{equ:cor22def}
\end{equation}
Now consider the coarsening dynamics after quenching the system 
from an equilibrium state at $T_{\upi} > 0$ to some temperature 
$T \geq 0$. The two-spin correlations in (\ref{equ:cor22def}) 
follow immediately from (\ref{equ:corquench}), i.e.\ 
$\langle \sigma_{i}(t) \, \sigma_{i+1}(t) \rangle = H_1''(2t)$.
The non-trivial term in (\ref{equ:cor22def}) is the two-time four-spin 
correlation function. Combining the results (\ref{equ:phicor02}), 
(\ref{equ:phicor22}) and (\ref{equ:cor22}) we may 
write any such correlation as 
\begin{eqnarray}
  C_{\ivec,\jvec}^{(2,2)}(t,t') & = & 
  \big[ H_{i_2-i_1}(2\dt) + \mathcal{F}_{i_1,i_2}^{\,\jvec}(t,t') \big] 
  H_{j_2-j_1}''(2t') 
  \label{equ:cor22general} \\
  & - & 
  \big[+\me^{-\dt} \, I_{i_1-j_1}(\gamma \dt) + 
    \mathcal{E}_{i_1,j_1}^{\,\jvec}(t,t') \big] 
  \big[-\me^{-\dt} \, I_{i_2-j_2}(\gamma \dt) + 
    \mathcal{E}_{i_2,j_2}^{\,\jvec}(t,t') \big] \nonumber \\
  & + & 
  \big[-\me^{-\dt} \, I_{i_1-j_2}(\gamma \dt) + 
    \mathcal{E}_{i_1,j_2}^{\,\jvec}(t,t') \big] 
  \big[+\me^{-\dt} \, I_{i_2-j_1}(\gamma \dt) + 
    \mathcal{E}_{i_2,j_1}^{\,\jvec}(t,t') \big]. \nonumber
\end{eqnarray}
To obtain the two-time correlation in the first term of (\ref{equ:cor22def}) 
we set $\ivec=(i,i+1)$ and $\jvec=(j,j+1)$. For these index vectors 
and bearing in mind
that $H_n''=-H_{-n}''$, we obtain from (\ref{equ:Eeven}) and
(\ref{equ:Feven}) the following
representations for the sums $\mathcal{E}$, $\mathcal{F}$ in 
(\ref{equ:cor22general}) 
\begin{eqnarray*}
  \mathcal{E}_{i_1,j_1}^{\,\jvec}(t,t') & = & H_{j-i}''(\dt,2t') + 
  \me^{-\dt} \, I_{i-j-1}(\gamma \dt) \, H_1''(2t') 
  \\
  \mathcal{E}_{i_1,j_2}^{\,\jvec}(t,t') & = & H_{j-i+1}''(\dt,2t') - 
  \me^{-\dt} \, I_{i-j}(\gamma \dt) \, H_1''(2t') 
  \\
  \mathcal{E}_{i_2,j_1}^{\,\jvec}(t,t') & = & H_{j-i-1}''(\dt,2t') + 
  \me^{-\dt} \, I_{i-j}(\gamma \dt) \, H_1''(2t') 
  \\
  \mathcal{E}_{i_2,j_2}^{\,\jvec}(t,t') & = & H_{j-i}''(\dt,2t') - 
  \me^{-\dt} \, I_{i-j+1}(\gamma \dt) \, H_1''(2t') 
  \\[1ex]
  \mathcal{F}_{i_1,i_2}^{\,\jvec}(t,t') & = & H_1''(2\dt,2t') 
  \\
  & - & \me^{-\dt} \left[ I_{i-j}(\gamma \dt) \, H_{i-j+1}''(2t') - 
    I_{i-j+1}(\gamma \dt) \, H_{i-j}''(2t') \right] \\
  & - & \me^{-\dt} \left[ I_{i-j-1}(\gamma \dt) \, H_{i-j}''(2t') - 
    I_{i-j}(\gamma \dt) \, H_{i-j-1}''(2t') \right] \\
  & + & \me^{-2\dt} I_{i-j}(\gamma \dt) \, I_{i-j}(\gamma \dt) \, 
    H_1''(2t') - \me^{-2\dt} I_{i-j-1}(\gamma \dt) \, 
    I_{i-j+1}(\gamma \dt) \, H_1''(2t')
\end{eqnarray*}
Substituting these expressions into (\ref{equ:cor22general}), using 
the identity (\ref{equ:Hpp_conversion}) to rewrite 
$H_n(2\dt)+H_n''(2\dt,2t')=H_n''(2t)$ and cancelling terms gives 
\begin{eqnarray}
  C_{\ivec,\jvec}^{(2,2)}(t,t') & = & 
  H_1''(2t) \, H_1''(2t') 
  \label{equ:cor22simplified} \\
  & + & 
  \big[ \makebox[5.2cm][l]{
  $\me^{-\dt} \, I_{i-j}(\gamma \dt)\hspace{2.3ex}+ H_{j-i}''(\dt,2t')$} \big]
  \big[ \makebox[5.2cm][l]{
  $\me^{-\dt} \, I_{i-j}(\gamma \dt)\hspace{2.3ex}- H_{j-i}''(\dt,2t')$} \big]
  \nonumber \\
  & - & 
  \big[ \makebox[5.2cm][l]{
  $\me^{-\dt} \, I_{i-j-1}(\gamma \dt) -  H_{j-i+1}''(\dt,2t')$} \big]
  \big[ \makebox[5.2cm][l]{
  $\me^{-\dt} \, I_{i-j+1}(\gamma \dt) +  H_{j-i-1}''(\dt,2t')$} \big]. 
  \nonumber 
\end{eqnarray}
Note that the first term on the r.h.s.\ of 
(\ref{equ:cor22simplified}), $H_1''(2t) \, H_1''(2t')$,
cancels with $\langle \sigma_i(t) \sigma_{i+1}(t) \rangle 
\langle \sigma_j(t') \sigma_{j+1}(t') \rangle$
when we substitute into (\ref{equ:cor22def}). 
The correlations (\ref{equ:cor22simplified}) are those 
in a finite ring of spins; since (\ref{equ:cor22simplified}) 
was obtained as the solutions of (\ref{equ:correl2}) on level $k=2$ 
the set $Q=Q_{\even}$ has to be used in the 
representation (\ref{equ:I}) for $I_n(x)$. The functions 
$H_n''(t_1,t_2)$ are given in (\ref{equ:H2prime2t}).
Now we focus on the quench from a random initial state 
$T_{\upi} = \infty$ and take the thermodynamic limit 
$N \to \infty$, thus replacing
$I_n(x)$ and $H_n''(t_1,t_2)$ 
in (\ref{equ:cor22simplified}) by $\In_n(x)$ and 
$\Hn_n(t_1,t_2)$, respectively. Substitution into 
(\ref{equ:cor22def}) then yields the two-time connected 
defect correlation function 
\begin{eqnarray}
  C_n(t,t') & = & 
  \big[ \makebox[4.7cm][l]{
  $\me^{-\dt} \, \In_n(\gamma \dt)\hspace{2.3ex}+ \Hn_n(\dt,2t')$} \big]
  \big[ \makebox[4.7cm][l]{
  $\me^{-\dt} \, \In_n(\gamma \dt)\hspace{2.3ex}- \Hn_n(\dt,2t')$} \big]
  \label{equ:cor22gamma} \\
  & - & 
  \big[ \makebox[4.7cm][l]{
  $\me^{-\dt} \, \In_{n+1}(\gamma \dt) -  \Hn_{n+1}(\dt,2t')$} \big]
  \big[ \makebox[4.7cm][l]{
  $\me^{-\dt} \, \In_{n-1}(\gamma \dt) +  \Hn_{n-1}(\dt,2t')$} \big].
  \nonumber 
\end{eqnarray}
For the $T=0$ quench (\ref{equ:cor22gamma}) may be simplified 
further using (\ref{equ:H2tzero}) 
\begin{eqnarray}
  C_n(t,t') & = & \me^{-(t+t')} \, [\In_{n-1}-\In_{n+1}](t+t') \, 
  \Hn_n(t-t',2t') \label{equ:cor22zero} \\ 
  & + &  
  \me^{-2t} \In_n(t-t') [ \In_{n-1}+2\In_n+\In_{n+1} ](t+t') - 
  \me^{-2(t+t')} [(\In_{n-1}+\In_n) (\In_n+\In_{n+1}) ](t+t'). 
  \nonumber 
\end{eqnarray}
For small $n$ the functions $\Hn_n(t_1,t_2)$ in (\ref{equ:cor22zero}) 
may be expressed purely in terms of modified Bessel functions 
$\In_n(x)$ via the recursion (\ref{equ:H2tzero}). For large $n$, 
on the other hand, it is more convenient to use
the representation (\ref{equ:Hn}), with $I_n(x)$ replaced by $\In_n(x)$ and
$\gamma=1$.

We have used the zero temperature result (\ref{equ:cor22zero}) 
in \cite{FDT} to study the non-equilibrium fluctuation-dissipation 
relations for defect observables $n_i$. Beyond this, we are 
not aware of any expressions equivalent to e.g.\ 
(\ref{equ:cor22gamma}) in the literature. 
As explained above, our results are of interest not only 
for the coarsening dynamics of the Glauber-Ising spin-chain, but also
directly give the corresponding particle-particle correlation and
response functions for the associated diffusion reaction process.

\begin{figure}[htb]
  \begin{picture}(15,5.5)
    \put(0,0){\epsfig{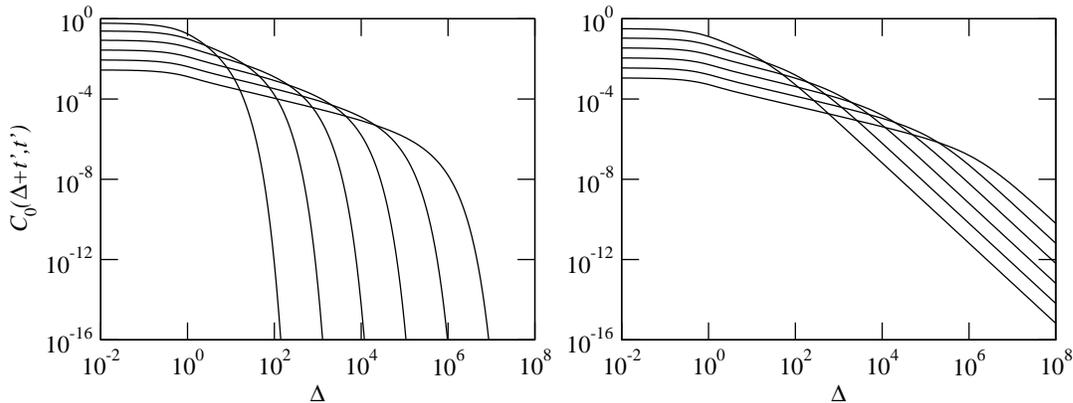}}
  \end{picture}
  \caption{\label{fig:autocor} Connected domain wall 
    autocorrelation functions $C_0(t,t')=4\left[\langle 
    n_i(t) n_i(t') \rangle-\langle n_i(t) \rangle 
    \langle n_i(t') \rangle \right]$ in equilibrium (left), i.e.\ 
    $t' \to \infty$,  and for the 
    coarsening dynamics after a zero temperature quench at $t=0$ 
    from a 
    random initial state (right). The different curves correspond 
    to decreasing temperature (left) and increasing system ``age'' $t'$ 
    (right) from top to bottom at small $\Delta=t-t'$. The plots 
    are obtained by numerical evaluation of 
    Eqs.~(\ref{equ:cor02teq},\ref{equ:cor02taging}) for 
    $\taueq,t'=10^1,10^2,\ldots,10^6$ and are discussed in the text.}
\end{figure}

From (\ref{equ:cor22gamma}) we can, for instance, derive an exact 
expression for the equilibrium domain-wall autocorrelation. This 
function corresponds to (four times) the particle autocorrelation 
in a diffusion limited reaction process $A + A \rightleftharpoons 
\emptyset$ at equilibrium. It follows from (\ref{equ:cor22gamma}) 
by setting $n=0$ and taking the limit $t' \to \infty$ with 
$\dt = t-t'$ fixed. This gives, using the representation (\ref{equ:Hn}) 
for $\Hn_n(\Delta,2t')$, 
\begin{equation}
  C_{\mathrm{eq}}(\dt) = \lim_{t' \to \infty} C_0(\dt+t',t') = 
  \me^{-2\dt} \In_0^{\,2}(\gamma \dt) - 
  \left\{ \me^{-\dt} \In_1(\gamma \dt) - \frac{\gamma}{2} 
  \int_{\dt}^{\infty} \dd \tau \, \me^{-\tau} [\In_0-\In_2](\gamma \tau) 
  \right\}^2.
  \label{equ:cor02teq}
\end{equation}
Expanding (\ref{equ:cor02teq}) asymptotically for 
$\dt,\taueq \to \infty$ at fixed
$\dt / \taueq$ gives the scaling forms 
\begin{eqnarray}
  1 \ll \dt \ll \taueq: & & C_{\mathrm{eq}}(\dt) \sim 
  \frac{2}{\sqrt{\pi \, \taueq \, \dt}} 
  \label{equ:cor02teqdiff} \\
  1 \ll \taueq \ll \dt: & & C_{\mathrm{eq}}(\dt) \sim 
  \frac{\taueq}{\pi \, \dt^2} \, \me^{-2 \, \dt/\taueq}
  \label{equ:cor02teqanni}
\end{eqnarray}
Plots of (\ref{equ:cor02teq}) for various $\taueq$ are 
shown in Fig.~\ref{fig:autocor}. In the regime $\Delta \ll \tau_{\eq}$ 
the particles may be considered as an ensemble of independent random 
walkers. Hence the decline of $C_{\mathrm{eq}}(\dt)$ corresponds 
to the return probability of a random walker. In the opposite 
regime, however, the particles are extremely likely to have been 
annihilated and replaced by new ones via the process $A+A 
\leftharpoondown \emptyset$. Since these new particles are
uncorrelated with the original ones, the connected correlation vanishes. 

For comparison we also consider the non-equilibrium case 
of diffusion limited pair-annihilation $A+A \rightarrow 
\emptyset$, for which the correlation function is given 
by the zero temperature formula (\ref{equ:cor22zero}). Setting 
$n=0$ again yields an exact expression for the particle 
autocorrelation 
\begin{equation}
  C_0(t,t') = 
  2 \, \me^{-2t} \, \In_0(\dt) \, [ \In_0+\In_1 ](t+t') - 
  \me^{-2(t+t')} \, [\In_0+\In_1]^2(t+t') 
  \label{equ:cor02taging}
\end{equation}
In (\ref{equ:cor02taging}) the time $t'$ plays a role analogous to
$\taueq$ in (\ref{equ:cor02teq}). So we expand 
for $\dt,t' \to \infty$ with $\dt/t'$ fixed. This gives 
the scaling forms 
\begin{eqnarray}
  1 \ll \dt \ll t': & & C_0(t,t') \sim \frac{2}{\pi \sqrt{2 \, t' \dt}} 
  \label{equ:cor02tagingdiff} \\
  1 \ll t' \ll \dt: & & C_0(t,t') \sim \frac{2 \, t'}{\pi \dt^2}
  \label{equ:cor02taginganni}
\end{eqnarray}
Plots of (\ref{equ:cor02taging}) are also shown in 
Fig.~\ref{fig:autocor}. For $\dt \ll t'$ we have a
situation that is exactly analogous to the equilibrium case, with
$\taueq$ replaced by $2\pi t'$.
In the opposite regime $\dt \gg t'$, however, the correlation function
decays much more slowly, now due to the annihilation of particles that
are very far apart rather than, as in equilibrium, the creation of new
particle pairs. We discuss in \cite{FDT} that the scaling form  
(\ref{equ:cor02taginganni}) is nontrivial and, for instance, gives 
rise to unanticipated nonequilibrium fluctuation-dissipation relations 
in the Glauber-Ising model.

The counterpart of the two-time connected domain wall correlation 
function (\ref{equ:cor22def}) is the two-time domain wall response 
function to local bond perturbations $h_{j,j+1}$ defined by
\begin{equation}
  R_{i-j}(t,t') = T 
  \left. \frac{\delta \, \langle \sigma_i(t) \, \sigma_{i+1}(t) \rangle }
  {\delta \, h_{j,j+1}(t')} \right|_{h_{j,j+1}=0}.
  \label{equ:res22def}
\end{equation}
In full analogy to the correlations we obtain $R_n(t,t')$ from
(\ref{equ:phires22}),(\ref{equ:res22}) and (\ref{equ:Eeven}), with
$\ivec=(i,i+1)$ and $\jvec=(j,j+1)$. For $N \to \infty$ and a quench 
from a random initial state $T_{\upi} = \infty$ we find
\begin{eqnarray}
  R_n(t,t') & = & \me^{-\dt} \, 
  \In_n \hspace{2.4ex} 
  \left[ - {\textstyle \left(1-\frac{\gamma^2}{2} \right) }
  \left( - \me^{-\dt} \, \In_n \hspace{2.3ex} + 
  \Hn_n \hspace{2.4ex} \right) + 
  {\textstyle \frac{\gamma^2}{2} }
  \left( + \me^{-\dt} \, \In_{n-2} + 
  \Hn_{n-2}  \right)
  \right] \nonumber \\ 
  & + & \me^{-\dt} \, 
  \In_{n-1} \left[ + {\textstyle \left(1-\frac{\gamma^2}{2} \right) }
  \left( - \me^{-\dt} \, \In_{n+1} + 
  \Hn_{n+1}  \right)  - 
  {\textstyle \frac{\gamma^2}{2} }
  \left( + \me^{-\dt} \, \In_{n-1} + 
  \Hn_{n-1}  \right)
  \right] \nonumber \\ 
  & - & \me^{-\dt} \, 
  \In_{n+1} \left[ + {\textstyle \left(1-\frac{\gamma^2}{2} \right) }
  \left( + \me^{-\dt} \, \In_{n-1} + 
  \Hn_{n-1}  \right) - 
  {\textstyle \frac{\gamma^2}{2} }
  \left( - \me^{-\dt} \, \In_{n+1} + 
  \Hn_{n+1}  \right)
  \right] \nonumber \\ 
  & - & \me^{-\dt} \, 
  \In_n \hspace{2.4ex} 
  \left[ - {\textstyle \left(1-\frac{\gamma^2}{2} \right)} 
  \left( + \me^{-\dt} \, \In_n \hspace{2.3ex} + 
  \Hn_n  \hspace{2.4ex} \right) + 
  {\textstyle \frac{\gamma^2}{2} }
  \left( - \me^{-\dt} \, \In_{n+2} + 
  \Hn_{n+2}  \right)
  \right]. 
  \label{equ:res22gamma}
\end{eqnarray}
All $\In_n$ in (\ref{equ:res22gamma}) have argument 
$\In_n(\gamma \dt)$ and all $\Hn_n=\Hn_n(\dt,2t')$. 
For the $T=0$ quench we may use the recursion (\ref{equ:H2tzero}) 
for $\Hn_n(t_1,t_2)$ to simplify this equation {\em drastically}. 
It turns out that the result may be written as
\begin{equation}
  R_n(t,t') = \frac{\partial}{\partial t'} \, \left\{ \me^{-2t} \, 
  \In_n(t-t') \left[ \In_{n-1}+2\In_n+\In_{n+1} \right](t+t') \right\}.
  \label{equ:res22zero}
\end{equation}

The functions $C_n$ (\ref{equ:cor22gamma}) and $R_n$ 
(\ref{equ:res22gamma}) are rather special: for
other index values, e.g.\ $(i,i+2)$, or in particular for higher 
orders $k,l$ the expressions become 
rather more complicated and are most efficiently evaluated
using symbolic software. As above, simplifications occur
for $N \to \infty$ and for a quench from $T_{\upi} = \infty$
to $T=0$. Any two-time multispin correlation and response 
function $C^{(k,l)}(t,t')$ and $R^{(k,l)}(t,t')$ with $k$ 
and $l$ even can then be 
expressed purely in terms of modified Bessel functions 
$\In_n(x)$, due to (\ref{equ:H2tzero}), (\ref{equ:Eeven}) and 
(\ref{equ:Feven}). It is clear from (\ref{equ:H2tzero}), of course,
that the number of 
$\In_n(x)$'s in the result grows with the index range covered 
by $\ivec,\jvec$.

\section{Conclusions}
\label{sec:conclusions}

We have presented a new approach for solving the full hierarchy of 
differential equations for mulitspin correlation and response functions in 
the finite one-dimensional Glauber-Ising model. Our result is 
the most explicit representation for these functions. The known 
results for equilibrium and dynamical correlation functions as 
well as two-time spin-spin correlation and response functions 
after a quench are easily recovered from our solution. Beyond 
that, however, we have derived closed expressions for arbitrary 
mutispin two-time correlation and response functions for the 
dynamics after a quench. We found that any such quantity 
can be expressed purely in terms of the four functions $I$, $H''$, 
$\mathcal{E}$ and $\mathcal{F}$, regardless of the number of 
spins involved. 

Our results for the two-time correlation and response functions formed
the basis for the study \cite{FDT}, where we gave a comprehensive
analysis of the nonequilibrium fluctuation-dissipation relations in
the Glauber-Ising model at zero temperature.  A number of other
applications of our results can be envisaged. For example, based on
the general expression for two-spin two-time correlations, we are
currently studying the existence of dynamical hereogeneities in the
Glauber-Ising model \cite{dynamicalhetero}.

Finally, we have emphasised the link between the Glauber-Ising chain
and diffusion-reaction processes. In particular, two-time
multi-particle correlation and response functions for a
one-dimensional diffusion-limited annihilation process with
appropriate rates follow immediately from our results. Using the exact
mapping between diffusion-annihilation and diffusion-coagulation
processes \cite{KrePfaWehHin95}, two-time solutions also follow for
the latter processes. We are currently exploring this link, with the
aim of e.g.\ obtaining exact scaling formulas for two-time correlation
and response functions in the one-spin facilitated
Fredrickson-Andersen model \cite{FreAnd84} at low temperatures. These
should help to clarify the meaning and extent of the apparent trivial
equilibrium fluctuation-dissipation behaviour found in numerical
simulations~\cite{BuhGar02b}.

\section{Acknowledgements}

We acknowledge financial support from the Austrian Academy of Sciences, 
the Wilhelm Macke Foundation and EPSRC Grant No.\ 00800822 (PM), and the 
Nuffield Foundation Grant No.\ NAL/00361/G (PS).

\begin{appendix}

\section{Projection of the Green's function}
\label{sec:greensproject}

Here we derive the identity
\begin{equation}
  \sum\limits_{\jvec \in N(k)} \Gk{k}{\ivec}{\jvec}{t} \, 
  b^{(k)}_{\jvec,\jvec'} = \sum\limits_{1\le\mu<\nu\le k} 
  (-1)^{\nu-\mu-1} \left(\frac{\dd}{\dd t} \, H_{i_\nu-i_\mu}(2t) \right) 
  \Gk{k-2}{\ivec \setminus (i_\mu,i_\nu)}{\jvec'}{t}
  \label{equ:Gb}
\end{equation}
that is used in Sec.~\ref{sec:recursion} for the inductive proof of 
(\ref{equ:Literations}). The functions $G^{(k)}(t)$ and $H_n(t)$ are 
given in (\ref{equ:greensfunction}) and (\ref{equ:H}), respectively, 
and $\ivec \in N(k)$, $\jvec' \in N(k-2)$ are ordered indices.
The matrix $b^{(k)}$, accounting for links between levels $k$ and 
$k-2$ in the hierarchy, is discussed in the text below 
(\ref{equ:matrixform}) and may be written as
\begin{equation}
  b^{(k)}_{\jvec,\jvec'}=\gamma \sum\limits_{\eta=1}^{k-1} 
  \delta_{j_{\eta+1},j_{\eta}+1} \, \delta_{\jvec \setminus 
  (j_{\eta},j_{\eta+1}),\jvec'} + \gamma \, \delta_{j_{1}+N,j_{k}+1} \, 
  \delta_{\jvec \setminus (j_{1},j_{k}),\jvec'}.
  \label{equ:bmatrix}
\end{equation}

Now we focus on (\ref{equ:Gb}): for any fixed $\jvec'$ the summation 
variable $\jvec$ on the l.h.s.\ of (\ref{equ:Gb}) must assume the 
particular values $\jvec=\jvec_{\eta,\ell}$ in order to hit a nonzero 
matrix element of $b^{(k)}$, where 
\begin{equation}
  \begin{array}{rll}
  1 \leq \eta \leq k-1: & 
  \jvec_{\eta,\ell} = (j_1',\ldots j_{\eta-1}',\ell,\ell+1,j_\eta',
  \ldots j_{k-2}') \\[1ex] 
  \eta = k : & 
  \jvec_{k,N} = (1,j_1',\ldots j_{k-2}',N) 
  \end{array}
  \label{equ:jetaell}
\end{equation}
The constraint $\jvec_{\eta,\ell} \in N(k)$ restricts the range of 
the parameters $\eta$ and $\ell$. If, for instance, $2 \leq \eta \leq 
k-2$ the conditions $j_{\eta-1}' < \ell$ and $\ell+1 < j_{\eta}'$ must 
be satisfied. So only if $j_{\eta-1}' \leq j_{\eta}'-3$ is there enough 
``space'' to insert the indices $(\ell, \ell+1)$ and retain strict 
ordering. Otherwise the corresponding vector $\jvec_{\eta,\ell}$ must 
be omitted. Nevertheless we may write
\begin{equation}
  \frac{1}{\gamma} \sum\limits_{\jvec \in N(k)} \Gk{k}{\ivec}{\jvec}{t} \, 
  b^{(k)}_{\jvec,\jvec'} = 
  \sum_{\ell=1}^{j_1'-1} G_{\ivec,\jvec_{1,\ell}}^{(k)}(t) + 
  \sum_{\eta=2}^{k-2} \sum_{\ell=j_{\eta-1}'}^{j_\eta'-1}  
  G_{\ivec,\jvec_{\eta,\ell}}^{(k)}(t) + 
  \!\!\!\!\sum_{\ell=j_{k-2}'}^{N-1} G_{\ivec,\jvec_{k-1,\ell}}^{(k)}(t) + 
  G_{\ivec,\jvec_{k,N}}^{(k)}(t). 
  \label{equ:jetaellsum}
\end{equation}
Note that in contrast to what we have just said we insert the indices 
$(\ell, \ell+1)$ between $j_{\eta-1}'$ and $j_\eta'$ in 
(\ref{equ:jetaellsum}) even if $j_{\eta-1}'=j_\eta'-1$. This is 
possible because the Green's function $G^{(k)}$ from
(\ref{equ:greensfunction})  
is identically zero if $j_{\eta,\ell}$ contains an index pair,  
as follows from its permutational antisymmetry discussed after
(\ref{equ:greensfunction}).
We may use permutational antisymmetry of $G^{(k)}$ once more to 
replace in (\ref{equ:jetaellsum}) all $\jvec_{\eta,\ell}$ with 
$1 \leq \eta < k$ by $\jvec_{1,\ell}$ since the corresponding 
permutations involve an even number of transpositions. The 
$\eta$-dependencies in (\ref{equ:jetaellsum}) via Green's functions 
then drop out and we can combine the sums 
\begin{equation}
  \sum\limits_{\jvec \in N(k)} \Gk{k}{\ivec}{\jvec}{t} \, 
  b^{(k)}_{\jvec,\jvec'} = 
  \gamma \sum_{\ell=1}^{N-1} G_{\ivec,\jvec_{1,\ell}}^{(k)}(t) + 
  \gamma \, G_{\ivec,\jvec_{k,N}}^{(k)}(t) = 
  \gamma \sum_{\ell=1}^N G_{\ivec,\jvec_{1,\ell}}^{(k)}(t).
  \label{equ:j1ellsum}
\end{equation}
The second equality in (\ref{equ:j1ellsum}) follows when using 
permutational antisymmetry of $G^{(k)}$ and $N$-(anti)\-periodicity 
(\ref{equ:I+N}) of the functions $I_n(x)$ contained in $G^{(k)}$. 
The reasoning applied is in full analogy to the discussion of 
$\bar{\Phi}^{(k)}$ in Sec.~\ref{sec:homogeneous}. 
Before proceeding, we discuss briefly why (\ref{equ:j1ellsum}) is
correct also for $j_1'=1$ or $j_{k-2}'=N$. One of the sum over
$\jvec_{1,\ell}$ and $\jvec_{k-1,\ell}$ in (\ref{equ:jetaellsum}) is
then empty and evaluates to zero, corresponding to the fact that
$\jvec_{1,\ell} \in N(k)$ or $\jvec_{k-1,\ell} \in N(k)$ cannot be
achieved for any $\ell$.  In (\ref{equ:j1ellsum}), on the other hand,
the corresponding terms drop out because $\jvec_{1,\ell}$ contains
index pairs.

The evaluation of (\ref{equ:j1ellsum}) requires us to keep track of 
the first two components of $\jvec_{1,\ell}$, viz.\ $(\ell,\ell+1)$, 
in the Green's function (\ref{equ:greensfunction}). To do so we 
focus on the subset of permutations $\Scal_{\mu,\nu}(k) \subset 
\Scal(k)$ that map these first two components onto components $\mu,\nu$, 
respectively. Any $\pi \in \Scal_{\mu,\nu}(k)$ may be factorized 
into $\pi = \pi_{\mu,\nu} \circ \bar{\pi}$ where $\bar{\pi}$ performs 
an arbitrary permutation on the components $3,\ldots k$ and 
$\pi_{\mu,\nu}$ then puts the components $1,2$ into positions 
$\mu,\nu$, respectively. We have $(-1)^\pi=(-1)^{\pi_{\mu,\nu}} \, 
(-1)^{\bar{\pi}}$, where $(-1)^{\pi_{\mu,\nu}}=(-1)^{\nu-\mu-1}$ 
for $\nu > \mu$ and $(-1)^{\pi_{\mu,\nu}}=-(-1)^{\mu-\nu-1}$ 
for $\nu < \mu$ follows by counting the number of transpositions 
in $\pi_{\mu,\nu}$. Using this factorisation for 
$\pi \in \Scal_{\mu,\nu}(k)$ and the fact $\bigcup_{\mu\neq\nu} 
\Scal_{\mu,\nu}(k) = \Scal(k)$ allows us to rewrite the Green's 
function (\ref{equ:greensfunction}) in the form
\begin{eqnarray}
  G_{\ivec,\jvec_{1,\ell}}^{(k)}(t) & = & \!\!\!\!\!\! 
  \sum_{1 \leq \mu < \nu \leq k} \!\!\!\!
  (-1)^{\nu-\mu-1} \, \me^{-2t} \, I_{i_\mu-l}(\gamma t) \, 
  I_{i_\nu-l-1}(\gamma t) \, 
  G_{\ivec \setminus (i_\mu,i_\nu),\jvec'}^{(k-2)}(t)
  \nonumber \\
  & - & \!\!\!\!\!\! 
  \sum_{1 \leq \nu < \mu \leq k} \!\!\!\!
  (-1)^{\mu-\nu-1} \, \me^{-2t} \, I_{i_\mu-l}(\gamma t) \, 
  I_{i_\nu-l-1}(\gamma t) \, 
  G_{\ivec \setminus (i_\nu,i_\mu),\jvec'}^{(k-2)}(t). 
  \label{equ:Gsplit}
\end{eqnarray}
Relabelling $\mu \leftrightarrow \nu$ in the second line of 
(\ref{equ:Gsplit}) and substitution into (\ref{equ:j1ellsum}) 
then gives
\begin{equation}
  \sum\limits_{\jvec \in N(k)} \!\! \Gk{k}{\ivec}{\jvec}{t} \, 
  b^{(k)}_{\jvec,\jvec'} = \gamma \!\!\!\!\!\!
  \sum_{1 \leq \mu < \nu \leq k} \!\!\!\!
  (-1)^{\nu-\mu-1} \, \me^{-2t} 
  \sum_{\ell=1}^N \left[ I_{i_\mu-l} \, 
  I_{i_\nu-l-1} - I_{i_\nu-l} \, 
  I_{i_\mu-l-1} \right](\gamma t) \, 
  G_{\ivec \setminus (i_\mu,i_\nu),\jvec'}^{(k-2)}(t).
  \label{equ:almostGb}
\end{equation}
Our derivation of the identity (\ref{equ:Gb}) is completed by using 
the convolution property (\ref{equ:Iconv}) of the functions $I_n(x)$ 
to evaluate the $\ell$-sum
\begin{equation}
  \gamma \, \me^{-2t} \sum_{\ell=1}^N \left[ I_{i_\mu-l} \, 
  I_{i_\nu-l-1} - I_{i_\nu-l} \, 
  I_{i_\mu-l-1} \right](\gamma t) = 
  \gamma \, \me^{-2t} \left[ I_{i_\nu-i_\mu-1} - 
  I_{i_\nu-i_\mu+1} \right](2\gamma t) 
  \label{equ:Hdiff}
\end{equation}
and identifying (\ref{equ:Hdiff}) as the $t$-derivative of $H_{i_\nu - 
i_\mu}(2t)$ given by (\ref{equ:H}).

\section{Properties of $I_n(x)$}
\label{sec:bessel}

The functions $I_n(x)$ appear in our analysis of the Glauber-Ising 
model when we solve (\ref{equ:homogeneousfourier}) for $\bar{\Phi}$ 
and invert the Fourier transforms. We define 
\begin{equation}
  I_n(x) = \frac{1}{N} \sum\limits_{q \in Q} \me^{i n q + x \cos q} = 
\frac{1}{N} \sum\limits_{q \in Q} \cos(n q) \, \me^{x \cos q}, 
  \label{equ:I}
\end{equation}
where we have to set $Q=Q_\mathrm{e}$ on even levels $k$ of the hierarchy
and $Q=Q_\mathrm{o}$ on odd ones. In the thermodynamic limit $N \to 
\infty$ the sets $Q_\mathrm{e}, Q_\mathrm{o}$ given in (\ref{equ:Qeven}, 
\ref{equ:Qodd}) become dense on $[0,2\pi]$. Hence for any fixed 
$n \in \mathbb{Z}$ and $x \in \mathbb{R}$ we have 
\begin{equation}
  \lim_{N \to \infty} I_n(x) = \int_0^{2 \pi} \frac{\dd q}{2 \pi} \, 
  \cos(n q) \, \me^{x \cos q} = \In_n(x)
  \label{equ:In}, 
\end{equation}
that is the $I_n(x)$ on all levels $k$ become modified Bessel functions 
$\In_n(x)$ as $N \to \infty$. For a comprehensive discussion 
of the $\In_n(x)$ see \cite{Mathbook}. Many properties of the $\In_n(x)$ 
in fact have an analogue for finite $N$. The following identities hold 
for $Q=Q_\mathrm{e}$ as well as $Q=Q_\mathrm{o}$:
\begin{align}
  \forall  0 \leq n < N: \quad & 
    I_n(0) = \delta_{n,0} 
  \label{equ:I0}\\[1ex]
  \forall x \in \mathbb{R} \quad \forall n \in \mathbb{Z}: \quad & 
    I_{-n}(x) = I_{n}(x) 
  \label{equ:Inneg}\\[1ex]
  \forall x \in \mathbb{R}^+ \quad \forall 
    -\lfloor {\textstyle \frac{N}{2}} \rfloor 
    \leq n \leq \lfloor {\textstyle \frac{N}{2}} \rfloor: \quad & 
    I_{n}(x) \geq 0
  \label{equ:Inpos}\\[1ex]
  \forall x \in \mathbb{R} \quad \forall n \in \mathbb{Z}: \quad &
    \frac{\dd}{\dd x} I_n(x) = \frac{1}{2} 
    \left[ I_{n-1}(x) + I_{n+1}(x) \right] 
  \label{equ:Idiff} \\[1ex]
  \forall x \in \mathbb{R} \quad \forall m,n \in \mathbb{Z} \quad 
  \forall q \in Q_{\even/\odd} : \quad &
    \sum_{k=m+1}^{m+N} \me^{ikq}\, I_{k+n}^{\even/\odd}(x) = 
    \me^{-inq+x \cos q} 
  \label{equ:Icossum} \\
  \forall x,y \in \mathbb{R} \quad \forall m,n \in \mathbb{Z}: \quad & 
    \sum_{k=m+1}^{m+N} I_k^{\even/\odd}(x) \, I_{n-k}^{\even/\odd}(y) = 
    I_n^{\even/\odd}(x+y) 
  \label{equ:Iconv}
\end{align}
In (\ref{equ:Icossum}, \ref{equ:Iconv}) the $\even/\odd$ symbols 
indicate that the identities hold on even as well as odd levels but 
not between even and odd ones. The functions $I_n(x)$ associated with 
even and odd levels differ in the following properties (here 
$m,n \in \mathbb{Z}$ and $x \in \mathbb{R}$): 
\begin{align}
  \mbox{Odd Levels ($Q=Q_\mathrm{o}$)} \quad \quad \quad  & 
  \quad \quad \quad
  \mbox{Even Levels ($Q=Q_\mathrm{e}$)} \nonumber \\[2ex]
  I_{n+N}(x) = I_n(x) \quad \quad \quad \quad & 
  \quad \quad \quad \quad 
  I_{n+N}(x) = -I_{n}(x) 
  \label{equ:I+N} \\[1ex]
  \sum_{k=m+1}^{m+N} I_k(x) = \me^x \quad \quad \quad \quad  & 
  \quad \quad \quad \quad 
  \sum_{k=m+1}^{m+N} |I_k(x)| \leq \me^x 
  \label{equ:Isum}
\end{align}

\begin{figure}[htb]
  \begin{picture}(15,5.5)
    \put(0,0){\epsfig{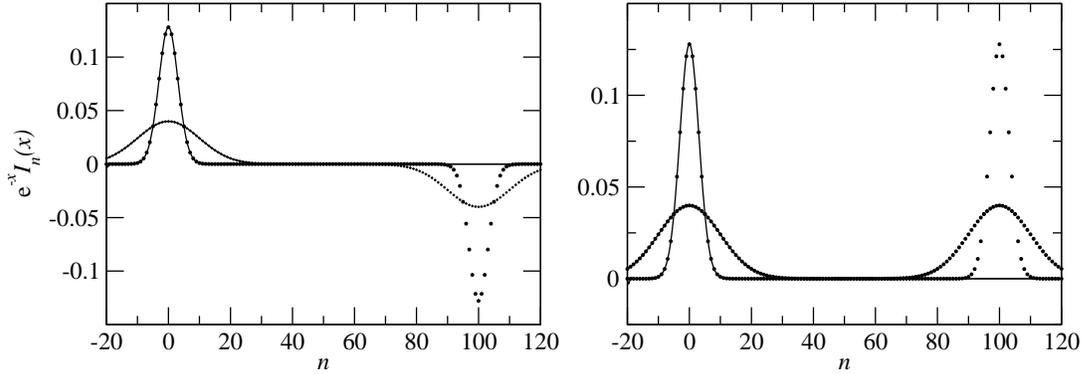}}
  \end{picture}
  \caption{\label{fig:I} Plots of the scaled functions $\me^{-x} I_n(x)$ 
    for a finite system of size $N=100$ (dots) and the $N \to \infty$ 
    limit $\me^{-x} \In_n(x)$ (lines). The (left) right figure shows 
    the $N$-(anti)\-periodic functions associated with (even) odd 
    levels, respectively. 
    The curves correspond to $x=10$ (narrow peaks) and $x=100$
    (broader peaks).}
\end{figure}

The identities (\ref{equ:I0}, \ref{equ:Inneg}, \ref{equ:Idiff}) 
follow immediately from the definition of $I_n(x)$. 
(\ref{equ:Icossum}, \ref{equ:Iconv}, \ref{equ:I+N}), on the other 
hand, are a consequence of the convolution property and 
$N$-(anti)\-periodicity of the discrete Fourier transforms 
(\ref{equ:fourier}, \ref{equ:fourierinv}). We have not managed 
to prove the seemingly trivial inequality (\ref{equ:Inpos}) in 
a rigorous way. Numerical evaluations of (\ref{equ:I}), however, 
leave no doubt that (\ref{equ:Inpos}) should be true in general. 
Assuming that (\ref{equ:Inpos}) is true, (\ref{equ:Isum}) can 
be verified as follows: on odd levels, where $I_n(x)$ is 
$N$-periodic, (\ref{equ:Inpos}) implies that $|I_n(x)|=I_n(x)$ 
for all $n \in \mathbb{Z}$ and hence (\ref{equ:Isum}) follows 
from (\ref{equ:Icossum}) by setting $q=0$; this is allowed 
since $0 \in Q_\odd$. On even 
levels, the $N$-antiperodicity of the $I_n(x)$ is cancelled by the 
modulus in (\ref{equ:Isum}). Hence we may shift the summation 
range to the region $-\lfloor \frac{N-1}{2} \rfloor \leq n \leq 
\lfloor \frac{N}{2} \rfloor$ where $I_n(x)$ is positive and drop 
the modulus. Then, by substituting (\ref{equ:I}) and exchanging 
sums, the bound (\ref{equ:Isum}) can be verified \cite{Thesis}. To illustrate 
the actual shape of the $I_n(x)$ we show some plots in Fig.~\ref{fig:I}.  

\section{Properties of $H_n(t)$ and Related Functions}
\label{sec:Hn}

The functions $H_n(t)$ originate from the projection of the Green's 
function (\ref{equ:Gb}), which is an essential ingredient for our 
solution (\ref{equ:gsolution}). Appendix~\ref{sec:greensproject} 
motivates the definition
\begin{equation}
  H_n(t) = \frac{\gamma}{2} \int_0^{t} 
  \dd\tau \, \me^{-\tau} \left[ I_{n-1}(\gamma\tau)-I_{n+1}(\gamma\tau) 
  \right]. 
  \label{equ:H}
\end{equation}
When substituting (\ref{equ:I}) for $I_n(x)$ the integral in (\ref{equ:H}) 
can be solved and we obtain the alternative representation 
\begin{equation}
  H_n(t)= \frac{1}{N} \sum\limits_{q \in Q} \sin (n \, q) \, 
  \frac{\gamma \sin q}{1-\gamma \cos q} 
  \left[ 1-\me^{-t \, (1-\gamma \cos q)} \right]
  \label{equ:Halt}
\end{equation}
In (\ref{equ:Halt}) one has to substitute $Q=Q_{\even/\odd}$ as 
appropriate. Like the $I_n(x)$, the $H_n(x)$ thus have slightly
different properties at even and odd levels of the hierarchy.  

Two obvious properties of $H_n(t)$ are $H_{-n}(t)=-H_n(t)$ and 
$H_n(0)=0$. In the thermodynamic limit $N \to \infty$ uniform 
convergence of $I_n(x)$ on any finite interval $x \in [0,a]$ 
and for fixed $n \in \mathbb{Z}$ 
allows us to replace $I_n(x)$ in (\ref{equ:H}) by its limit 
$\In_n(x)$. In (\ref{equ:Halt}), on the other hand, the limit 
$N \to \infty$ amounts to replacing $\frac{1}{N} \sum_q$ by 
$\frac{1}{2 \pi} \int \dd q$. In analogy with 
Appendix~\ref{sec:bessel} we use the notation $\Hn_n(t)$ for the 
$N \to \infty$ limit of $H_n(t)$. Further properties of the 
functions $H_n(t)$ may be obtained directly from (\ref{equ:H}) 
using the features of $I_n(x)$ given in Appendix~\ref{sec:bessel}. 
It is interesting to note that, for $\gamma=1$, $\Hn_n(2t)$ 
is essentially the
probability that two independent, continuous time random 
walkers on a one-dimensional lattice, which are a distance $n>0$ apart 
at time $t=0$, meet before time $t$ \cite{DerZei96}.

In the context of 
Sec.~\ref{sec:correl2quench},~\ref{sec:responsequench} the 
following sum, which we use to extend the definition of $H_n$ 
to two time arguments, is of relevance 
\begin{equation}
  H_{j-i}^{\even/\odd}(t_1,t_2) = \sum_n \me^{-t_1} 
  I_{i-n}^{\even/\odd}(\gamma \, t_1) H_{j-n}^{\even/\odd}(t_2).
  \label{equ:H2tdef}
\end{equation}
Here $\even/\odd$ emphasises that both $I$ and $H$ on the r.h.s.\ 
of (\ref{equ:H2tdef}) must be associated either with even or odd 
levels. From (\ref{equ:Iconv}) and (\ref{equ:H}) we immediately 
obtain 
\begin{equation}
  H_n(t_1,t_2)=\frac{\gamma}{2} \int\limits_{t_1}^{t_1+t_2} 
  \dd\tau \, \me^{-\tau} \left[ I_{n-1}(\gamma\tau)-I_{n+1}(\gamma\tau) 
  \right] \label{equ:Hn}
\end{equation}
and thence, or from (\ref{equ:Icossum}) and (\ref{equ:Halt}),
\begin{equation}
  H_n(t_1,t_2)= \frac{1}{N} \sum\limits_{q \in Q} \sin (n \, q) \, 
  \frac{\gamma \sin q}{1-\gamma \cos q} \, \me^{-t_1 \, (1-\gamma \cos q)} 
  \, \left[ 1-\me^{-t_2 \, (1-\gamma \cos q)} \right].
  \label{equ:Hn1}
\end{equation}
According to e.g.\ (\ref{equ:H},\ref{equ:Hn}) we have the obvious 
link $H_n(t)=H_n(0,t)$ between functions with one and two time 
arguments. But it is quite remarkable that we may conversely write  
\begin{equation}
  H_n(t_1,t_2)=H_n(t_1+t_2)-H_n(t_1),
  \label{equ:H_conversion}
\end{equation}
when considering the definition (\ref{equ:H2tdef}). The expressions 
(\ref{equ:Hn},\ref{equ:Hn1}) are particularly useful in the thermodynamic 
limit $N \to \infty$ where (\ref{equ:H2tdef}) becomes an infinite 
summation. In (\ref{equ:Hn},\ref{equ:Hn1}), however, this limit is 
straightforward. As for (\ref{equ:H},\ref{equ:Halt}) we then use the 
symbol $\Hn_n(t_1,t_2)$.

In Sec.~\ref{sec:correlquench} we obtained another function 
$H_n''(t)$ as given by (\ref{equ:Hn2prime}) that is closely 
related to $H_n(t)$. The physical relevance of $H_n''(t)$ is 
discussed in the text below (\ref{equ:Hn2prime}). In 
mathematical terms we may simply consider $H_n''(t)$ as a 
generalisation of $H_n(t)$ since we have $H_n''(t)=H_n(t)$ 
for $\gamma_{\upi} = 0$. For $\gamma_{\upi} \neq 0$, a 
representation of $H_n''(t)$ in the form (\ref{equ:H}) is 
not possible. Note that while $H_n(t)$ is defined on even 
as well as odd levels of the hierarchy, $H_n''(t)$ is 
always associated with even levels, i.e.\ the sum in 
(\ref{equ:Hn2prime}) runs over $Q_{\even}$. In analogy 
with (\ref{equ:H2tdef}) we now extend the definition of 
$H_n''(t)$ to two time arguments. The following 
sums are relevant for Appendix~\ref{sec:EF}
\begin{eqnarray}
  H_{j-i}''(t_1,t_2) & = & 
    \sum_n \me^{-t_1} I_{i-n}^{\even}(\gamma \, t_1) H_{j-n}''(t_2) 
    \label{equ:H2prime2tdef}\\
  \widetilde{H}_{j-i}''(t_1,t_2) & = & 
    \sum_n \sgn(j-n) \, \me^{-t_1} I_{i-n}^{\odd}(\gamma \, t_1) 
    H_{j-n}''(t_2) 
    \label{equ:H2prime2ttildedef}
\end{eqnarray}
where $1 \leq i,j \leq N$ and the sum runs over $n=1,\ldots N$ for 
a finite system. 
The superscripts $\even$ and $\odd$ of $I_n(x)$ indicate that these 
functions must be associated with even and odd levels of the hierarchy, 
respectively. The sum (\ref{equ:H2prime2tdef}) may be evaluated easily 
using (\ref{equ:Hn2prime}) and (\ref{equ:Icossum}) 
\begin{eqnarray}
  H_n''(t_1,t_2)
  & = & \frac{1}{N} \sum_{q \in Q_\mathrm{e}} 
  \sin (n \, q) \, \me^{-t_1 (1-\gamma 
  \cos q)} \times \nonumber \\
  && \left\{ 
  \frac{\gamma_\mathrm{i} \sin 
  q}{1-\gamma_\mathrm{i} \cos q} \, 
  \me^{-t_2 (1-\gamma 
  \cos q)}+\frac{\gamma \sin 
  q}{1-\gamma \cos q} \left[ 1-\me^{-t_2 
  (1-\gamma \cos q)} \right] \right\}.
  \label{equ:H2prime2t}
\end{eqnarray}
In analogy to $H$ we obtain from (\ref{equ:Hn2prime}), (\ref{equ:Halt}) 
and (\ref{equ:H2prime2t}) the link $H_n''(t)=H_n''(0,t)$ and the 
non-trivial identity
\begin{equation}
  H_n''(t_1,t_2)=H_n''(t_1+t_2)-H_n(t_1) 
  \label{equ:Hpp_conversion}
\end{equation}
for the sum (\ref{equ:H2prime2tdef}). For $\gamma_{\upi}=0$ the 
equations (\ref{equ:Hn2prime}), (\ref{equ:H2prime2tdef}), 
(\ref{equ:H2prime2t}) and (\ref{equ:Hpp_conversion}) reduce to 
(\ref{equ:Halt}), (\ref{equ:H2tdef}), (\ref{equ:Hn1}) and 
(\ref{equ:H_conversion}), respectively, with $H''$ replaced by
$H$. Consistent with our notation elsewhere we 
will denote the $N \to \infty$ limit of $H''$ by $\Hn''$.

In contrast to (\ref{equ:H2prime2tdef}), the sum 
(\ref{equ:H2prime2ttildedef}) cannot be evaluated explicitly. 
But it may be rewritten in terms of the integral representation 
\begin{equation}
  \widetilde{H}_n''(t_1,t_2) = \sgn(n) \, H_n''(t_1, 
  t_2) + \gamma \int_0^{t_1} \dd \tau \, \me^{-\tau} \, 
  I_n^{\odd}(\gamma \, \tau) \, H_1''(t_1-\tau,t_2).
  \label{equ:H2prime2ttilde}
\end{equation}
Note that, in contrast to $H_n''$, the function $\tilde{H}_n''$ is even 
in $n$. As for (\ref{equ:H2prime2tdef}), (\ref{equ:H2prime2t}) 
we will drop the double-primes in (\ref{equ:H2prime2ttildedef}), 
(\ref{equ:H2prime2ttilde}) for $\gamma_{\upi} = 0$, and denote
the $N \to \infty$ 
limit of $\tilde{H}_n''(t_1,t_2)$ by $\tilde{\Hn}_n''(t_1,t_2)$.
Due to uniform convergence of the integrand, this limit is obtained by
replacing $I$,$H$ with $\In$,$\Hn$ in (\ref{equ:H2prime2ttilde}).

Equivalence of (\ref{equ:H2prime2ttildedef}) and (\ref{equ:H2prime2ttilde}) 
may be verified as follows: substitute the expression (\ref{equ:Hn2prime}) 
for $H_n''(t)$ into (\ref{equ:H2prime2ttildedef}) and exchange the
sums over $n$ and $q$. Now, the sum over $n$ is of the form  
\begin{equation}
  f_{i,j}(x,q)=\sum\limits_{n=1}^N \sgn(j-n) \, I_{i-n}^{\odd}(x) \,
  \sin \left[ q \, (j-n) \right], 
  \label{equ:fij}
\end{equation}
where $q$ is the summation variable in $H''$ and thus $q \in 
Q_\mathrm{e}$. Differentiating (\ref{equ:fij}) w.r.t.\ $x$ 
using (\ref{equ:Idiff}), shifting the summation variable $n \to 
n \pm 1$ as appropriate and rewriting the result in terms of sums 
over the range $n=1,\ldots N$ yields 
\begin{eqnarray}
  \frac{\partial}{\partial x} \, f_{i,j}(x,q) & = & 
  \frac{1}{2} \sum\limits_{n=1}^N \sgn(j-n+1) \, I_{i-n}^\odd(x) \, 
  \sin[q \, (j-n+1)] \label{equ:fij1}\\
  & + & \frac{1}{2} \, \sgn(j-N) \, I_{i-N-1}^\odd(x) \, 
  \sin[q \, (j-N)] - 
  \frac{1}{2} \, \sgn(j) \, I_{i-1}^\odd(x) \, \sin(qj) \nonumber \\
  & + & \frac{1}{2} \sum\limits_{n=1}^N \sgn(j-n-1) \, I_{i-n}^\odd(x) \, 
  \sin[q \, (j-n-1)] \nonumber \\
  & + & \frac{1}{2} \, \sgn(j-1) \, I_i^\odd(x) \, 
  \sin[q \, (j-1)] - 
  \frac{1}{2} \, \sgn(j-N-1) \, I_{i-N}^\odd(x) \, \sin[q \, (j-N-1)]. 
  \nonumber 
\end{eqnarray}
Next focus on the second line in (\ref{equ:fij1}). Due to 
$N$-periodicity (\ref{equ:I+N}) of the functions $I_n(x)$ 
on odd levels we have $I_{i-N-1}^\odd(x)=I_{i-1}^\odd(x)$. 
Also, since $q \in Q_\even$ as given by (\ref{equ:Qeven}), 
the identity $\sin[q \, (j-N)]=-\sin(qj)$ holds. Recalling, 
finally, that $1 \leq j \leq N$ is assumed and noting that 
$\sgn(j-N)=-\sgn(j)$ for $1 \leq j < N$ thus shows that 
the two terms in the second line of (\ref{equ:fij1}) cancel 
each other for $1 \leq j < N$. When $j=N$ 
both terms are zero because $\sin(0)=\sin(Nq)=0$. It follows 
similarly that the fourth line in (\ref{equ:fij1}) drops 
out. Expressing $\sgn(j-n \pm 1)$ via $\sgn(j-n)$ by adding
compensating terms for $j=n,n\pm1$ and using the 
trigonometric identity $\sin(a+b)+\sin(a-b)=2 \sin(a) \cos(b)$ 
allows us to combine the sums in (\ref{equ:fij1}) in the 
form (\ref{equ:fij}). Consequently the $f_{i,j}(x,q)$ 
satisfy the inhomogeneous differential equations 
\begin{equation}
  \frac{\partial}{\partial x} \, f_{i,j}(x,q) - \cos(q) \, f_{i,j}(x,q) = 
  I_{i-j}^{\odd}(x) \, \sin(q).
  \label{equ:fijdiff}
\end{equation}
From (\ref{equ:I0}) and (\ref{equ:fij}) we also have $f_{i,j}(0,q) = 
\sgn(j-i) \, \sin[q \, (j-i)]$. Thus we obtain $f_{i,j}(x,q)$ by 
integrating (\ref{equ:fijdiff}) for this initial condition, i.e.\  
\begin{equation}
  f_{i,j}(x,q)=\sgn(j-i) \, \sin[q \, (j-i)] \, \me^{x \, \cos q} + 
  \sin q \int_0^x \dd x' \, \me^{(x-x') \, \cos q} \, I_{i-j}^{\odd}(x').
  \label{equ:fijint}
\end{equation}
Using (\ref{equ:fijint}) in (\ref{equ:H2prime2ttildedef}) immediately 
produces (\ref{equ:H2prime2ttilde}) if one bears in mind
(\ref{equ:H2prime2t}).

We conclude this section by showing that the general expressions 
(\ref{equ:H2prime2t}), (\ref{equ:H2prime2ttilde}) simplify quite 
significantly if one considers a quench from the random initial 
configuration corresponding to $T_{\upi} = \infty$ to $T = 0$, i.e.\ 
$\gamma_{\upi} = 0$ and $\gamma=1$. Let us first focus on 
$H_n''(t_1,t_2)$ as given by (\ref{equ:H2prime2t}), which for 
$\gamma_{\upi}=0$ we write as $H_n(t_1,t_2)$. Since also
$\gamma=1$, we may use the trigonometric identity 
\begin{equation}
  \sin [(n+1) \, x] \, \frac{\sin x}{1-\cos x} = 
  \sin (n \, x) \, \frac{\sin x}{1-\cos x} + \cos (n \, x) + \cos [(n+1) \, x]
\end{equation}
in (\ref{equ:Hn1}). This yields the zero temperature recursion formula 
\begin{equation}
  H_{n+1}(t_1,t_2) = H_n(t_1,t_2) + \me^{-t_1} \left[ 
  I_n + I_{n+1} \right](t_1) - \me^{-(t_1+t_2)} \left[ 
  I_n + I_{n+1} \right](t_1+t_2)
  \label{equ:H2tzero}
\end{equation}
for $H_n(t_1,t_2)$. Starting from $H_0(t_1,t_2)=0$, any $H_n(t_1,t_2)$ 
can thus be expressed purely in terms of functions $I_n(x)$. By setting 
$t_1=0$ in (\ref{equ:H2tzero}) and using (\ref{equ:I0}) we also obtain 
a recursion for $H_n(t)=H_n(0,t)$, that is 
\begin{equation}
  H_1(t)=1-\me^{-t} \left[ I_0 + I_1 \right](t) \quad \mbox{and} \quad 
  H_{n+1}(t) = H_n(t) - \me^{-t} \left[ I_n + I_{n+1} \right](t),  
  \label{equ:Hzero}
\end{equation}
where $1<n+1<N$. Outside that range we have $H_{-n}(t)=-H_n(t)$ and 
$N$-(anti)\-periodicity in $n$ according to e.g.\ (\ref{equ:H}). Equations 
(\ref{equ:H2tzero},\ref{equ:Hzero}) apply in the thermodynamic 
limit, too, if we replace $H$ by $\Hn$ and $I_n(x)$ by modified Bessel 
functions $\In_n(x)$. 

Now we consider $\tilde{H}_n''(t_1,t_2)$ at $\gamma_{\upi}=0$, $\gamma=1$ 
and, for the sake of simplicity, also $N \to \infty$. As explained
above, we drop the double-primes in (\ref{equ:H2prime2ttilde}) for 
$\gamma_{\upi}=0$ and replace $H$,$I$ by $\Hn$,$\In$ in the 
thermodynamic limit. This gives 
\begin{equation}
  \widetilde{\Hn}_n(t_1,t_2) = \sgn(n) \, \Hn_n(t_1, 
  t_2) + \int_0^{t_1} \dd \tau \, \me^{-\tau} \, 
  \In_n(\tau) \, \Hn_1(t_1-\tau,t_2). 
  \label{equ:H2ttilde}
\end{equation}
Expressing $\Hn_1(t_1-\tau,t_2)$ via (\ref{equ:H2tzero}) allows us 
to rewrite (\ref{equ:H2ttilde}) as 
\begin{eqnarray}
  \widetilde{\Hn}_n(t_1,t_2) & = & \sgn(n) \, \Hn_n(t_1, t_2) + 
  \me^{-t_1} \, f_n(t_1) - \me^{-(t_1+t_2)} \, f_n(t_1+t_2) 
  \nonumber \\
  & + & \me^{-(t_1+t_2)} \int_0^{t_2} \!\! \dd \tau \, 
  \In_n(t_1+t_2-\tau) \, \left[ \In_0 + \In_1 \right](\tau), 
  \label{equ:H2ttildeint}
\end{eqnarray}
where the functions $f_n(t)$ are given by the single-sided convolution 
integral 
\begin{equation}
  f_n(t)=\int_0^t \dd \tau \, \In_n(\tau) \, 
  \left[ \In_0 + \In_1 \right](t-\tau).
  \label{equ:fn}
\end{equation}
By Laplace transforming (\ref{equ:fn}) it is easy to verify that 
$f_0(t)=\me^t-\In_0(t)$ and $f_{n+1}(t)=f_n(t)-2 \In_{n+1}(t)$. 
This recursion formula produces a very similar expression for 
$\me^{-t} f_n(t)$ as (\ref{equ:Hzero}) for $H_n(t)$ which in fact 
amounts to the link 
\begin{equation}
  \me^{-t} \, \In_n(t) + \me^{-t} \, f_n(t) = 
  \delta_{n,0} + \sgn(n) \, \Hn_n(t).
  \label{equ:fnexpression}
\end{equation}
Substituting (\ref{equ:fnexpression}) into (\ref{equ:H2ttildeint}) 
and using (\ref{equ:H_conversion}) reduces the expression 
(\ref{equ:H2ttilde}) for $\tilde{\Hn}_n(t_1,t_2)$ to 
\begin{equation}
  \me^{-t_1} \, \In_n(t_1) + \widetilde{\Hn}_n(t_1,t_2) = 
  \me^{-(t_1+t_2)} \left\{ \In_n(t_1+t_2) + 
  \!\! \int_0^{t_2} \!\! \dd \tau \, 
  \In_n(t_1+t_2-\tau) \, \left[ \In_0 + \In_1 \right](\tau)
  \right\}.
  \label{equ:H2ttildezero}
\end{equation}

\section{The Sums $\mathcal{E}$ and $\mathcal{F}$}
\label{sec:EF}

Here we focus on the sums $\mathcal{E}$, $\mathcal{F}$ given 
in (\ref{equ:Esum}), (\ref{equ:Fsum}). These appear in 
the expressions for the homogeneous solutions of the two-time 
correlation and response functions discussed in 
Sec.~\ref{sec:correl2quench},~\ref{sec:responsequench}. 
Our intention in this section is to reduce the sums in
(\ref{equ:Esum}), (\ref{equ:Fsum}) over ranges
scaling with $N$ to sums over a finite number of terms, which
therefore remain manageable even for $N\to\infty$.

Let us first consider the one-dimensional sum $\mathcal{E}$ 
given in (\ref{equ:Esum}), where we use the abbreviation 
$p=\dim(\jvec)$. For a finite system we have $\jvec \in 
N(p)$ and the summation range is $m=1,2,\ldots N$. If $p$ 
is even, then the product over the $\sgn$'s in (\ref{equ:Esum}) 
is $+1$ for $m<j_1$ and $m>j_p$. Therefore the expression 
\begin{equation}
  \boldsymbol{\delta}_{\jvec,m} = 1-\prod_{\lambda=1}^{\mathrm{dim}(\jvec)} 
  \sgn(j_\lambda-m)
  \label{equ:deltajm}
\end{equation}
can be non-zero only for $j_1 \leq m \leq j_p$. In fact we have 
$\boldsymbol{\delta}_{\jvec,j_\nu}=1$ for $\nu=1,\ldots p$ and 
$\boldsymbol{\delta}_{\jvec,m}=2$ for $j_{2\nu-1} < m < j_{2\nu}$ 
with $\nu=1,\ldots p/2$. Otherwise $\boldsymbol{\delta}_{\jvec,m}$ 
is zero. In terms of (\ref{equ:deltajm}) we may rewrite 
(\ref{equ:Esum}) as
\begin{equation}
  \mathcal{E}_{i_\eps,j_\nu}^{\,\jvec}(t,t') = 
  \sum_m \left[1-\boldsymbol{\delta}_{\jvec,m} \right] \me^{-\dt} \, 
  I_{i_\eps-m}(\gamma \dt) \, H_{j_\nu-m}''(2t'). 
  \label{equ:Esumsplit}
\end{equation}
Now recall that $C^{(k,l)}(t,t')$ 
and $R^{(k,l)}(t,t')$ are non-vanishing only if $k+l$ is even. Hence if
$p=l$ (or $p=l\pm2$ 
for responses where from (\ref{equ:jnus}) the argument $\jvec$ in
(\ref{equ:Esumsplit})
might be a vector $\jvec^{\nu,\mathrm{s}}$) is even, then so must $k$
be. Consequently the functions $I_n(x)$ in (\ref{equ:Esumsplit}) are
associated with even levels. This allows us to rewrite (\ref{equ:Esumsplit}) 
\begin{eqnarray}
  \mathcal{E}_{i_\eps,j_\nu}^{\,\jvec}(t,t') = 
  H_{j_\nu-i_\eps}''(\dt,2t') - 
  \sum_m \boldsymbol{\delta}_{\jvec,m} \, 
  \me^{-\dt} \, I_{i_\eps-m}(\gamma \dt) \,
  H_{j_\nu-m}''(2t') 
  \label{equ:Eeven}
\end{eqnarray}
using (\ref{equ:H2prime2tdef}). The remaining summation in 
(\ref{equ:Eeven}) only produces non-zero terms in the 
$N$-independent range $j_1 \leq m \leq j_p$. Consequently 
the thermodynamic limit $N \to \infty$ of (\ref{equ:Eeven}) 
is straightforward.

Following a similar strategy we now rewrite 
the sum $\mathcal{E}$ for $p=\dim(\jvec)$ odd. In this 
case the product over the $\sgn$'s in (\ref{equ:Esum}) 
gives $+1$ for $m<j_1$ but $-1$ for $m>j_p$. Hence we 
introduce 
\begin{equation}
  \boldsymbol{\sigma}_{\jvec,m}^a = \sgn(a-m)-
  \prod_{\lambda=1}^{\mathrm{dim}(\jvec)} \sgn(j_\lambda-m),
  \label{equ:sigmajm}
\end{equation}
where $a$ is some reference site that should be chosen in 
the range $j_1 \leq a \leq j_p$. Then (\ref{equ:sigmajm}) 
again only produces non-zero values for $j_1 \leq m \leq 
j_p$. In terms of (\ref{equ:sigmajm}) and with the choice $a=j_{\nu}$ the
sum (\ref{equ:Esum}) becomes 
\begin{equation}
  \mathcal{E}_{i_\eps,j_\nu}^{\,\jvec}(t,t') = 
  \sum_m \left[\sgn(j_\nu-m) - \boldsymbol{\sigma}_{\jvec,m}^{j_\nu} 
  \right] \me^{-\dt} \, I_{i_\eps-m}(\gamma \dt) \,
  H_{j_\nu-m}''(2t').
  \label{equ:Esumtildesplit}
\end{equation}
Applying the same reasoning as below (\ref{equ:Esumsplit}) we 
conclude that in (\ref{equ:Esumtildesplit}) the functions $I_n(x)$ 
must be associated with odd levels. Hence we utilise 
(\ref{equ:H2prime2ttildedef}) to rewrite (\ref{equ:Esumtildesplit}) 
as
\begin{equation}
  \mathcal{E}_{i_\eps,j_\nu}^{\,\jvec}(t,t') = 
  \widetilde{H}_{j_\nu-i_\eps}''(\dt,2t') - 
  \sum_m \boldsymbol{\sigma}_{\jvec,m}^{j_\nu} \, 
  \me^{-\dt} \, I_{i_\eps-m}(\gamma \dt) \,
  H_{j_\nu-m}''(2t'). 
  \label{equ:Eodd}
\end{equation}
Because the summation range in (\ref{equ:Eodd}) that produces non-zero 
terms is again $N$-independent and bounded by $j_1 \leq m \leq 
j_p$, the thermodynamic limit of (\ref{equ:Eodd}) is obvious.

For the two-dimensional sum $\mathcal{F}$ given in (\ref{equ:Fsum}) 
with even $p=\dim{\jvec}$ we may proceed in full analogy with
$\mathcal{E}$. Expressing the products over the $\sgn$'s 
in (\ref{equ:Fsum}) by (\ref{equ:deltajm}) yields 
\begin{equation}
  \mathcal{F}_{i_\eps,i_\delta}^{\,\jvec}(t,t') = 
  \sum_{m,n} \left[1-\boldsymbol{\delta}_{\jvec,m}-
  \boldsymbol{\delta}_{\jvec,n} + \boldsymbol{\delta}_{\jvec,m} \, 
  \boldsymbol{\delta}_{\jvec,n} \right] 
  \me^{-2 \dt} \, I_{i_\eps-m}(\gamma \dt) \, I_{i_\delta-n}(\gamma \dt) \,
  H_{n-m}''(2t'). 
  \label{equ:Fsumsplit}
\end{equation}
The unrestricted one and two-dimensional sums in (\ref{equ:Fsumsplit}) 
may be evaluated using (\ref{equ:Iconv}) and (\ref{equ:H2prime2tdef}). 
We are then left with bounded summation ranges $j_1 \leq m,n \leq j_p$ 
\begin{eqnarray}
  \mathcal{F}_{i_\eps,i_\delta}^{\,\jvec}(t,t') & = & 
  H_{i_\delta-i_\eps}''(2\dt,2t') \nonumber \\[1ex]
  & - & \sum_{m} \boldsymbol{\delta}_{\jvec,m} \, 
  \me^{-\dt} \left[ I_{i_\eps-m}(\gamma \dt) \, H_{i_\delta-m}''(\dt,2t')-
  I_{i_\delta-m}(\gamma \dt) \, H_{i_\eps-m}''(\dt,2t') \right] \nonumber \\
  & + & \sum_{m,n} \boldsymbol{\delta}_{\jvec,m} \, 
  \boldsymbol{\delta}_{\jvec,n} \, 
  \me^{-2 \dt} \, I_{i_\eps-m}(\gamma \dt) \, I_{i_\delta-n}(\gamma \dt) \,
  H_{n-m}''(2t'). 
  \label{equ:Feven}
\end{eqnarray}

Rewriting the two-dimensional sum $\mathcal{F}$ for $p=\dim(\jvec)$ 
odd is somewhat problematic. If we express the products over the 
$\sgn$'s in (\ref{equ:Fsum}) using (\ref{equ:sigmajm}) it is 
not possible to choose reference points $a$ s.t.\ we arrive at 
sums of the form (\ref{equ:H2prime2ttildedef}). It turns out that 
the most compact -- yet still bulky -- result follows when choosing 
\begin{eqnarray}
  \mathcal{F}_{i_\eps,i_\delta}^{\,\jvec}(t,t') & = & 
  \sum_{m,n} \left[\sgn(i_\eps-m) \, \sgn(i_\delta-n) - 
  \sgn(i_\eps-m) \, \boldsymbol{\sigma}_{\jvec,n}^{i_\delta} - 
  \sgn(i_\delta-n) \, \boldsymbol{\sigma}_{\jvec,m}^{i_\eps} + 
  \boldsymbol{\sigma}_{\jvec,m}^{i_\eps} \, 
  \boldsymbol{\sigma}_{\jvec,n}^{i_\delta} \right] \nonumber \\
  & \times & \me^{-2 \dt} \, I_{i_\eps-m}(\gamma \dt) \, 
  I_{i_\delta-n}(\gamma \dt) \, H_{n-m}''(2t'). 
  \label{equ:Fsumtildesplit}
\end{eqnarray}
By generalising (\ref{equ:fij}) integral representations for 
the sums in (\ref{equ:Fsumtildesplit}) may be derived. One 
should find
\begin{eqnarray}
  \mathcal{F}_{i_\eps,i_\delta}^{\,\jvec}(t,t') & = & 
  \sum_{m,n} 
  \boldsymbol{\sigma}_{\jvec,m}^{i_\eps}   \, 
  \boldsymbol{\sigma}_{\jvec,n}^{i_\delta} \, 
  \me^{-2 \dt} \, I_{i_\eps-m}(\gamma \dt) \, 
  I_{i_\delta-n}(\gamma \dt) \, H_{n-m}''(2t') 
  \nonumber \\
  & - & \frac{\gamma}{2} \sum_m 
  \boldsymbol{\sigma}_{\jvec,m}^{i_\eps} \,
  \me^{-\dt} \, I_{i_\eps-m}(\gamma \dt) 
  \int_0^{\dt} \dd \tau \, 
  \me^{-\tau} \, I_{0}(\gamma \tau) \left[ 
  H_{i_\delta-m-1}''- H_{i_\delta-m+1}'' \right](\dt-\tau,2t') 
  \nonumber \\
  & + & \frac{\gamma}{2} \sum_m 
  \boldsymbol{\sigma}_{\jvec,m}^{i_\delta} \,
  \me^{-\dt} \, I_{i_\delta-m}(\gamma \dt) 
  \int_0^{\dt} \dd \tau \, 
  \me^{-\tau} \, I_{0}(\gamma \tau) \left[ 
  H_{i_\eps-m-1}''- H_{i_\eps-m+1}'' \right](\dt-\tau,2t') 
  \nonumber \\
  & + & \frac{\gamma^2}{4} 
  \int_0^{\dt} \dd \tau_1 \, \me^{-\tau_1} \, I_0(\gamma \tau_1) \,
  \int_0^{\dt} \dd \tau_2 \, \me^{-\tau_2} \, I_0(\gamma \tau_2) 
  \nonumber \\
  && \quad\quad \times \left[ 
  -H_{i_\delta-i_\eps-2}'' +2 H_{i_\delta-i_\eps}''- H_{i_\delta-i_\eps+2}'' 
  \right](2 \dt-\tau_1-\tau_2,2t').
  \label{equ:Fodd}
\end{eqnarray}
As the reference points in (\ref{equ:Fsumtildesplit}) are $i_\eps$ and 
$i_\delta$ the bounds on the summation ranges in (\ref{equ:Fodd}) are 
of the form $\min(i_\eps,j_1) \leq m \leq \max(i_\eps,j_l)$ etc. For 
$\gamma_{\upi}=0$ and $\gamma=1$ the expression (\ref{equ:Fodd}) may 
be simplified using (\ref{equ:H2tzero}).

\section{The Notation of Equation (\ref{equ:responsephisolution})}
\label{sec:res1122}

In this Appendix we illustrate by means of two simple examples how 
arbitrary homogeneous solutions for two-time multispin response 
functions may be obtained from (\ref{equ:responsephisolution}). 

Let us first consider the simplest possible case of the spin response 
function to a local magnetic field. Here we have $k=l=1$ and 
$\ivec=(i)$, $\jvec=(j)$. According to the formalism presented in 
Sec.~\ref{sec:responsequench} the first step in the evaluation of 
(\ref{equ:responsephisolution}) is to determine the squashed index 
vectors (\ref{equ:jnus}). In the case at hand there is only one 
such vector since $l=1$, viz.\ $\jvec^{1,\mathrm{s}}=\jvec^{1}=
(j-1,j,j+1)$ with $l^{1,\mathrm{s}}=3$. Then, by setting $k=l=1$ in 
(\ref{equ:responsephisolution}) and substituting $l^{1,\mathrm{s}}$, 
we obtain
\begin{eqnarray}
  \Phi_{i,j}^{(1,1)}(t,t') & = & \me^{-\dt} \, I_{i-j}(\gamma \dt) 
  \left[ \left(1-\frac{\gamma^2}{2} \right) \sum_{\pi \in \mathcal{P}'(1)} 
  \sign \prod_{\lambda=1}^1 {\mathcal{V}'}_{(i \cup j)_{\pi(2\lambda-1)}, 
  (i \cup j)_{\pi(2\lambda)}}^{\, j}(t,t') \right. \nonumber \\
  & + & \left. \frac{\gamma^2}{2} 
  \sum_{\pi \in \mathcal{P}'(2)} \sign \prod_{\lambda=1}^2 
  {\mathcal{V}'}_{(i \cup \jvec^{1,\mathrm{s}})_{\pi(2\lambda-1)}, 
  (i \cup \jvec^{1,\mathrm{s}})_{\pi(2\lambda)}}^{\, \jvec^{1,\mathrm{s}}}
  (t,t') \right]. \nonumber
\end{eqnarray}
The summations over pairings in this equation are restricted to 
those, $\mathcal{P}'$, that contain the pair $(i,j)$, where we set 
$\mathcal{V}_{i,j}'=1$. In the first 
sum we have $\mathcal{P}(1)=\{ \mathrm{Id} \}$ and pairs are drawn 
from $i \cup j = (i,j)$. Hence $\mathcal{P}'(1)=\{ \mathrm{Id} \}$ 
and the only term in this sum is $\mathcal{V}_{i,j}'=1$. In the 
second sum, however, we have to consider the 3 pairings 
$(1,2,3,4)$, $(1,3,2,4)$ and $(1,4,2,3)$ contained in $\mathcal{P}(2)$. 
As pairs are drawn from $i \cup \jvec^{1,\mathrm{s}}=(i,j-1,j,j+1)$ 
the only pairing contained in $\mathcal{P}'(2)$ is $(1,3,2,4)$; the 
sign of the corresponding permutation $\pi$ is $(-1)^\pi = -1$. So 
there is again only one term in this sum, namely 
$-\mathcal{V}_{i,j}' \mathcal{V}_{j-1,j+1}'=-\mathcal{V}_{j-1,j+1}$, 
and therefore 
\begin{equation}
  \Phi_{i,j}^{(1,1)}(t,t') = \me^{-\dt} \, I_{i-j}(\gamma \dt) 
  \left[ {\textstyle \left( 1-\frac{\gamma^2}{2} \right)} - 
  {\textstyle \frac{\gamma^2}{2} }
  \mathcal{V}_{j-1,j+1}^{\, \jvec^{1,\mathrm{s}}}(t,t')
  \right]. \nonumber
\end{equation}
Since both indices of $\mathcal{V}$ are components of 
$\jvec^{1,\mathrm{s}}$ we have to use the $(a,b)=(j_\mu,j_\nu)$ 
case of (\ref{equ:Vcal}) to express this function. Making the 
corresponding substitution in the above expression produces the 
result (\ref{equ:phires11}) given in Sec.~\ref{sec:responsequench}.

The next higher nontrivial response functions are those with $k+l=4$. 
As a second example we consider the case $k=l=2$ where $\ivec=(i_1,i_2)$ 
and $\jvec=(j_1,j_2)$. We restrict ourselves further to vectors $\jvec$ 
that satisfy $j_2=j_1+1$, corresponding to a perturbation that couples
to adjacent
sites of the lattice. In this situation we have two index vectors 
$\jvec^\nu$ given by $\jvec^1=(j_1-1,j_1,j_1+1,j_2)$ and 
$\jvec^2=(j_1,j_2-1,j_2,j_2+1)$. From (\ref{equ:jnus}) we obtain the 
corresponding squashed versions $\jvec^{1,\mathrm{s}}=(j_1-1,j_1)$ 
and $\jvec^{2,\mathrm{s}}=(j_2,j_2+1)$ using $j_2=j_1+1$. Their 
dimensions are obviously $l^{1,\mathrm{s}}=l^{2,\mathrm{s}}=2$. Now, 
by substituting $k=l=2$ in (\ref{equ:responsephisolution}), explicitly 
writing out the summations over $\mu$ and $\nu$ and plugging in 
$l^{\nu,\mathrm{s}}$ we obtain 
\begin{eqnarray}
  \Phi_{\ivec,\jvec}^{(2,2)} & = & \me^{-\dt} \, 
  I_{i_1-j_1}(\gamma \dt) \left[ \left(1-\frac{\gamma^2}{2} \right) 
  \sum_{\pi \in \mathcal{P}'(2)} \sign \prod_{\lambda=1}^2 
  {\mathcal{V}'}_{(\ivec \cup \jvec)_{\pi(2\lambda-1)}, 
  (\ivec \cup \jvec)_{\pi(2\lambda)}}^{\, \jvec}(t,t') \right. \nonumber \\
  &  & \quad + \left. \frac{\gamma^2}{2} 
  \sum_{\pi \in \mathcal{P}'(2)} \sign \prod_{\lambda=1}^2 
  {\mathcal{V}'}_{(\ivec \cup \jvec^{1,\mathrm{s}})_{\pi(2\lambda-1)}, 
  (\ivec \cup \jvec^{1,\mathrm{s}})_{\pi(2\lambda)}}^{\, \jvec^{1,\mathrm{s}}} 
  (t,t') \right] \nonumber \\ 
  & + & \me^{-\dt} \, 
  I_{i_2-j_1}(\gamma \dt) \left[ \left(1-\frac{\gamma^2}{2} \right) 
  \sum_{\pi \in \mathcal{P}'(2)} \sign \prod_{\lambda=1}^2 
  {\mathcal{V}'}_{(\ivec \cup \jvec)_{\pi(2\lambda-1)}, 
  (\ivec \cup \jvec)_{\pi(2\lambda)}}^{\, \jvec}(t,t') \right. \nonumber \\
  &  & \quad + \left. \frac{\gamma^2}{2} 
  \sum_{\pi \in \mathcal{P}'(2)} \sign \prod_{\lambda=1}^2 
  {\mathcal{V}'}_{(\ivec \cup \jvec^{1,\mathrm{s}})_{\pi(2\lambda-1)}, 
  (\ivec \cup \jvec^{1,\mathrm{s}})_{\pi(2\lambda)}}^{\, \jvec^{1,\mathrm{s}}} 
  (t,t') \right] \nonumber \\
  & - & \me^{-\dt} \, 
  I_{i_1-j_2}(\gamma \dt) \left[ \left(1-\frac{\gamma^2}{2} \right) 
  \sum_{\pi \in \mathcal{P}'(2)} \sign \prod_{\lambda=1}^2 
  {\mathcal{V}'}_{(\ivec \cup \jvec)_{\pi(2\lambda-1)}, 
  (\ivec \cup \jvec)_{\pi(2\lambda)}}^{\, \jvec}(t,t') \right. \nonumber \\
  &  & \quad + \left. \frac{\gamma^2}{2} 
  \sum_{\pi \in \mathcal{P}'(2)} \sign \prod_{\lambda=1}^2 
  {\mathcal{V}'}_{(\ivec \cup \jvec^{2,\mathrm{s}})_{\pi(2\lambda-1)}, 
  (\ivec \cup \jvec^{2,\mathrm{s}})_{\pi(2\lambda)}}^{\, \jvec^{2,\mathrm{s}}} 
  (t,t') \right] \nonumber \\
  & - & \me^{-\dt} \, 
  I_{i_2-j_2}(\gamma \dt) \left[ \left(1-\frac{\gamma^2}{2} \right) 
  \sum_{\pi \in \mathcal{P}'(2)} \sign \prod_{\lambda=1}^2 
  {\mathcal{V}'}_{(\ivec \cup \jvec)_{\pi(2\lambda-1)}, 
  (\ivec \cup \jvec)_{\pi(2\lambda)}}^{\, \jvec}(t,t') \right. \nonumber \\
  &  & \quad + \left. \frac{\gamma^2}{2} 
  \sum_{\pi \in \mathcal{P}'(2)} \sign \prod_{\lambda=1}^2 
  {\mathcal{V}'}_{(\ivec \cup \jvec^{2,\mathrm{s}})_{\pi(2\lambda-1)}, 
  (\ivec \cup \jvec^{2,\mathrm{s}})_{\pi(2\lambda)}}^{\, \jvec^{2,\mathrm{s}}} 
  (t,t') \right]. \nonumber 
\end{eqnarray}
Within the 4 square brackets the pairings are restricted to those, 
$\mathcal{P}'$, that contain the pairs $(i_1,j_1)$, $(i_2,j_1)$, 
$(i_1,j_2)$ and $(i_2,j_2)$ from top to bottom, respectively. 
The functions $\mathcal{V}'$ carrying these indices are to be 
replaced by $1$. In all pairing sums the sets $\mathcal{P}'(2)$ 
are of course subsets of $\mathcal{P}(2)$; the latter contains the pairings 
$(1,2,3,4)$, $(1,3,2,4)$ and $(1,4,2,3)$. We now focus on the 
first square bracket; the remaining ones follow similarly. In 
the first sum, drawing pairs from $\ivec \cup \jvec = 
(i_1,i_2,j_1,j_2)$, the only pairing containing $(i_1,j_1)$ is 
$(1,3,2,4)$ and has the sign $-1$. So this sum reduces to 
$-\mathcal{V}_{i_1,j_1}' \mathcal{V}_{i_2,j_2}' = 
-\mathcal{V}_{i_2,j_2}$. The second sum, on the other hand, draws 
pairs from $\ivec \cup \jvec^{1,\mathrm{s}}=(i_1,i_2,j_1-1,j_1)$. 
Hence the only pairing that contains $(i_1,j_1)$ is $(1,4,2,3)$ 
having the sign $+1$. This sum therefore collapses to 
$+\mathcal{V}_{i_1,j_1}' \mathcal{V}_{i_2,j_1-1}' = 
+\mathcal{V}_{i_2,j_1-1}$. Applying this reasoning to all summations 
yields
\begin{eqnarray}
  \Phi_{\ivec,\jvec}^{(2,2)} \!\! & = & \me^{-\dt} \, 
  I_{i_1-j_1}(\gamma \dt) \left[ - 
  {\textstyle \left(1-\frac{\gamma^2}{2} \right) }
  \mathcal{V}_{i_2,j_2}^{\, \jvec}(t,t') +
  {\textstyle \frac{\gamma^2}{2} }
  \mathcal{V}_{i_2,j_1-1}^{\, \jvec^{1,\mathrm{s}}}(t,t') 
  \right] \nonumber \\ 
  & + & \me^{-\dt} \, 
  I_{i_2-j_1}(\gamma \dt) \left[ + 
  {\textstyle \left(1-\frac{\gamma^2}{2} \right) }
  \mathcal{V}_{i_1,j_2}^{\, \jvec}(t,t') -
  {\textstyle \frac{\gamma^2}{2} }
  \mathcal{V}_{i_1,j_1-1}^{\, \jvec^{1,\mathrm{s}}}(t,t') 
  \right] \nonumber \\ 
  & - & \me^{-\dt} \, 
  I_{i_1-j_2}(\gamma \dt) \left[ + 
  {\textstyle \left(1-\frac{\gamma^2}{2} \right) }
  \mathcal{V}_{i_2,j_1}^{\, \jvec}(t,t') -
  {\textstyle \frac{\gamma^2}{2} }
  \mathcal{V}_{i_2,j_2+1}^{\, \jvec^{2,\mathrm{s}}}(t,t') 
  \right] \nonumber \\ 
  & - & \me^{-\dt} \, 
  I_{i_2-j_2}(\gamma \dt) \left[ - 
  {\textstyle \left(1-\frac{\gamma^2}{2} \right) }
  \mathcal{V}_{i_1,j_1}^{\, \jvec}(t,t') +
  {\textstyle \frac{\gamma^2}{2} }
  \mathcal{V}_{i_1,j_2+1}^{\, \jvec^{2,\mathrm{s}}}(t,t')
  \right]. \nonumber 
\end{eqnarray}
Note that the indices of all functions $\mathcal{V}$ comprise components 
of $\ivec$ and $\jvec$ or $\ivec$ and $\jvec^{\nu,\mathrm{s}}$. So all 
of them must be expressed via the $(a,b)=(i_\eps,j_\nu)$ case of 
(\ref{equ:Vcal}). We remind the reader that for $\nu$ in 
$(-1)^{\nu-1}$ in (\ref{equ:Vcal}) one has to substitute the 
{\em actual component number} of the index $j_\nu=(\jvec)_\nu$ in
$\jvec$. This  
means, for instance, $j_1-1=(\jvec^{1,\mathrm{s}})_1$ is component 
$\nu=1$ in $\jvec^{1,\mathrm{s}}$ while $j_1=(\jvec^{1,\mathrm{s}})_2$ 
is component $\nu=2$ in $\jvec^{1,\mathrm{s}}$. Accurately applying 
(\ref{equ:Vcal}) to the above expression yields the result 
(\ref{equ:phires22}) given in Sec.~\ref{sec:responsequench}.

\end{appendix}

\bibliography{1D_Ising_Refs}

\begin{thebibliography}{26}
\expandafter\ifx\csname natexlab\endcsname\relax\def\natexlab#1{#1}\fi
\expandafter\ifx\csname bibnamefont\endcsname\relax
  \def\bibnamefont#1{#1}\fi
\expandafter\ifx\csname bibfnamefont\endcsname\relax
  \def\bibfnamefont#1{#1}\fi
\expandafter\ifx\csname citenamefont\endcsname\relax
  \def\citenamefont#1{#1}\fi
\expandafter\ifx\csname url\endcsname\relax
  \def\url#1{\texttt{#1}}\fi
\expandafter\ifx\csname urlprefix\endcsname\relax\def\urlprefix{URL }\fi
\providecommand{\bibinfo}[2]{#2}
\providecommand{\eprint}[2][]{\url{#2}}

\bibitem[{\citenamefont{Glauber}(1963)}]{Glauber63}
\bibinfo{author}{\bibfnamefont{R.~J.} \bibnamefont{Glauber}},
  \bibinfo{journal}{J.\ Math.\ Phys.} \textbf{\bibinfo{volume}{4}},
  \bibinfo{pages}{294} (\bibinfo{year}{1963}).

\bibitem[{\citenamefont{Bedeaux et~al.}(1970)\citenamefont{Bedeaux, Shuler, and
  Oppenheim}}]{BedShuOpp70}
\bibinfo{author}{\bibfnamefont{D.}~\bibnamefont{Bedeaux}},
  \bibinfo{author}{\bibfnamefont{K.~E.} \bibnamefont{Shuler}},
  \bibnamefont{and}
  \bibinfo{author}{\bibfnamefont{I.}~\bibnamefont{Oppenheim}},
  \bibinfo{journal}{Journal of Statistical Physics}
  \textbf{\bibinfo{volume}{2}}, \bibinfo{pages}{1} (\bibinfo{year}{1970}).

\bibitem[{\citenamefont{Brey and Prados}(1993)}]{BrePra93b}
\bibinfo{author}{\bibfnamefont{J.~J.} \bibnamefont{Brey}} \bibnamefont{and}
  \bibinfo{author}{\bibfnamefont{A.}~\bibnamefont{Prados}},
  \bibinfo{journal}{Physica A} \textbf{\bibinfo{volume}{197}},
  \bibinfo{pages}{569} (\bibinfo{year}{1993}).

\bibitem[{\citenamefont{Brey and Prados}(1996)}]{BrePra96}
\bibinfo{author}{\bibfnamefont{J.~J.} \bibnamefont{Brey}} \bibnamefont{and}
  \bibinfo{author}{\bibfnamefont{A.}~\bibnamefont{Prados}},
  \bibinfo{journal}{Phys.\ Rev.\ E} \textbf{\bibinfo{volume}{53}},
  \bibinfo{pages}{458} (\bibinfo{year}{1996}).

\bibitem[{\citenamefont{Prados et~al.}(1997)\citenamefont{Prados, Brey, and
  {S{\'a}nchez-Rey}}}]{PraBreSan97}
\bibinfo{author}{\bibfnamefont{A.}~\bibnamefont{Prados}},
  \bibinfo{author}{\bibfnamefont{J.~J.} \bibnamefont{Brey}}, \bibnamefont{and}
  \bibinfo{author}{\bibfnamefont{B.}~\bibnamefont{{S{\'a}nchez-Rey}}},
  \bibinfo{journal}{Europhys.\ Lett.} \textbf{\bibinfo{volume}{40}},
  \bibinfo{pages}{13} (\bibinfo{year}{1997}).

\bibitem[{\citenamefont{Spouge}(1988)}]{Spouge88}
\bibinfo{author}{\bibfnamefont{J.~L.} \bibnamefont{Spouge}},
  \bibinfo{journal}{Phys.\ Rev.\ Lett.} \textbf{\bibinfo{volume}{60}},
  \bibinfo{pages}{871} (\bibinfo{year}{1988}).

\bibitem[{\citenamefont{Amar and Family}(1990)}]{AmaFam90}
\bibinfo{author}{\bibfnamefont{J.~G.} \bibnamefont{Amar}} \bibnamefont{and}
  \bibinfo{author}{\bibfnamefont{F.}~\bibnamefont{Family}},
  \bibinfo{journal}{Phys.\ Rev.\ A} \textbf{\bibinfo{volume}{41}},
  \bibinfo{pages}{3258} (\bibinfo{year}{1990}).

\bibitem[{\citenamefont{Santos}(1997)}]{Santos97}
\bibinfo{author}{\bibfnamefont{J.~E.} \bibnamefont{Santos}},
  \bibinfo{journal}{J.\ Phys.\ A-Math.\ Gen.} \textbf{\bibinfo{volume}{30}},
  \bibinfo{pages}{3249} (\bibinfo{year}{1997}).

\bibitem[{\citenamefont{Brey and Prados}(2002)}]{BrePra02b}
\bibinfo{author}{\bibfnamefont{J.~J.} \bibnamefont{Brey}} \bibnamefont{and}
  \bibinfo{author}{\bibfnamefont{A.}~\bibnamefont{Prados}},
  \bibinfo{journal}{Europhys.\ Lett} \textbf{\bibinfo{volume}{57}},
  \bibinfo{pages}{171} (\bibinfo{year}{2002}).

\bibitem[{\citenamefont{Godr{\`e}che and Luck}(2000)}]{GodLuc00}
\bibinfo{author}{\bibfnamefont{C.}~\bibnamefont{Godr{\`e}che}}
  \bibnamefont{and} \bibinfo{author}{\bibfnamefont{J.~M.} \bibnamefont{Luck}},
  \bibinfo{journal}{J.\ Phys.\ A} \textbf{\bibinfo{volume}{33}},
  \bibinfo{pages}{1151} (\bibinfo{year}{2000}).

\bibitem[{\citenamefont{Lippiello and Zannetti}(2000)}]{LipZan00}
\bibinfo{author}{\bibfnamefont{E.}~\bibnamefont{Lippiello}} \bibnamefont{and}
  \bibinfo{author}{\bibfnamefont{M.}~\bibnamefont{Zannetti}},
  \bibinfo{journal}{Phys.\ Rev.\ E} \textbf{\bibinfo{volume}{61}},
  \bibinfo{pages}{3369} (\bibinfo{year}{2000}).

\bibitem[{\citenamefont{Mayer et~al.}(2003)\citenamefont{Mayer, Berthier,
  Garrahan, and Sollich}}]{FDT}
\bibinfo{author}{\bibfnamefont{P.}~\bibnamefont{Mayer}},
  \bibinfo{author}{\bibfnamefont{L.}~\bibnamefont{Berthier}},
  \bibinfo{author}{\bibfnamefont{J.~P.} \bibnamefont{Garrahan}},
  \bibnamefont{and} \bibinfo{author}{\bibfnamefont{P.}~\bibnamefont{Sollich}},
  \bibinfo{journal}{cond-mat/0301493}  (\bibinfo{year}{2003}).

\bibitem[{\citenamefont{Felderhof}(1971)}]{Felderhof71}
\bibinfo{author}{\bibfnamefont{B.~U.} \bibnamefont{Felderhof}},
  \bibinfo{journal}{Rep.\ Math.\ Phys.} \textbf{\bibinfo{volume}{1}},
  \bibinfo{pages}{215} (\bibinfo{year}{1971}).

\bibitem[{\citenamefont{Aliev}(1998)}]{Aliev98}
\bibinfo{author}{\bibfnamefont{M.~A.} \bibnamefont{Aliev}},
  \bibinfo{journal}{Phys.\ Lett.\ A} \textbf{\bibinfo{volume}{241}},
  \bibinfo{pages}{19} (\bibinfo{year}{1998}).

\bibitem[{\citenamefont{Derrida and Zeitak}(1996)}]{DerZei96}
\bibinfo{author}{\bibfnamefont{B.}~\bibnamefont{Derrida}} \bibnamefont{and}
  \bibinfo{author}{\bibfnamefont{R.}~\bibnamefont{Zeitak}},
  \bibinfo{journal}{Phys.\ Rev.\ E} \textbf{\bibinfo{volume}{54}},
  \bibinfo{pages}{2513} (\bibinfo{year}{1996}).

\bibitem[{\citenamefont{Mayer}(unpublished)}]{Thesis}
\bibinfo{author}{\bibfnamefont{P.}~\bibnamefont{Mayer}},
  \emph{\bibinfo{title}{{\rm Ph.D. thesis, King's College London}}}
  (\bibinfo{year}{unpublished}).

\bibitem[{\citenamefont{Fontes et~al.}(2001)\citenamefont{Fontes, Isopi,
  Newman, and Stein}}]{FonIsoNewStein01}
\bibinfo{author}{\bibfnamefont{L.~R.} \bibnamefont{Fontes}},
  \bibinfo{author}{\bibfnamefont{M.}~\bibnamefont{Isopi}},
  \bibinfo{author}{\bibfnamefont{C.~M.} \bibnamefont{Newman}},
  \bibnamefont{and} \bibinfo{author}{\bibfnamefont{D.~L.} \bibnamefont{Stein}},
  \bibinfo{journal}{Phys.\ Rev.\ Lett.} \textbf{\bibinfo{volume}{87}},
  \bibinfo{pages}{110201} (\bibinfo{year}{2001}).

\bibitem[{\citenamefont{Cugliandolo and Kurchan}(1993)}]{CugKur93}
\bibinfo{author}{\bibfnamefont{L.~F.} \bibnamefont{Cugliandolo}}
  \bibnamefont{and} \bibinfo{author}{\bibfnamefont{J.}~\bibnamefont{Kurchan}},
  \bibinfo{journal}{Phys.\ Rev.\ Lett.} \textbf{\bibinfo{volume}{71}},
  \bibinfo{pages}{173} (\bibinfo{year}{1993}).

\bibitem[{\citenamefont{Cugliandolo et~al.}(1997)\citenamefont{Cugliandolo,
  Kurchan, and Peliti}}]{CugKurPel97}
\bibinfo{author}{\bibfnamefont{L.~F.} \bibnamefont{Cugliandolo}},
  \bibinfo{author}{\bibfnamefont{J.}~\bibnamefont{Kurchan}}, \bibnamefont{and}
  \bibinfo{author}{\bibfnamefont{L.}~\bibnamefont{Peliti}},
  \bibinfo{journal}{Phys.\ Rev.\ E} \textbf{\bibinfo{volume}{55}},
  \bibinfo{pages}{3898} (\bibinfo{year}{1997}).

\bibitem[{\citenamefont{Crisanti and Ritort}(2003)}]{CriRit03}
\bibinfo{author}{\bibfnamefont{A.}~\bibnamefont{Crisanti}} \bibnamefont{and}
  \bibinfo{author}{\bibfnamefont{F.}~\bibnamefont{Ritort}},
  \bibinfo{journal}{Journal of Physics A: Mathematical and General}
  \textbf{\bibinfo{volume}{36}}, \bibinfo{pages}{181} (\bibinfo{year}{2003}).

\bibitem[{\citenamefont{Sollich et~al.}(2002)\citenamefont{Sollich, Fielding,
  and Mayer}}]{SolFieMay02}
\bibinfo{author}{\bibfnamefont{P.}~\bibnamefont{Sollich}},
  \bibinfo{author}{\bibfnamefont{S.}~\bibnamefont{Fielding}}, \bibnamefont{and}
  \bibinfo{author}{\bibfnamefont{P.}~\bibnamefont{Mayer}},
  \bibinfo{journal}{J.\ Phys.\ Cond.\ Matt.} \textbf{\bibinfo{volume}{14}},
  \bibinfo{pages}{1683} (\bibinfo{year}{2002}).

\bibitem[{\citenamefont{Mayer et~al.}(unpublished)\citenamefont{Mayer,
  Berthier, Garrahan, and Sollich}}]{dynamicalhetero}
\bibinfo{author}{\bibfnamefont{P.}~\bibnamefont{Mayer}},
  \bibinfo{author}{\bibfnamefont{L.}~\bibnamefont{Berthier}},
  \bibinfo{author}{\bibfnamefont{J.~P.} \bibnamefont{Garrahan}},
  \bibnamefont{and} \bibinfo{author}{\bibfnamefont{P.}~\bibnamefont{Sollich}}
  (\bibinfo{year}{unpublished}).

\bibitem[{\citenamefont{Krebs et~al.}(1995)\citenamefont{Krebs,
  Pfannm{\"{u}}ller, Wehefritz, and Hinrichsen}}]{KrePfaWehHin95}
\bibinfo{author}{\bibfnamefont{K.}~\bibnamefont{Krebs}},
  \bibinfo{author}{\bibfnamefont{M.~P.} \bibnamefont{Pfannm{\"{u}}ller}},
  \bibinfo{author}{\bibfnamefont{B.}~\bibnamefont{Wehefritz}},
  \bibnamefont{and}
  \bibinfo{author}{\bibfnamefont{H.}~\bibnamefont{Hinrichsen}},
  \bibinfo{journal}{J.\ Stat.\ Phys.} \textbf{\bibinfo{volume}{78}},
  \bibinfo{pages}{1429} (\bibinfo{year}{1995}).

\bibitem[{\citenamefont{Fredrickson and Andersen}(1984)}]{FreAnd84}
\bibinfo{author}{\bibfnamefont{G.~H.} \bibnamefont{Fredrickson}}
  \bibnamefont{and} \bibinfo{author}{\bibfnamefont{H.~C.}
  \bibnamefont{Andersen}}, \bibinfo{journal}{Phys.\ Rev.\ Lett.}
  \textbf{\bibinfo{volume}{53}}, \bibinfo{pages}{1244} (\bibinfo{year}{1984}).

\bibitem[{\citenamefont{Buhot and Garrahan}(2002)}]{BuhGar02b}
\bibinfo{author}{\bibfnamefont{A.}~\bibnamefont{Buhot}} \bibnamefont{and}
  \bibinfo{author}{\bibfnamefont{J.~P.} \bibnamefont{Garrahan}},
  \bibinfo{journal}{Phys.\ Rev.\ Lett.} \textbf{\bibinfo{volume}{88}},
  \bibinfo{pages}{225702} (\bibinfo{year}{2002}).

\bibitem[{\citenamefont{Gradshteyn and Ryzhik}(2000)}]{Mathbook}
\bibinfo{author}{\bibfnamefont{L.~S.} \bibnamefont{Gradshteyn}}
  \bibnamefont{and} \bibinfo{author}{\bibfnamefont{I.~M.}
  \bibnamefont{Ryzhik}}, \emph{\bibinfo{title}{Table of Integrals, Series, and
  Products}} (\bibinfo{publisher}{Academic Press}, \bibinfo{address}{New York},
  \bibinfo{year}{2000}).

\end{thebibliography}

\end{document}